%Paper: astro-ph/9307022
%From: JAP@STARLINK.ROE.AC.UK
%Date: Wed, 14 Jul 93 15:10 BST

% MND.TEX (Computer Modern version)
% Copyright (c) 1992 Cambridge University Press
% Last Modification : 28.11.1992, M. Reed

\catcode `\@=11 % @ signs are letters

\def\@version{1.3}
\def\@verdate{28.11.1992}

%%% \input cmndfon.tex
%%% \input mnddef.tex
%%% \input twocol.tex   % This \inputs setup.tex

% CMNDFON.TEX v0.7a
% Computer Modern font file for MNS/D designs (not CUP Lasercomp compatible)
% Last Modification : 1.8.92, M. Reed
%
% Font family sizes available:
%   8pt, 9pt, 10pt, 11pt, 14pt and 17pt.
%
% Faces available:
%   \rm, math italic, symbol, \it, \bf, \sl, \tt, \sc, \sf, \cal and \em.
%
%   \em gives EMPHASIS, so :
%       \rm text text text {\em hello} fdsjhg sjgfj
%   gives \rm .... {\it ..} ....
%
%   and
%       \it text text text {\em hello} fdsjhg sjgfj
%   gives \it .... {\rm ..} ....
%
%   math italic bold - use via \mathchardef (see fonttest.tex).
%   bold symbol - use via \mathchardef (see fonttest.tex).
%
% Odd sizes:
%   none

\font\fiverm=cmr5
\font\fivei=cmmi5	\skewchar\fivei='177
\font\fivesy=cmsy5	\skewchar\fivesy='60
\font\fivebf=cmbx5

\font\sevenrm=cmr7
\font\seveni=cmmi7	\skewchar\seveni='177
\font\sevensy=cmsy7	\skewchar\sevensy='60
\font\sevenbf=cmbx7

\font\eightrm=cmr8
\font\eightbf=cmbx8
\font\eightit=cmti8
\font\eighti=cmmi8			\skewchar\eighti='177
\font\eightmib=cmmib10 at 8pt	\skewchar\eightmib='177
\font\eightsy=cmsy8			\skewchar\eightsy='60
\font\eightsyb=cmbsy10 at 8pt	\skewchar\eightsyb='60
\font\eightsl=cmsl8
\font\eighttt=cmtt8			\hyphenchar\eighttt=-1
\font\eightcsc=cmcsc10 at 8pt
\font\eightsf=cmss8

\font\ninerm=cmr9
\font\ninebf=cmbx9
\font\nineit=cmti9
\font\ninei=cmmi9			\skewchar\ninei='177
\font\ninemib=cmmib10 at 9pt	\skewchar\ninemib='177
\font\ninesy=cmsy9			\skewchar\ninesy='60
\font\ninesyb=cmbsy10 at 9pt	\skewchar\ninesyb='60
\font\ninesl=cmsl9
\font\ninett=cmtt9			\hyphenchar\ninett=-1
\font\ninecsc=cmcsc10 at 9pt
\font\ninesf=cmss9

\font\tenrm=cmr10
\font\tenbf=cmbx10
\font\tenit=cmti10
\font\teni=cmmi10		\skewchar\teni='177
\font\tenmib=cmmib10	\skewchar\tenmib='177
\font\tensy=cmsy10		\skewchar\tensy='60
\font\tensyb=cmbsy10	\skewchar\tensyb='60
\font\tenex=cmex10
\font\tensl=cmsl10
\font\tentt=cmtt10		\hyphenchar\tentt=-1
\font\tencsc=cmcsc10
\font\tensf=cmss10

\font\elevenrm=cmr10 scaled \magstephalf
\font\elevenbf=cmbx10 scaled \magstephalf
\font\elevenit=cmti10 scaled \magstephalf
\font\eleveni=cmmi10 scaled \magstephalf	\skewchar\eleveni='177
\font\elevenmib=cmmib10 scaled \magstephalf	\skewchar\elevenmib='177
\font\elevensy=cmsy10 scaled \magstephalf	\skewchar\elevensy='60
\font\elevensyb=cmbsy10 scaled \magstephalf	\skewchar\elevensyb='60
\font\elevensl=cmsl10 scaled \magstephalf
\font\eleventt=cmtt10 scaled \magstephalf	\hyphenchar\eleventt=-1
\font\elevencsc=cmcsc10 scaled \magstephalf
\font\elevensf=cmss10 scaled \magstephalf

\font\fourteenrm=cmr10 scaled \magstep2
\font\fourteenbf=cmbx10 scaled \magstep2
\font\fourteenit=cmti10 scaled \magstep2
\font\fourteeni=cmmi10 scaled \magstep2		\skewchar\fourteeni='177
\font\fourteenmib=cmmib10 scaled \magstep2	\skewchar\fourteenmib='177
\font\fourteensy=cmsy10 scaled \magstep2	\skewchar\fourteensy='60
\font\fourteensyb=cmbsy10 scaled \magstep2	\skewchar\fourteensyb='60
\font\fourteensl=cmsl10 scaled \magstep2
\font\fourteentt=cmtt10 scaled \magstep2	\hyphenchar\fourteentt=-1
\font\fourteencsc=cmcsc10 scaled \magstep2
\font\fourteensf=cmss10 scaled \magstep2

\font\seventeenrm=cmr10 scaled \magstep3
\font\seventeenbf=cmbx10 scaled \magstep3
\font\seventeenit=cmti10 scaled \magstep3
\font\seventeeni=cmmi10 scaled \magstep3	\skewchar\seventeeni='177
\font\seventeenmib=cmmib10 scaled \magstep3	\skewchar\seventeenmib='177
\font\seventeensy=cmsy10 scaled \magstep3	\skewchar\seventeensy='60
\font\seventeensyb=cmbsy10 scaled \magstep3	\skewchar\seventeensyb='60
\font\seventeensl=cmsl10 scaled \magstep3
\font\seventeentt=cmtt10 scaled \magstep3	\hyphenchar\seventeentt=-1
\font\seventeencsc=cmcsc10 scaled \magstep3
\font\seventeensf=cmss10 scaled \magstep3

\def\@typeface{Computer Modern} % define the typeface in use

\def\hexnumber@#1{\ifnum#1<10 \number#1\else
 \ifnum#1=10 A\else\ifnum#1=11 B\else\ifnum#1=12 C\else
 \ifnum#1=13 D\else\ifnum#1=14 E\else\ifnum#1=15 F\fi\fi\fi\fi\fi\fi\fi}

\def\mib{\hexnumber@\mibfam}
\def\syb{\hexnumber@\sybfam}

\def\makestrut{%
  \setbox\strutbox=\hbox{%
    \vrule height.7\baselineskip depth.3\baselineskip width 0pt}%
}

\def\bls#1{%
  \normalbaselineskip=#1%
  \normalbaselines%
  \makestrut%
}

% families \itfam, \slfam, \bffam, \ttfam defined in PLAIN.
%
% \itfam is \fam4
% \slfam is \fam5
% \bffam is \fam6
% \ttfam is \fam7

\newfam\mibfam % \fam8
\newfam\sybfam % \fam9
\newfam\scfam  % \fam10
\newfam\sffam  % \fam11

\def\em{\ifdim\fontdimen1\font>0 \rm\else\it\fi}

\textfont3=\tenex
\scriptfont3=\tenex
\scriptscriptfont3=\tenex

\def\eightpoint{% 8^7^5
  \def\rm{\fam0\eightrm}%
  \textfont0=\eightrm \scriptfont0=\sevenrm \scriptscriptfont0=\fiverm%
  \textfont1=\eighti  \scriptfont1=\seveni  \scriptscriptfont1=\fivei%
  \textfont2=\eightsy \scriptfont2=\sevensy \scriptscriptfont2=\fivesy%
  \textfont\itfam=\eightit\def\it{\fam\itfam\eightit}%
  \textfont\bffam=\eightbf%
    \scriptfont\bffam=\sevenbf%
      \scriptscriptfont\bffam=\fivebf%
  \def\bf{\fam\bffam\eightbf}%
  \textfont\slfam=\eightsl\def\sl{\fam\slfam\eightsl}%
  \textfont\ttfam=\eighttt\def\tt{\fam\ttfam\eighttt}%
  \textfont\scfam=\eightcsc\def\sc{\fam\scfam\eightcsc}%
  \textfont\sffam=\eightsf\def\sf{\fam\sffam\eightsf}%
  \textfont\mibfam=\eightmib%
  \textfont\sybfam=\eightsyb%
  \bls{10pt}%
}

\def\ninepoint{% 9^7^5
  \def\rm{\fam0\ninerm}%
  \textfont0=\ninerm \scriptfont0=\sevenrm \scriptscriptfont0=\fiverm%
  \textfont1=\ninei  \scriptfont1=\seveni  \scriptscriptfont1=\fivei%
  \textfont2=\ninesy \scriptfont2=\sevensy \scriptscriptfont2=\fivesy%
  \textfont\itfam=\nineit\def\it{\fam\itfam\nineit}%
  \textfont\bffam=\ninebf%
    \scriptfont\bffam=\sevenbf%
      \scriptscriptfont\bffam=\fivebf%
  \def\bf{\fam\bffam\ninebf}%
  \textfont\slfam=\ninesl\def\sl{\fam\slfam\ninesl}%
  \textfont\ttfam=\ninett\def\tt{\fam\ttfam\ninett}%
  \textfont\scfam=\ninecsc\def\sc{\fam\scfam\ninecsc}%
  \textfont\sffam=\ninesf\def\sf{\fam\sffam\ninesf}%
  \textfont\mibfam=\ninemib%
  \textfont\sybfam=\ninesyb%
  \bls{12pt}%
}

\def\tenpoint{% 10^7^5
  \def\rm{\fam0\tenrm}%
  \textfont0=\tenrm \scriptfont0=\sevenrm \scriptscriptfont0=\fiverm%
  \textfont1=\teni  \scriptfont1=\seveni  \scriptscriptfont1=\fivei%
  \textfont2=\tensy \scriptfont2=\sevensy \scriptscriptfont2=\fivesy%
  \textfont\itfam=\tenit\def\it{\fam\itfam\tenit}%
  \textfont\bffam=\tenbf%
    \scriptfont\bffam=\sevenbf%
      \scriptscriptfont\bffam=\fivebf%
  \def\bf{\fam\bffam\tenbf}%
  \textfont\slfam=\tensl\def\sl{\fam\slfam\tensl}%
  \textfont\ttfam=\tentt\def\tt{\fam\ttfam\tentt}%
  \textfont\scfam=\tencsc\def\sc{\fam\scfam\tencsc}%
  \textfont\sffam=\tensf\def\sf{\fam\sffam\tensf}%
  \textfont\mibfam=\tenmib%
  \textfont\sybfam=\tensyb%
  \bls{12pt}%
}

\def\elevenpoint{% 11^8^5
  \def\rm{\fam0\elevenrm}%
  \textfont0=\elevenrm \scriptfont0=\eightrm \scriptscriptfont0=\fiverm%
  \textfont1=\eleveni  \scriptfont1=\eighti  \scriptscriptfont1=\fivei%
  \textfont2=\elevensy \scriptfont2=\eightsy \scriptscriptfont2=\fivesy%
  \textfont\itfam=\elevenit\def\it{\fam\itfam\elevenit}%
  \textfont\bffam=\elevenbf%
    \scriptfont\bffam=\eightbf%
      \scriptscriptfont\bffam=\fivebf%
  \def\bf{\fam\bffam\elevenbf}%
  \textfont\slfam=\elevensl\def\sl{\fam\slfam\elevensl}%
  \textfont\ttfam=\eleventt\def\tt{\fam\ttfam\eleventt}%
  \textfont\scfam=\elevencsc\def\sc{\fam\scfam\elevencsc}%
  \textfont\sffam=\elevensf\def\sf{\fam\sffam\elevensf}%
  \textfont\mibfam=\elevenmib%
  \textfont\sybfam=\elevensyb%
  \bls{13pt}%
}

\def\fourteenpoint{% 14^10^7
  \def\rm{\fam0\fourteenrm}%
  \textfont0\fourteenrm  \scriptfont0\tenrm  \scriptscriptfont0\sevenrm%
  \textfont1\fourteeni   \scriptfont1\teni   \scriptscriptfont1\seveni%
  \textfont2\fourteensy  \scriptfont2\tensy  \scriptscriptfont2\sevensy%
  \textfont\itfam=\fourteenit\def\it{\fam\itfam\fourteenit}%
  \textfont\bffam=\fourteenbf%
    \scriptfont\bffam=\tenbf%
      \scriptscriptfont\bffam=\sevenbf%
  \def\bf{\fam\bffam\fourteenbf}%
  \textfont\slfam=\fourteensl\def\sl{\fam\slfam\fourteensl}%
  \textfont\ttfam=\fourteentt\def\tt{\fam\ttfam\fourteentt}%
  \textfont\scfam=\fourteencsc\def\sc{\fam\scfam\fourteencsc}%
  \textfont\sffam=\fourteensf\def\sf{\fam\sffam\fourteensf}%
  \textfont\mibfam=\fourteenmib%
  \textfont\sybfam=\fourteensyb%
  \bls{17pt}%
}

\def\seventeenpoint{% 17^11^9
  \def\rm{\fam0\seventeenrm}%
  \textfont0\seventeenrm  \scriptfont0\elevenrm  \scriptscriptfont0\ninerm%
  \textfont1\seventeeni   \scriptfont1\eleveni   \scriptscriptfont1\ninei%
  \textfont2\seventeensy  \scriptfont2\elevensy  \scriptscriptfont2\ninesy%
  \textfont\itfam=\seventeenit\def\it{\fam\itfam\seventeenit}%
  \textfont\bffam=\seventeenbf%
    \scriptfont\bffam=\elevenbf%
      \scriptscriptfont\bffam=\ninebf%
  \def\bf{\fam\bffam\seventeenbf}%
  \textfont\slfam=\seventeensl\def\sl{\fam\slfam\seventeensl}%
  \textfont\ttfam=\seventeentt\def\tt{\fam\ttfam\seventeentt}%
  \textfont\scfam=\seventeencsc\def\sc{\fam\scfam\seventeencsc}%
  \textfont\sffam=\seventeensf\def\sf{\fam\sffam\seventeensf}%
  \textfont\mibfam=\seventeenmib%
  \textfont\sybfam=\seventeensyb%
  \bls{20pt}%
}

\lineskip=1pt      \normallineskip=\lineskip
\lineskiplimit=0pt \normallineskiplimit=\lineskiplimit

%%% \endinput

% MNSDEF.TEX v8.04
% Last modification : 23.11.1992, M. Reed

% NUMBER THE DESIGN ELEMENTS

\def\Nulle{0}  % null element
\def\Aue{1}    % article author(s)
\def\Afe{2}    % author affiliation
    % acceptance date
\def\Sue{4}    % summary
\def\Hae{5}    % heading A
\def\Hbe{6}    % heading B
\def\Hce{7}    % heading C
\def\Hde{8}    % heading D
    % keywords
\def\Txe{10}   % text
\def\Lie{11}   % list
\def\Bbe{12}   % bibliography

% TEMPORARY REGISTERS

\newdimen\DimenA
\newbox\BoxA

\newcount\LastMac \LastMac=\Nulle
\newcount\HeaderNumber \HeaderNumber=0
\newcount\DefaultHeader \DefaultHeader=\HeaderNumber
\newskip\Indent

\newskip\half      \half=5.5pt plus 1.5pt minus 2.25pt
\newskip\one       \one=11pt plus 3pt minus 5.5pt
\newskip\onehalf   \onehalf=16.5pt plus 5.5pt minus 8.25pt
\newskip\two       \two=22pt plus 5.5pt minus 11pt

\def\Half{\vskip-\lastskip\vskip\half}
\def\One{\vskip-\lastskip\vskip\one}
\def\OneHalf{\vskip-\lastskip\vskip\onehalf}
\def\Two{\vskip-\lastskip\vskip\two}

%%% The next two macros define non changing dimensions which don't
%%% enlarge when \Rereree mode is in use. Used in \table and \figure.

\def\rTenPT{10pt plus \Feathering}

\def\TenPT{10pt plus \Feathering} % leading
\def\ElevenPT{11pt plus \Feathering}

\def\Raggedright{% set lines unjustified
 \rightskip=0pt plus \hsize
% \spaceskip=.3333em\xspaceskip=.5em
}

\def\Fullout{% set lines justified
\rightskip=0pt
% \spaceskip=0pt\xspaceskip=0pt
}

\def\Hang#1#2{% set hanging indentation
 \hangindent=#1
 \hangafter=#2
}

\def\EveryMac{% called at start of most macros
 \Fullout
 \everypar{}
}

% DESIGN ELEMENT DEFINITIONS

% Article opening

\def\title#1{% article title
 \EveryMac
 \LastMac=\Nulle
 \global\HeaderNumber=0
 \global\DefaultHeader=1
 \vbox to 1pc{\vss}
 \seventeenpoint% \baselineskip=20pt
 \Raggedright
 \noindent \bf #1
}

\def\author#1{% article author(s)
 \EveryMac
 \ifnum\LastMac=\Afe \OneHalf
  \else \Two
 \fi
 \LastMac=\Aue
 \fourteenpoint% \baselineskip=17pt
 \Raggedright
 \noindent \rm #1\par
 \vskip 3pt\relax
}

\def\affiliation#1{% author(s) affiliation
 \EveryMac
 \LastMac=\Afe
 \eightpoint\bls{\TenPT}% \baselineskip=\TenPT
 \Raggedright
 \noindent \it #1\par
}

\def\abstract{%
 \EveryMac
 \Two
 \LastMac=\Sue
 \everypar{\Hang{11pc}{0}}
 \noindent\ninebf ABSTRACT\par
 \tenpoint\bls{\ElevenPT}% \baselineskip=\ElevenPT
 \Fullout
 \noindent\rm
}

% The \maketitle macro ensures that the two spanning material appears
% at the top of the first page, and that it has two lines of space
% underneath it. If you forget this in you input, no output will be produced.
% The \BeginOpening (alias \begintopmatter) macro should be called at the
% very start of the input file, so that it is in force when the document
% starts. This ensures that when \maketitle is called that the group is
% closed, and the material gets printed.

\def\maketitle{%
  \Two%
  \EndOpening%
  \MakePage%
}

% Page offset

\def\pageoffset#1#2{\hoffset=#1\relax\voffset=#2\relax}

% Headings

\def\Autonumber{% set header auto numbering
 \global\AutoNumbertrue  %  ensure this declaration is GLOBAL
}

\newif\ifAutoNumber \AutoNumberfalse
\newcount\Sec        %  heading auto number counters
\newcount\SecSec
\newcount\SecSecSec

\Sec=0

\def\:{\let\@sptoken= } \:  % this makes \@sptoken a space token
\def\:{\@xifnch} \expandafter\def\: {\futurelet\@tempc\@ifnch}

\def\@ifnextchar#1#2#3{%
  \let\@tempMACe #1%
  \def\@tempMACa{#2}%
  \def\@tempMACb{#3}%
  \futurelet \@tempMACc\@ifnch%
}

\def\@ifnch{%
\ifx \@tempMACc \@sptoken%
  \let\@tempMACd\@xifnch%
\else%
  \ifx \@tempMACc \@tempMACe%
    \let\@tempMACd\@tempMACa%
  \else%
    \let\@tempMACd\@tempMACb%
  \fi%
\fi%
\@tempMACd%
}

\def\@ifstar#1#2{\@ifnextchar *{\def\@tempMACa*{#1}\@tempMACa}{#2}}

\def\section{\@ifstar{\@ssection}{\@section}}

\def\@section#1{% heading A (\section{....})
 \EveryMac
 \Two
 \LastMac=\Hae
 \ninepoint\bls{\ElevenPT}% \baselineskip=\ElevenPT
 \bf
 \Raggedright
 \ifAutoNumber
  \advance\Sec by 1
  \noindent\number\Sec\hskip 1pc \uppercase{#1}
  \SecSec=0
 \else
  \noindent \uppercase{#1}
 \fi
 \nobreak
}

\def\@ssection#1{%  main section heading (\section*{....})
 \EveryMac
 \ifnum\LastMac=\Hae \Half
  \else \OneHalf
 \fi
 \LastMac=\Hae
 \tenpoint\bls{\ElevenPT}% \baselineskip=\ElevenPT
 \bf
 \Raggedright
 \noindent\uppercase{#1}
}

\def\subsection#1{% heading B
 \EveryMac
 \ifnum\LastMac=\Hae \Half
  \else \OneHalf
 \fi
 \LastMac=\Hbe
 \tenpoint\bls{\ElevenPT}% \baselineskip=\ElevenPT
 \bf
 \Raggedright
 \ifAutoNumber
  \advance\SecSec by 1
  \noindent\number\Sec.\number\SecSec
  \hskip 1pc #1
  \SecSecSec=0
 \else
  \noindent #1
 \fi
 \nobreak
}

\def\subsubsection#1{% heading C
 \EveryMac
 \ifnum\LastMac=\Hbe \Half
  \else \OneHalf
 \fi
 \LastMac=\Hce
 \ninepoint\bls{\ElevenPT}% \baselineskip=\ElevenPT
 \it
 \Raggedright
 \ifAutoNumber
  \advance\SecSecSec by 1
  \noindent\number\Sec.\number\SecSec.\number\SecSecSec
  \hskip 1pc #1
 \else
  \noindent #1
 \fi
 \nobreak
}

\def\paragraph#1{% heading D
 \EveryMac
 \One
 \LastMac=\Hde
 \ninepoint\bls{\ElevenPT}% \baselineskip=\ElevenPT
 \noindent \it #1
 \rm
}

% Text

\def\tx{% text
 \EveryMac
 \ifnum\LastMac=\Lie \Half\fi
 \ifnum\LastMac=\Hae \nobreak\Half\fi
 \ifnum\LastMac=\Hbe \nobreak\Half\fi
 \ifnum\LastMac=\Hce \nobreak\Half\fi
 \ifnum\LastMac=\Lie \else \noindent\fi
 \LastMac=\Txe
 \ninepoint\bls{\ElevenPT}% \baselineskip=\ElevenPT
 \rm
}

% Lists

\def\item{% list item
 \par
 \EveryMac
 \ifnum\LastMac=\Lie
  \else \Half
 \fi
 \LastMac=\Lie
 \ninepoint\bls{\ElevenPT}% \baselineskip=\ElevenPT
 \rm
}

% References

\def\bibitem{% bibliography
 \par
 \EveryMac
 \ifnum\LastMac=\Bbe
  \else \Half
 \fi
 \LastMac=\Bbe
 \Hang{1.5em}{1}
 \eightpoint\bls{\TenPT}% \baselineskip=\TenPT
 \Raggedright
 \noindent \rm
}

% Page heads

\newtoks\CatchLine

\def\@journal{Mon.\ Not.\ R.\ Astron.\ Soc.\ }  % The journal title string
\def\@pubyear{1993}        % Assign a default publication year
\def\@pagerange{000--000}  % Assign a default page-range
\def\@volume{000}          % Assign a default volume number
\def\@microfiche{}         %

\def\pubyear#1{\gdef\@pubyear{#1}\@makecatchline}
\def\pagerange#1{\gdef\@pagerange{#1}\@makecatchline}
\def\volume#1{\gdef\@volume{#1}\@makecatchline}
\def\microfiche#1{\gdef\@microfiche{and Microfiche\ #1}\@makecatchline}

\def\@makecatchline{%
  \global\CatchLine{%
    {\rm \@journal {\bf \@volume},\ \@pagerange\ (\@pubyear)\ \@microfiche}}%
}

\@makecatchline % Assign a catchline, using the above defaults

\newtoks\LeftHeader
\def\shortauthor#1{% left page head
 \global\LeftHeader{#1}
}

\newtoks\RightHeader
\def\shorttitle#1{% right page head
 \global\RightHeader{#1}
}

\def\PageHead{% recto/verso running heads
 \EveryMac
 \ifnum\HeaderNumber=1 \Pagehead
  \else \Catchline
 \fi
}

\def\Catchline{%
 \vbox to 0pt{\vskip-22.5pt
  \hbox to \PageWidth{\vbox to8.5pt{}\noindent
  \eightpoint\the\CatchLine\hfill}\vss}
 \nointerlineskip
}

\def\Pagehead{%
 \ifodd\pageno
   \vbox to 0pt{\vskip-22.5pt
   \hbox to \PageWidth{\vbox to8.5pt{}\elevenpoint\it\noindent
    \hfill\the\RightHeader\hskip1.5em\rm\folio}\vss}
 \else
   \vbox to 0pt{\vskip-22.5pt
   \hbox to \PageWidth{\vbox to8.5pt{}\elevenpoint\rm\noindent
   \folio\hskip1.5em\it\the\LeftHeader\hfill}\vss}
 \fi
 \nointerlineskip
}

\def\PageFoot{} % No page footer as default

\def\authorcomment#1{%
  \gdef\PageFoot{%
    \nointerlineskip%
    \vbox to 22pt{\vfil%
      \hbox to \PageWidth{\elevenpoint\rm\noindent \hfil #1 \hfil}}%
  }%
}

\everydisplay{\displaysetup}

\newif\ifeqno
\newif\ifleqno

\def\displaysetup#1$${%
 \displaytest#1\eqno\eqno\displaytest
}

\def\displaytest#1\eqno#2\eqno#3\displaytest{%
 \if!#3!\ldisplaytest#1\leqno\leqno\ldisplaytest
 \else\eqnotrue\leqnofalse\def\eqn{#2}\def\eq{#1}\fi
 \generaldisplay$$}

\def\ldisplaytest#1\leqno#2\leqno#3\ldisplaytest{%
 \def\eq{#1}%
 \if!#3!\eqnofalse\else\eqnotrue\leqnotrue
  \def\eqn{#2}\fi}

\def\generaldisplay{%
\ifeqno \ifleqno
   \hbox to \hsize{\noindent
     $\displaystyle\eq$\hfil$\displaystyle\eqn$}
  \else
    \hbox to \hsize{\noindent
     $\displaystyle\eq$\hfil$\displaystyle\eqn$}
  \fi
 \else
 \hbox to \hsize{\vbox{\noindent
  $\displaystyle\eq$\hfil}}
 \fi
}

\def\@notice{%
  \par\Two%
  \bls{12pt}%
  \noindent\tenrm This paper has been produced using the Blackwell
                  Scientific Publications \TeX\ macros.%
}

%  redefine \bye to output our identification notice :
\outer\def\bye{\@notice\par\vfill\supereject\end}

%  define a sign on :
\everyjob{%
  \Warn{Monthly notices of the RAS journal style (\@typeface)\space
        v\@version,\space \@verdate.}\Warn{}%
}

%%% \endinput

% TWOCOL.TEX v7.2
% Last modified : 22.11.1992, M. Reed

%--------------------------------------------------------%
%                     INITIALISATION                     %
%--------------------------------------------------------%

\newif\if@debug \@debugfalse  %  when false, only warnings displayed

\def\Print#1{\if@debug\immediate\write16{#1}\else \fi}
\def\Warn#1{\immediate\write16{#1}}
\def\wlog#1{}

\newcount\Iteration % temporary loop counter

\newif\ifFigureBoxes  % Figure spaces will be outlined if true
\FigureBoxestrue

\def\Single{0} \def\Double{1}                 % ItemSPAN's
\def\Figure{0} \def\Table{1}                  % ItemTYPE's

\def\InStack{0}  % ItemSTATUS
\def\InZoneA{1}
\def\InZoneB{2}
\def\InZoneC{3}

\newcount\TEMPCOUNT % temporary count register
\newdimen\TEMPDIMEN % temporary dimen register
\newbox\TEMPBOX     % temporary box register
\newbox\VOIDBOX     % a box which is permenately void

\newcount\LengthOfStack % number of items currently in stack
\newcount\MaxItems      % maximum number of items allowed in stack
\newcount\StackPointer
\newcount\Point         % used in calculation for generating the
                        % physical address of an item in the stack.
\newcount\NextFigure    % number of next figure to be output
\newcount\NextTable     % number of next table to be output
\newcount\NextItem      % Next item to consider by order in stack

\newcount\StatusStack   % set to point to top of STATUS stack
\newcount\NumStack      % set to point to top of NUMBER stack
\newcount\TypeStack     % set to point to top of TYPE stack
\newcount\SpanStack     % set to point to top of SPAN stack
\newcount\BoxStack      % set to point to top of BOX stack

\newcount\ItemSTATUS    % status of present item
\newcount\ItemNUMBER    % number of present item
\newcount\ItemTYPE      % type of present item
\newcount\ItemSPAN      % span of present item
\newbox\ItemBOX         % box of present item
\newdimen\ItemSIZE      % size of present item
                        % (calculated by GetItemBOX)

\newdimen\PageHeight    % vertical measure of full page
\newdimen\TextLeading   % distance between baselines of body text
\newdimen\Feathering    % amount of interline stretch
\newcount\LinesPerPage  % height of page in text lines
\newdimen\ColumnWidth   % width of 1 column of text
\newdimen\ColumnGap     % size of gap between columns
\newdimen\PageWidth     % = \ColumnWidth * 2 + \ColumnGap
\newdimen\BodgeHeight   % Bodge to align figures and tables with text
\newcount\Leading       % Set to same as \TextLeading above

\newdimen\ZoneBSize  % size of items in ZoneB
\newdimen\TextSize   % size of text in ZoneB
\newbox\ZoneABOX     % contains Zone A material
\newbox\ZoneBBOX     % contains Zone B material
\newbox\ZoneCBOX     % contains Zone C material

\newif\ifFirstSingleItem
\newif\ifFirstZoneA
\newif\ifMakePageInComplete
\newif\ifMoreFigures \MoreFiguresfalse % set true in join stack
\newif\ifMoreTables  \MoreTablesfalse  % set true in join stack

\newif\ifFigInZoneB % used to determine in which zone an item
\newif\ifFigInZoneC % will be placed based on what is in other
\newif\ifTabInZoneB % zones already for a given page.
\newif\ifTabInZoneC

\newif\ifZoneAFullPage

\newbox\MidBOX    % = LeftBOX+gap+RightBOX
\newbox\LeftBOX
\newbox\RightBOX
\newbox\PageBOX   % complete made-up page

\newif\ifLeftCOL  % flags first pass through output routine
\LeftCOLtrue

\newdimen\ZoneBAdjust

\newcount\ItemFits
\def\Yes{1}
\def\No{2}

% Read setup file.

%%% \input SETUP.TEX

% SETUP.TEX v7.1b
% Last modified : 22.11.1992, M. Reed

\MaxItems=15
\NextFigure=0        % used for article opening
\NextTable=1

\BodgeHeight=6pt
\TextLeading=11pt    % baselineskip of body text
\Leading=11
\Feathering=0pt      % amount of interline stretch
\LinesPerPage=61     % number of text lines per full page -1
\topskip=\TextLeading
\ColumnWidth=20pc    % width of text columns
\ColumnGap=2pc       % gap between columns

\def\ItemSep{\vskip \TextLeading plus \TextLeading minus 4pt}

\FigureBoxesfalse %  No figure boxes

\parskip=0pt
\parindent=18pt
\widowpenalty=0
\clubpenalty=10000
\tolerance=1500
\hbadness=1500
\abovedisplayskip=6pt plus 2pt minus 2pt
\belowdisplayskip=6pt plus 2pt minus 2pt
\abovedisplayshortskip=6pt plus 2pt minus 2pt
\belowdisplayshortskip=6pt plus 2pt minus 2pt

\PageHeight=\TextLeading % calculate height of page
\multiply\PageHeight by \LinesPerPage
\advance\PageHeight by \topskip

\PageWidth=2\ColumnWidth
\advance\PageWidth by \ColumnGap

%--------------------------------------------------------%
%                         STACKS                         %
%--------------------------------------------------------%

% THE ITEM STACK
% The item stack contains contains figures and tables
% in the order in which they appear in the article source
% code.

% allocate stack space

\newcount\DUMMY \StatusStack=\allocationnumber
\newcount\DUMMY \newcount\DUMMY \newcount\DUMMY
\newcount\DUMMY \newcount\DUMMY \newcount\DUMMY
\newcount\DUMMY \newcount\DUMMY \newcount\DUMMY
\newcount\DUMMY \newcount\DUMMY \newcount\DUMMY
\newcount\DUMMY \newcount\DUMMY \newcount\DUMMY

\newcount\DUMMY \NumStack=\allocationnumber
\newcount\DUMMY \newcount\DUMMY \newcount\DUMMY
\newcount\DUMMY \newcount\DUMMY \newcount\DUMMY
\newcount\DUMMY \newcount\DUMMY \newcount\DUMMY
\newcount\DUMMY \newcount\DUMMY \newcount\DUMMY
\newcount\DUMMY \newcount\DUMMY \newcount\DUMMY

\newcount\DUMMY \TypeStack=\allocationnumber
\newcount\DUMMY \newcount\DUMMY \newcount\DUMMY
\newcount\DUMMY \newcount\DUMMY \newcount\DUMMY
\newcount\DUMMY \newcount\DUMMY \newcount\DUMMY
\newcount\DUMMY \newcount\DUMMY \newcount\DUMMY
\newcount\DUMMY \newcount\DUMMY \newcount\DUMMY

\newcount\DUMMY \SpanStack=\allocationnumber
\newcount\DUMMY \newcount\DUMMY \newcount\DUMMY
\newcount\DUMMY \newcount\DUMMY \newcount\DUMMY
\newcount\DUMMY \newcount\DUMMY \newcount\DUMMY
\newcount\DUMMY \newcount\DUMMY \newcount\DUMMY
\newcount\DUMMY \newcount\DUMMY \newcount\DUMMY

\newbox\DUMMY   \BoxStack=\allocationnumber
\newbox\DUMMY   \newbox\DUMMY \newbox\DUMMY
\newbox\DUMMY   \newbox\DUMMY \newbox\DUMMY
\newbox\DUMMY   \newbox\DUMMY \newbox\DUMMY
\newbox\DUMMY   \newbox\DUMMY \newbox\DUMMY
\newbox\DUMMY   \newbox\DUMMY \newbox\DUMMY

\def\wlog{\immediate\write-1}

% \GetItemSTATUS, \GetItemNUMBER, \GetItemTYPE, \GetItemSPAN,
% \GetItemBox
% are used to get details of a particular item from the item
% stack. The argument to each of these is the items position
% in the stack (usually \StackPointer)...not the items number.

\def\GetItemAll#1{%
 \GetItemSTATUS{#1}
 \GetItemNUMBER{#1}
 \GetItemTYPE{#1}
 \GetItemSPAN{#1}
 \GetItemBOX{#1}
}

% Note: \LeaveStack uses this routine. Do not destroy \Point
\def\GetItemSTATUS#1{%
 \Point=\StatusStack
 \advance\Point by #1
 \global\ItemSTATUS=\count\Point
}

% Note: \LeaveStack uses this routine. Do not destroy \Point
\def\GetItemNUMBER#1{%
 \Point=\NumStack
 \advance\Point by #1
 \global\ItemNUMBER=\count\Point
}

% Note: \LeaveStack uses this routine. Do not destroy \Point
\def\GetItemTYPE#1{%
 \Point=\TypeStack
 \advance\Point by #1
 \global\ItemTYPE=\count\Point
}

% Note: \LeaveStack uses this routine. Do not destroy \Point
\def\GetItemSPAN#1{%
 \Point\SpanStack
 \advance\Point by #1
 \global\ItemSPAN=\count\Point
}

% Note: \LeaveStack uses this routine. Do not destroy \Point
\def\GetItemBOX#1{%
 \Point=\BoxStack
 \advance\Point by #1
 \global\setbox\ItemBOX=\vbox{\copy\Point}
 \global\ItemSIZE=\ht\ItemBOX
 \global\advance\ItemSIZE by \dp\ItemBOX
 \TEMPCOUNT=\ItemSIZE
 \divide\TEMPCOUNT by \Leading
 \divide\TEMPCOUNT by 65536
 \advance\TEMPCOUNT by 1
 \ItemSIZE=\TEMPCOUNT pt
 \global\multiply\ItemSIZE by \Leading
}

% item joins stack

\def\JoinStack{%
 \ifnum\LengthOfStack=\MaxItems % stack is full of items
  \Warn{WARNING: Stack is full...some items will be lost!}
 \else
  \Point=\StatusStack
  \advance\Point by \LengthOfStack
  \global\count\Point=\ItemSTATUS
  \Point=\NumStack
  \advance\Point by \LengthOfStack
  \global\count\Point=\ItemNUMBER
  \Point=\TypeStack
  \advance\Point by \LengthOfStack
  \global\count\Point=\ItemTYPE
  \Point\SpanStack
  \advance\Point by \LengthOfStack
  \global\count\Point=\ItemSPAN
  \Point=\BoxStack
  \advance\Point by \LengthOfStack
  \global\setbox\Point=\vbox{\copy\ItemBOX}
  \global\advance\LengthOfStack by 1
  \ifnum\ItemTYPE=\Figure % used in \MakePage
   \global\MoreFigurestrue
  \else
   \global\MoreTablestrue
  \fi
 \fi
}

% item leaves stack
% #1=physical position of the item to be removed

\def\LeaveStack#1{%
 {\Iteration=#1
 \loop
 \ifnum\Iteration<\LengthOfStack
  \advance\Iteration by 1
  \GetItemSTATUS{\Iteration}
   \advance\Point by -1
   \global\count\Point=\ItemSTATUS
  \GetItemNUMBER{\Iteration}
   \advance\Point by -1
   \global\count\Point=\ItemNUMBER
  \GetItemTYPE{\Iteration}
   \advance\Point by -1
   \global\count\Point=\ItemTYPE
  \GetItemSPAN{\Iteration}
   \advance\Point by -1
   \global\count\Point=\ItemSPAN
  \GetItemBOX{\Iteration}
   \advance\Point by -1
   \global\setbox\Point=\vbox{\copy\ItemBOX}
 \repeat}
 \global\advance\LengthOfStack by -1
}

% clean stack
% This routine scans through the stack and removes anything
% that does not have STATUS=\InStack.

\newif\ifStackNotClean

\def\CleanStack{%
 \StackNotCleantrue
 {\Iteration=0
  \loop
   \ifStackNotClean
    \GetItemSTATUS{\Iteration}
    \ifnum\ItemSTATUS=\InStack
     \advance\Iteration by 1
     \else
      \LeaveStack{\Iteration}
    \fi
   \ifnum\LengthOfStack<\Iteration
    \StackNotCleanfalse
   \fi
 \repeat}
}

% Find item.
% This macro searches from the top to the bottom of the
% stack for an item of a specified type and number.
% #1=type, #2=number
% If the specified item is found, then \StackPointer is set
% to point to it, else \StackPointer=-1.
% This routine is used to find the next figure or table
% by number.

\def\FindItem#1#2{%
 \global\StackPointer=-1 % assume item isn't in stack for now
 {\Iteration=0
  \loop
  \ifnum\Iteration<\LengthOfStack
   \GetItemSTATUS{\Iteration}
   \ifnum\ItemSTATUS=\InStack
    \GetItemTYPE{\Iteration}
    \ifnum\ItemTYPE=#1
     \GetItemNUMBER{\Iteration}
     \ifnum\ItemNUMBER=#2
      \global\StackPointer=\Iteration
      \Iteration=\LengthOfStack % terminate loop
     \fi
    \fi
   \fi
  \advance\Iteration by 1
 \repeat}
}

% Find next type
% This macro searches from the top to the bottom of the stack
% looking for the first item which has STATUS=\InStack.
% If it is a figure then a figure is what will be considered
% next by \MakePage else table.

\def\FindNext{%
 \global\StackPointer=-1 % assume stack is empty for now
 {\Iteration=0
  \loop
  \ifnum\Iteration<\LengthOfStack
   \GetItemSTATUS{\Iteration}
   \ifnum\ItemSTATUS=\InStack
    \GetItemTYPE{\Iteration}
   \ifnum\ItemTYPE=\Figure
    \ifMoreFigures
      \global\NextItem=\Figure
      \global\StackPointer=\Iteration
      \Iteration=\LengthOfStack % terminate loop
    \fi
   \fi
   \ifnum\ItemTYPE=\Table
    \ifMoreTables
      \global\NextItem=\Table
      \global\StackPointer=\Iteration
      \Iteration=\LengthOfStack % terminate loop
    \fi
   \fi
  \fi
  \advance\Iteration by 1
 \repeat}
}

% Change status
% Macro to change the status of a specified item in stack.
% #1=item, #2=new status

\def\ChangeStatus#1#2{%
 \Point=\StatusStack
 \advance\Point by #1
 \global\count\Point=#2
}

%--------------------------------------------------------%
%                       MAKEPAGE                         %
%--------------------------------------------------------%

% This macro is called at the start of every new page
% including the first. It scans through the stack picking
% out items which should be placed on this page. It then
% leaves space for the items to be placed later. The routine
% terminates when either there is no room on the page to
% fit the next figure or table, or there are no more items
% in the stack.

\def\Zone{\InZoneA}

\ZoneBAdjust=0pt

\def\MakePage{% allocate space on this page for stack items
 \global\ZoneBSize=\PageHeight
 \global\TextSize=\ZoneBSize
 \global\ZoneAFullPagefalse
 \global\topskip=\TextLeading
 \MakePageInCompletetrue
 \MoreFigurestrue
 \MoreTablestrue
 \FigInZoneBfalse
 \FigInZoneCfalse
 \TabInZoneBfalse
 \TabInZoneCfalse
 \global\FirstSingleItemtrue
 \global\FirstZoneAtrue
 \global\setbox\ZoneABOX=\box\VOIDBOX
 \global\setbox\ZoneBBOX=\box\VOIDBOX
 \global\setbox\ZoneCBOX=\box\VOIDBOX
 \loop
  \ifMakePageInComplete
 \FindNext
 \ifnum\StackPointer=-1
  \NextItem=-1
  \MoreFiguresfalse
  \MoreTablesfalse
 \fi
 \ifnum\NextItem=\Figure
   \FindItem{\Figure}{\NextFigure}
   \ifnum\StackPointer=-1 \global\MoreFiguresfalse
   \else
    \GetItemSPAN{\StackPointer}
    \ifnum\ItemSPAN=\Single \def\Zone{\InZoneB}\relax
     \ifFigInZoneC \global\MoreFiguresfalse\fi
    \else
     \def\Zone{\InZoneA}
     \ifFigInZoneB \def\Zone{\InZoneC}\fi
    \fi
   \fi
   \ifMoreFigures\Print{}\FigureItems\fi
 \fi
\ifnum\NextItem=\Table
   \FindItem{\Table}{\NextTable}
   \ifnum\StackPointer=-1 \global\MoreTablesfalse
   \else
    \GetItemSPAN{\StackPointer}
    \ifnum\ItemSPAN=\Single\relax
     \ifTabInZoneC \global\MoreTablesfalse\fi
    \else
     \def\Zone{\InZoneA}
     \ifTabInZoneB \def\Zone{\InZoneC}\fi
    \fi
   \fi
   \ifMoreTables\Print{}\TableItems\fi
 \fi
   \MakePageInCompletefalse % assume page is complete
   \ifMoreFigures\MakePageInCompletetrue\fi
   \ifMoreTables\MakePageInCompletetrue\fi
 \repeat
%\Print{TextSize=\the\TextSize}
%\Print{ZoneBSize=\the\ZoneBSize}
 \ifZoneAFullPage
  \global\TextSize=0pt
  \global\ZoneBSize=0pt
  \global\vsize=0pt\relax
  \global\topskip=0pt\relax
  \vbox to 0pt{\vss}
  \eject
 \else
 \global\advance\ZoneBSize by -\ZoneBAdjust
 \global\vsize=\ZoneBSize
 \global\hsize=\ColumnWidth
 \global\ZoneBAdjust=0pt
 \ifdim\TextSize<23pt
 \Warn{}
 \Warn{* Making column fall short: TextSize=\the\TextSize *}
 \vskip-\lastskip\eject\fi
 \fi
}

\def\MakeRightCol{% allocate space for the right column of text
 \global\TextSize=\ZoneBSize
 \MakePageInCompletetrue
 \MoreFigurestrue
 \MoreTablestrue
 \global\FirstSingleItemtrue
 \global\setbox\ZoneBBOX=\box\VOIDBOX
 \def\Zone{\InZoneB}
 \loop
  \ifMakePageInComplete
 \FindNext
 \ifnum\StackPointer=-1
  \NextItem=-1
  \MoreFiguresfalse
  \MoreTablesfalse
 \fi
 \ifnum\NextItem=\Figure
   \FindItem{\Figure}{\NextFigure}
   \ifnum\StackPointer=-1 \MoreFiguresfalse
   \else
    \GetItemSPAN{\StackPointer}
    \ifnum\ItemSPAN=\Double\relax
     \MoreFiguresfalse\fi
   \fi
   \ifMoreFigures\Print{}\FigureItems\fi
 \fi
 \ifnum\NextItem=\Table
   \FindItem{\Table}{\NextTable}
   \ifnum\StackPointer=-1 \MoreTablesfalse
   \else
    \GetItemSPAN{\StackPointer}
    \ifnum\ItemSPAN=\Double\relax
     \MoreTablesfalse\fi
   \fi
   \ifMoreTables\Print{}\TableItems\fi
 \fi
   \MakePageInCompletefalse % assume page is complete
   \ifMoreFigures\MakePageInCompletetrue\fi
   \ifMoreTables\MakePageInCompletetrue\fi
 \repeat
 \ifZoneAFullPage
  \global\TextSize=0pt
  \global\ZoneBSize=0pt
  \global\vsize=0pt\relax
  \global\topskip=0pt\relax
  \vbox to 0pt{\vss}
  \eject
 \else
 \global\vsize=\ZoneBSize
 \global\hsize=\ColumnWidth
 \ifdim\TextSize<23pt
 \Warn{}
 \Warn{* Making column fall short: TextSize=\the\TextSize *}
 \vskip-\lastskip\eject\fi
\fi
}

\def\FigureItems{% Stack pointer points to next figure
 \Print{Considering...}
 \ShowItem{\StackPointer}
 \GetItemBOX{\StackPointer} % auto calculates ItemSIZE
 \GetItemSPAN{\StackPointer}
  \CheckFitInZone % check to see if item fits
  \ifnum\ItemFits=\Yes
   \ifnum\ItemSPAN=\Single
     \ChangeStatus{\StackPointer}{\InZoneB} % flag to be output
     \global\FigInZoneBtrue
     \ifFirstSingleItem
      \hbox{}\vskip-\BodgeHeight
     \global\advance\ItemSIZE by \TextLeading
     \fi
     \unvbox\ItemBOX\ItemSep
     \global\FirstSingleItemfalse
     \global\advance\TextSize by -\ItemSIZE% allocate space
     \global\advance\TextSize by -\TextLeading
   \else
    \ifFirstZoneA
     \global\advance\ItemSIZE by \TextLeading
     \global\FirstZoneAfalse\fi
    \global\advance\TextSize by -\ItemSIZE
    \global\advance\TextSize by -\TextLeading
    \global\advance\ZoneBSize by -\ItemSIZE
    \global\advance\ZoneBSize by -\TextLeading
    \ifFigInZoneB\relax
     \else
     \ifdim\TextSize<3\TextLeading
     \global\ZoneAFullPagetrue
     \fi
    \fi
    \ChangeStatus{\StackPointer}{\Zone}
    \ifnum\Zone=\InZoneC \global\FigInZoneCtrue\fi
  \fi
   \Print{TextSize=\the\TextSize}
   \Print{ZoneBSize=\the\ZoneBSize}
  \global\advance\NextFigure by 1
   \Print{This figure has been placed.}
  \else
   \Print{No space available for this figure...holding over.}
   \Print{}
   \global\MoreFiguresfalse
  \fi
}

\def\TableItems{% Stack pointer points to next table
 \Print{Considering...}
 \ShowItem{\StackPointer}
 \GetItemBOX{\StackPointer} % auto calculates ItemSIZE
 \GetItemSPAN{\StackPointer}
  \CheckFitInZone % check to see of item fits in Zone
  \ifnum\ItemFits=\Yes
   \ifnum\ItemSPAN=\Single
    \ChangeStatus{\StackPointer}{\InZoneB}
     \global\TabInZoneBtrue
     \ifFirstSingleItem
      \hbox{}\vskip-\BodgeHeight
     \global\advance\ItemSIZE by \TextLeading
     \fi
     \unvbox\ItemBOX\ItemSep
     \global\FirstSingleItemfalse
     \global\advance\TextSize by -\ItemSIZE
     \global\advance\TextSize by -\TextLeading
   \else
    \ifFirstZoneA
    \global\advance\ItemSIZE by \TextLeading
    \global\FirstZoneAfalse\fi
    \global\advance\TextSize by -\ItemSIZE
    \global\advance\TextSize by -\TextLeading
    \global\advance\ZoneBSize by -\ItemSIZE
    \global\advance\ZoneBSize by -\TextLeading
    \ifFigInZoneB\relax
     \else
     \ifdim\TextSize<3\TextLeading
     \global\ZoneAFullPagetrue
     \fi
    \fi
    \ChangeStatus{\StackPointer}{\Zone}
    \ifnum\Zone=\InZoneC \global\TabInZoneCtrue\fi
   \fi
%   \Print{TextSize=\the\TextSize}
%   \Print{ZoneBSize=\the\ZoneBSize}
  \global\advance\NextTable by 1
   \Print{This table has been placed.}
  \else
  \Print{No space available for this table...holding over.}
   \Print{}
   \global\MoreTablesfalse
  \fi
}

% These macros check to see if an item of ItemSIZE will
% fit in a particular zone. If it will, then ItemFits
% will be set true else false.

\def\CheckFitInZone{%
{\advance\TextSize by -\ItemSIZE
 \advance\TextSize by -\TextLeading
 \ifFirstSingleItem
  \advance\TextSize by \TextLeading
 \fi
 \ifnum\Zone=\InZoneA\relax
  \else \advance\TextSize by -\ZoneBAdjust
 \fi
 \ifdim\TextSize<3\TextLeading \global\ItemFits=\No
 \else \global\ItemFits=\Yes\fi}
}

\def\BF#1#2{% Blank figure (with no caption to be pasted in by hand)
 \ItemSTATUS=\InStack
 \ItemNUMBER=#1
 \ItemTYPE=\Figure
 \if#2S \ItemSPAN=\Single
  \else \ItemSPAN=\Double
 \fi
 \setbox\ItemBOX=\vbox{}
}

\def\BT#1#2{% Blank table (to be pasted in by hand)
 \ItemSTATUS=\InStack
 \ItemNUMBER=#1
 \ItemTYPE=\Table
 \if#2S \ItemSPAN=\Single
  \else \ItemSPAN=\Double
 \fi
 \setbox\ItemBOX=\vbox{}
}

\def\BeginOpening{%
 \hsize=\PageWidth
 \global\setbox\ItemBOX=\vbox\bgroup
}

\let\begintopmatter=\BeginOpening  %  alias for \BeginOpening

\def\EndOpening{%
 \egroup
 \ItemNUMBER=0
 \ItemTYPE=\Figure
 \ItemSPAN=\Double
 \ItemSTATUS=\InStack
 \JoinStack
}

%%% \def\FC#1#2#3#4{%
%%%  \ItemSTATUS=\InStack
%%%  \ItemNUMBER=#1
%%%  \ItemTYPE=\Figure
%%%  \if#2S \ItemSPAN=\Single \TEMPDIMEN=\ColumnWidth
%%%   \else \ItemSPAN=\Double \TEMPDIMEN=\PageWidth
%%%  \fi
%%% {\hsize=\TEMPDIMEN
%%%  \global\setbox\ItemBOX=\vbox{%
%%%  \ifFigureBoxes\B{\TEMPDIMEN}{#3}
%%%  \else \vbox to #3{\vfil}\fi%
%%% %\eightpoint\baselineskip=\rTenPT\vskip 5.5pt plus 6pt\rm\noindent #4}
%%%  \eightpoint\bls{\rTenPT}\vskip 5.5pt plus 6pt\rm\noindent #4}
%%% }
%%%  \JoinStack
%%%  \Print{Processing source for figure {\the\ItemNUMBER}}
%%% % \ifnum\ItemNUMBER=\NextFigure
%%% %  \ifnum\ItemSPAN=\Single
%%% %   \ItemSIZE=\ht\ItemBOX
%%% %   \advance\ItemSIZE by \dp\ItemBOX
%%% %   \advance\ItemSIZE by 12pt
%%% %   \global\TextSize=\pagegoal
%%% %   \global\advance\TextSize by -\pagetotal
%%% %   \CheckFitInZone
%%% %   \ifnum\ItemFits=\Yes
%%% %    \TEMPDIMEN=\pagegoal
%%% %    \advance\TEMPDIMEN by -\ItemSIZE
%%% %    \global\pagegoal=\TEMPDIMEN
%%% %    \global\setbox\ZoneBBOX=\vbox{\box\ZoneBBOX\ItemSep\box\ItemBOX}
%%% %    \global\advance\NextFigure by 1
%%% %    \global\advance\TextSize by -\ItemSIZE
%%% %   \else \global\setbox\ItemBOX=\vbox{\box\ItemBOX\ItemSep}\JoinStack\fi
%%% %  \else \global\setbox\ItemBOX=\vbox{\box\ItemBOX\ItemSep}\JoinStack\fi
%%% % \else \global\setbox\ItemBOX=\vbox{\box\ItemBOX\ItemSep}\JoinStack\fi
%%% % \JoinStack
%%% % \Print{Figure {\the\ItemNUMBER} has joined stack}
%%% }

\newbox\tmpbox

\def\FC#1#2#3#4{%
  \ItemSTATUS=\InStack
  \ItemNUMBER=#1
  \ItemTYPE=\Figure
  \if#2S
    \ItemSPAN=\Single \TEMPDIMEN=\ColumnWidth
  \else
    \ItemSPAN=\Double \TEMPDIMEN=\PageWidth
  \fi
  {\hsize=\TEMPDIMEN
   \global\setbox\ItemBOX=\vbox{%
     \ifFigureBoxes
       \B{\TEMPDIMEN}{#3}
     \else
       \vbox to #3{\vfil}%
     \fi%
     \eightpoint\rm\bls{\rTenPT}%
     \vskip 5.5pt plus 6pt%
     \setbox\tmpbox=\vbox{#4\par}%
     \ifdim\ht\tmpbox>10pt %  1 line in eightpoint is approx 5.5pt
       \noindent #4\par%
     \else
       \hbox to \hsize{\hfil #4\hfil}%
     \fi%
   }%
  }%
  \JoinStack%
  \Print{Processing source for figure {\the\ItemNUMBER}}%
}

  %  alias for \FC

\def\TH#1#2#3#4{%
 \ItemSTATUS=\InStack
 \ItemNUMBER=#1
 \ItemTYPE=\Table
 \if#2S \ItemSPAN=\Single \TEMPDIMEN=\ColumnWidth
  \else \ItemSPAN=\Double \TEMPDIMEN=\PageWidth
 \fi
{\hsize=\TEMPDIMEN
%\eightpoint\baselineskip=\rTenPT\rm
\eightpoint\bls{\rTenPT}\rm
\global\setbox\ItemBOX=\vbox{\noindent#3\vskip 5.5pt plus5.5pt\noindent#4}}
 \JoinStack
 \Print{Processing source for table {\the\ItemNUMBER}}
}

  %  alias for \TH

\def\UnloadZoneA{%
\FirstZoneAtrue
 \Iteration=0
  \loop
   \ifnum\Iteration<\LengthOfStack
    \GetItemSTATUS{\Iteration}
    \ifnum\ItemSTATUS=\InZoneA
     \GetItemBOX{\Iteration}
     \ifFirstZoneA \vbox to \BodgeHeight{\vfil}%
     \FirstZoneAfalse\fi
     \unvbox\ItemBOX\ItemSep
     \LeaveStack{\Iteration}
     \else
     \advance\Iteration by 1
   \fi
 \repeat
}

\def\UnloadZoneC{%
\Iteration=0
  \loop
   \ifnum\Iteration<\LengthOfStack
    \GetItemSTATUS{\Iteration}
    \ifnum\ItemSTATUS=\InZoneC
     \GetItemBOX{\Iteration}
     \ItemSep\unvbox\ItemBOX
     \LeaveStack{\Iteration}
     \else
     \advance\Iteration by 1
   \fi
 \repeat
}

%--------------------------------------------------------%
%                         DIAGNOSTICS                    %
%--------------------------------------------------------%

\def\ShowItem#1{% Show details of on item entry in stack
  {\GetItemAll{#1}
  \Print{\the#1:
  {TYPE=\ifnum\ItemTYPE=\Figure Figure\else Table\fi}
  {NUMBER=\the\ItemNUMBER}
  {SPAN=\ifnum\ItemSPAN=\Single Single\else Double\fi}
  {SIZE=\the\ItemSIZE}}}
}

\def\ShowStack{%
 \Print{}
 \Print{LengthOfStack = \the\LengthOfStack}
 \ifnum\LengthOfStack=0 \Print{Stack is empty}\fi
 \Iteration=0
 \loop
 \ifnum\Iteration<\LengthOfStack
  \ShowItem{\Iteration}
  \advance\Iteration by 1
 \repeat
}

\def\B#1#2{%
\hbox{\vrule\kern-0.4pt\vbox to #2{%
\hrule width #1\vfill\hrule}\kern-0.4pt\vrule}
}

%% FOLLOWING LINE CANNOT BE BROKEN BEFORE 80 CHAR
\def\Ref#1{\begingroup\global\setbox\TEMPBOX=\vbox{\hsize=2in\noindent#1}\endgroup
\ht1=0pt\dp1=0pt\wd1=0pt\vadjust{\vtop to 0pt{\advance
\hsize0.5pc\kern-10pt\moveright\hsize\box\TEMPBOX\vss}}}

\def\MarkRef#1{\leavevmode\thinspace\hbox{\vrule\vtop
{\vbox{\hrule\kern1pt\hbox{\vphantom{\rm/}\thinspace{\rm#1}%
\thinspace}}\kern1pt\hrule}\vrule}\thinspace}%

%--------------------------------------------------------%
%                     OUTPUT ROUTINE                     %
%--------------------------------------------------------%

\output{%
 \ifLeftCOL
  \global\setbox\LeftBOX=\vbox to \ZoneBSize{\box255\unvbox\ZoneBBOX}
  \global\LeftCOLfalse
  \MakeRightCol
 \else
  \setbox\RightBOX=\vbox to \ZoneBSize{\box255\unvbox\ZoneBBOX}
  \setbox\MidBOX=\hbox{\box\LeftBOX\hskip\ColumnGap\box\RightBOX}
  \setbox\PageBOX=\vbox to \PageHeight{%
  \UnloadZoneA\box\MidBOX\UnloadZoneC}
  \shipout\vbox{\PageHead\box\PageBOX\PageFoot}
  \global\advance\pageno by 1
  \global\HeaderNumber=\DefaultHeader
  \global\LeftCOLtrue
  \CleanStack
  \MakePage
 \fi
}

%%% \endinput

\catcode `\@=12 % @ signs are non-letters

%%% \dump
\message{*** process twice to get section numbers etc. correct ***}
%
% Marginal adjustments using \pageoffset maybe required when printing
% proofs on a Laserprinter (this is usually not needed).
% Syntax: \pageoffset{ +/- hor. offset}{ +/- vert. offset}
% e.g.    \pageoffset{-3pc}{-4pc}

\pageoffset{-0.8cm}{0.2cm}

\def\bigstrut{\vrule width0pt height0.6truecm}
\font\japit = cmti10 at 11truept
\def\name#1{{\it #1\/}}
%
% JAP macros
%
\def\japitem#1{\par\hang\textindent{#1}}
\def\ref{\parskip =0pt\par\noindent\hangindent\parindent
    \parskip =2ex plus .5ex minus .1ex}
\def\gs{\mathrel{\raise1.16pt\hbox{$>$}\kern-7.0pt
\lower3.06pt\hbox{{$\scriptstyle \sim$}}}}
\def\ls{\mathrel{\raise1.16pt\hbox{$<$}\kern-7.0pt
\lower3.06pt\hbox{{$\scriptstyle \sim$}}}}
\def\pmb#1{\setbox0 =\hbox{#1}%
  \kern-.025em\copy0\kern-\wd0
  \kern.05em\copy0\kern-\wd0
  \kern-.025em\raise.0433em\box0 }

\def\ss{\rm\scriptscriptstyle}

% Edited down version of thesis macros suitable for Paper I.

%
% Fonts & Sizes
%

\font\sf    = cmss10

\font\csc   = cmcsc10 scaled 833

 at 20.7truept
%
% Symbols
%
\def\gs{\mathrel{\raise0.27ex\hbox{$>$}\kern-0.70em % Greater/squiggles
\lower0.71ex\hbox{{$\scriptstyle \sim$}}}}
\def\ls{\mathrel{\raise0.27ex\hbox{$<$}\kern-0.70em % Less than/squiggles
\lower0.71ex\hbox{{$\scriptstyle \sim$}}}}
\def\micron{\hbox{$\mu$m}}
\def\hrs{\hbox{$\rm^h$}}
\def\mins{\hbox{$\rm^m$}}
\def\secs{\hbox{$\rm^s$}}
\def\deg{{\hbox{\twelvept$^{\circ}$}}}
\def\amin{\hbox{$'$}}
\def\asec{\hbox{$''$}}

\def\sqr#1#2{\vbox{\hrule height#2        % Square symbol of size #1 and line
    \hbox{\vrule width#2 height#1 \kern#1 % thickness #2.
      \vrule width#2}
    \hrule height#2}}

\def\msqr#1#2{\vcenter{\sqr{#1}{#2}}}  % Square in maths mode centred on
                                       % math formula axis

% Universal square symbol - horizontal, maths, superscripts, etc...

\def\square{{\ifmmode\mathchoice{\msqr{0.54em}{0.04em}}{\msqr{0.54em}{0.04em}}%
{\msqr{0.38em}{0.03em}}{\msqr{0.27em}{0.03em}}\else
\raise 0.04em\sqr{0.54em}{0.04em}\fi}}

\def\sqdeg{\hbox{\square{\twelvept$\,^{\circ}$}}}   % Square degrees.
\def\sqamin{\hbox{\square$\,'$}}         % Square arcminutes.
\def\sqasec{\hbox{\square$\,''$}}        % Square arcseconds.
\def\electron{\hbox{$\rm e^{\hbox{-}}$}} % Electron (e minus).

%
% Sectioning and layout commands
%
 % Old maths use of \sec

%
% Boxing commands
%

%
% Define a box of #1 by #2 truecm to hold a PGPLOT figure.
%
\def\pgplot#1#2#3{\hbox to #1 truecm{
     \vbox to #2 truecm{\special{#3}\vfill} \hfill}}

% \pgtweak allows the box containing the graph to be tweaked - a margin
% of #3 truecm is stripped from the TOP of the box and #4 truecm from the
% RIGHT. This is especially useful when their is no top title. Then the top
% space can be reclaimed from PGPAPER. It works be creating the new smaller box
% and then letting the old box (containing the figure) to stick out to the top
% and right.

\def\pgtweak#1#2#3#4#5{{\dimen1=#1 truecm \dimen2=#2 truecm
   \advance\dimen1 by -#3 truecm
   \advance\dimen2 by -#4 truecm
   \vbox to \dimen2{\vss
   \hbox to \dimen1{\pgplot{#1}{#2}{#5}\hss}}}}
%
%========================================================================
%
% "Shadow box macro"
%
%a temporary box and dimension register are need for this implementation
\newdimen\dropht \newbox\dropbox

\def\dropshadow#1#2{\setbox\dropbox=%
%first take what you are given and outline it
%This is the boxit routine in the TeXbook, ex. 21.3
\vbox{\hrule\hbox{\vrule\kern3pt
        \vbox{\kern3pt#1\kern3pt}\kern3pt\vrule}\hrule}
%Now grab put the shadow on, using the dimensions from the
%box you just made to force the size of the rules.
\dropht=\ht\dropbox\advance\dropht by -#2
       \vbox{\baselineskip0pt\lineskip0pt
             \hbox{\copy\dropbox\vrule width#2 height\dropht}
             \hbox{\kern#2\vrule height#2 width \wd\dropbox}}}

%Now test it.
%\setbox4=\vbox {\hsize 23pc This is a test of the dropshadow routine.}
%\dropshadow{\box4}{4pt}
%========================================================================
%
% Miscelleanous macros
%
\let\maththinspace=\, % Save old definition...
\def\,{\relax\ifmmode\maththinspace\else\thinspace\fi} % ... and redefine it.

      % Make math default \rm font.

\def\etal{{\it et~al}}
      % emission line e.g. %emline{S}{II}
 % e.g. %emline{S}{II}{6717}

%
% Journals and referances
%
\def\jfont{\it}
\def\vol#1{{\bf#1}}
\def\MNRAS{{\jfont Mon. Not. R. Astr. Soc.}}
\def\ApJ{{\jfont Astrophys. J.}}
\def\AstrAst{{\jfont Astr. Astrophys.}}
\def\AstrAstSup{{\jfont Astr. Astrophys. Suppl. Ser.}}
\def\AstrJ{{\jfont Astr. J.}}
%
% Hyphenation
%
\hyphenation{infra-red}
\hyphenation{inter-stellar}
\hyphenation{flat-field}
\hyphenation{near-star}
\hyphenation{gal-axy}
%
% file: TXStags.tex             TeXsis                  version 2.11
%=======================================================================
% TAGS - create lables for things                       source: TechRpt
\message{Labels and tags.}

\catcode`@=11

% ================
% Error Reporting:
% \emsg writes the message "#1", ON A NEW LINE, on the terminal and in the log

\def\emsg#1{\begingroup%
   \def\TeX{TeX}\def\label##1{}%
   \immediate\write16{#1}
   \endgroup}

% \@errmark writes a short error message in the left margin
\newif\ifmarkerrors     \markerrorsfalse        % default is off

\def\@errmark#1{\ifmarkerrors           % only if this is on
    \vadjust{\vtop to 0pt{\vss          % box with no height
     \kern-\baselineskip                % insert text next to line
     \llap{{\tt #1->\hskip 0.5cm }}%    %  in left margin
    }}%                                 % end \vbox and \vadjust
    \fi}

% I/O and switches:
\newread\auxfilein      % input for jobname.aux file
\newwrite\auxfileout    % output for jobname.aux file
\newif\ifauxswitch      % enable writing to .aux file?
\auxswitchtrue          % default is "yes"
\let\XA=\expandafter    % shorthand for \expandafter
\let\NX=\noexpand       % shorthand for \noexpand

%-------------------------
% Initialization for TAGS

\def\auxinit{%          % first look for old file jobname.aux and read it
    \openin\auxfilein=\jobname.aux      % open old .aux file for input
    \ifeof\auxfilein{\closein\auxfilein}% no: just close it
    \else\closein\auxfilein             % yes: close it so to \input
       \input \jobname.aux                      %   and read it in
    \fi                                         % else ignore it
  \ifauxswitch                                  % if auxswitch true:
    \immediate\openout\auxfileout=\jobname.aux  % open new .aux file for output
  \else                                         % if auxswitch false:
    \gdef\auxout##1##2{}%               % turn off \tag 's writing to .aux file
  \fi}

%-------------------------
% \tag{name}{value} defines "\name" as "value", and writes the definition
% to the .aux file.  To eliminate the output to the .aux file, just
% \def\auxout#1#2{}.
% This uses \csname...\endcsname everywhere to allow label names to
% include practically anything, including _, -, 0-9, (), ...
% Spaces may be included in a name, but they are ignored.

\def\tag#1#2{\relax\edef\@@ttt{#2}%             % put value in \@@ttt
            \stripblanks #1\endlist             % remove blanks from label
            \XA\let\csname \tok\endcsname=\z@   % remove old definition
            \auxout{\tok}{\@@ttt}%              % write def to .aux file
            \XA\xdef\csname\tok\endcsname{\@@ttt}}      % define it now

% \auxout{name}{value} writes the definition to .aux file

\def\auxout#1#2{\immediate\write\auxfileout{\NX\expandafter\gdef
                \NX\csname #1\NX\endcsname{#2}}}

%-------------------------
%  \label{name} tags name with the current value of \ttt, which is set
%  to the proper value by various macros, such as \chapter, \section, etc...

\def\label#1{\tag{#1}{\ttt}}

%-------------------------
%  \use{name} uses the value of \name if \name is defined; otherwise it
%  puts {\bf name} into the text and writes an error message to the .LOG

\def\use#1{\stripblanks #1\endlist                      % remove all blanks
  \XA\ifx\csname\tok\endcsname\relax                    % is "\#1" undefined?
    \emsg{> UNDEFINED TAG `\tok' ON PAGE \folio.}        % yes: error message
    {\bf\tok}\@errmark{UNDEF}%                          %   and mark in output
  \else{\csname\tok\endcsname}\fi\relax}                % no: evaluate it

%--------------------------
% The following macro is used to see if a control sequence has been
% defined, using the fact that \csname makes a default assignment of
% undefined control sequences to \relax.
% Use is:
%    \ifundefined{name}<true text>\else<false text>\fi
% name must NOT contain the leading \
% (This does NOT work if nested in  another \if !  See pg 211 of TeXbook)

\def\ifundefined#1{\XA\ifx\csname#1\endcsname\relax}

% \stripblanks removes extraneous blanks from a token list
% Format:
%   \stripblanks text\endlist
% defines \tok as the text with ALL blanks removed.
% \tok is \empty if text is ALL blank.

\def\stripblanks{\let\tok=\empty\@stripblanks}
\def\@stripblanks#1{\def\next{#1}\@striplist}
\def\@striplist{\ifx\next\endlist \let\next\relax
               \else \@stripspace\let\next\@stripblanks \fi
               \next}
\def\endlist{\endlist}          % \endlist is undefined on purpose
\def\@stripspace{\XA\if\space\next\else\edef\tok{\tok\next}\fi}

\catcode`@=12

% >>> EOF TXStags.tex <<<
%
% The following macros switch sizes automatically for the standard
% TeX fonts - roman,bold,italic,slanted,san-serif,typewriter and
% math symbols.
%
%       \tenpt               10pt type
%       \elevenpt            11pt type
%       \twelvept            12pt type
%       \fourteenpt          14pt type
%       \seventeenpt         17pt type
%
% After a size change you are in \rm by default.

\def\myfont{\font}

\newskip\ttglue\newfam\ssfam

%+++ 5 pt fonts already set up by PLAIN TeX.

%+++ 7 pt fonts - most already set up by PLAIN TeX.
\myfont\sevenit=cmti7

%+++ 8 pt fonts (some preloaded):
\myfont\eightrm=cmr8
\myfont\eighti=cmmi8
\myfont\eightsy=cmsy8
\myfont\eightbf=cmbx8
\myfont\eighttt=cmtt8
\myfont\eightit=cmti8
\myfont\eightsl=cmsl8
\myfont\eightss=cmss8

%+++ 9 pt fonts (some preloaded):
\myfont\ninerm=cmr9
\myfont\ninei=cmmi9
\myfont\ninesy=cmsy9
\myfont\ninebf=cmbx9
\myfont\ninett=cmtt9
\myfont\nineit=cmti9
\myfont\ninesl=cmsl9
\myfont\niness=cmss9

%+++ 10 pt fonts:               % most are already loaded by Plain TeX
\myfont\tenss=cmss10                              % 10 pt sans serif
\myfont\tencsc=cmcsc10                            % 10 pt caps & small caps

%+++ 11 pt fonts:
\myfont\elevenrm=cmr10 scaled \magstephalf        % 11 pt  Roman
\myfont\eleveni=cmmi10 scaled \magstephalf        % 11 pt  math italic
\myfont\elevensy=cmsy10 scaled \magstephalf       % 11 pt  math symbol
\myfont\elevenex=cmex10 scaled \magstephalf       % 11 pt  math extended
%%symbols
\myfont\elevenbf=cmbx10 scaled \magstephalf       % 11 pt  Roman bold extended
\myfont\elevensl=cmsl10 scaled \magstephalf       % 11 pt  slanted
\myfont\eleventt=cmtt10 scaled \magstephalf       % 11 pt  typewriter
\myfont\elevenit=cmti10 scaled \magstephalf       % 11 pt  Roman text italic
\myfont\elevenss=cmss10 scaled \magstephalf       % 11 pt  sans serif
\myfont\elevencsc=cmcsc10 scaled \magstephalf     % 11 pt caps & small caps
\skewchar\eleveni='177                          % see p 414 of the TeXbook
\skewchar\elevensy='60                          % see p 414 of the TeXbook
\hyphenchar\eleventt=-1                         % see p 414 of the TeXbook

%+++ 12 pt fonts:
\myfont\twelverm=cmr10 scaled \magstep1           % 12 pt  Roman
\myfont\twelvei=cmmi10 scaled \magstep1           % 12 pt  math italic
\myfont\twelvesy=cmsy10 scaled \magstep1          % 12 pt  math symbol
\myfont\twelveex=cmex10 scaled \magstep1          % 12 pt  math extended
\myfont\twelvebf=cmbx10 scaled \magstep1          % 12 pt  Roman bold extended
\myfont\twelvesl=cmsl10 scaled \magstep1          % 12 pt  slanted
\myfont\twelvett=cmtt10 scaled \magstep1          % 12 pt  typewriter
\myfont\twelveit=cmti10 scaled \magstep1          % 12 pt  Roman text itali
\myfont\twelvess=cmss10 scaled \magstep1          % 12 pt  sans serif
\myfont\twelvecsc=cmcsc10 scaled \magstep1        % 12 pt caps & small caps
\skewchar\twelvei='177                          % see p 414 of the TeXbook
\skewchar\twelvesy='60                          % see p 414 of the TeXbook
\hyphenchar\twelvett=-1                         % see p 414 of the TeXbook

%+++ 14 pt fonts:
\myfont\fourteenrm=cmr10 scaled \magstep2           % 14 pt  Roman
\myfont\fourteeni=cmmi10 scaled \magstep2           % 14 pt  math italic
\myfont\fourteensy=cmsy10 scaled \magstep2          % 14 pt  math symbol
\myfont\fourteenex=cmex10 scaled \magstep2          % 14 pt  math extended
\myfont\fourteenbf=cmbx10 scaled \magstep2          % 14 pt  Roman bold
%%extended
\myfont\fourteensl=cmsl10 scaled \magstep2          % 14 pt  slanted
\myfont\fourteentt=cmtt10 scaled \magstep2          % 14 pt  typewriter
\myfont\fourteenit=cmti10 scaled \magstep2          % 14 pt  Roman text itali
\myfont\fourteenss=cmss10 scaled \magstep2          % 14 pt  sans serif
\myfont\fourteencsc=cmcsc10 scaled \magstep2        % 14 pt caps & small caps
\skewchar\fourteeni='177                          % see p 414 of the TeXbook
\skewchar\fourteensy='60                          % see p 414 of the TeXbook
\hyphenchar\fourteentt=-1                         % see p 414 of the TeXbook

%+++ 17 pt fonts:
\myfont\seventeenrm=cmr10 scaled \magstep3           % 17 pt  Roman
\myfont\seventeeni=cmmi10 scaled \magstep3           % 17 pt  math italic
\myfont\seventeensy=cmsy10 scaled \magstep3          % 17 pt  math symbol
\myfont\seventeenex=cmex10 scaled \magstep3          % 17 pt  math extended
\myfont\seventeenbf=cmbx10 scaled \magstep3          % 17 pt  Roman bold
%%extended
\myfont\seventeensl=cmsl10 scaled \magstep3          % 17 pt  slanted
\myfont\seventeentt=cmtt10 scaled \magstep3          % 17 pt  typewriter
\myfont\seventeenit=cmti10 scaled \magstep3          % 17 pt  Roman text itali
\myfont\seventeenss=cmss12 scaled \magstep2          % 17 pt  sans serif
\myfont\seventeencsc=cmcsc10 scaled \magstep3        % 17 pt caps & small caps
\skewchar\seventeeni='177                          % see p 414 of the TeXbook
\skewchar\seventeensy='60                          % see p 414 of the TeXbook
\hyphenchar\seventeentt=-1                         % see p 414 of the TeXbook

\def\fontname#1{\ifcase#1 zero\or%
one\or two\or three\or four\or five\or six\or seven\or eight\or nine\or ten%
\or eleven\or twelve\or thirteen\or fourteen\or fifteen\or sixteen%
\or seventeen\or eighteen\or nineteen\or twenty\or twentyone\or twentytwo%
\or twentythree\or twentyfour\or twentyfive\or twentysix\or twentyseven%
\or twentyeight\or twentynine\or thirty\else unknown\fi}

\def\setfam#1#2#3#4#5{%
\textfont#1=\expandafter\csname\fontname{#2}#5\endcsname%
\scriptfont#1=\expandafter\csname\fontname{#3}#5\endcsname%
\scriptscriptfont#1=\expandafter\csname\fontname{#4}#5\endcsname}

\def\setdimens#1#2#3{\normalbaselineskip=#1%
\setbox\strutbox=\hbox{\vrule height#2 depth#3 width0pt}%
\normalbaselines\rm}

%
% Switch to 10pt type
%
\message{10pt,}
\def\tenpt{\setfam1{10}75i\setfam2{10}75{sy}%
\setfam0{10}75{rm}\def\rm{\fam0\tenrm}%
\setfam\bffam{10}87{bf}\def\bf{\fam\bffam\tenbf}%
\setfam\itfam{10}87{it}\def\it{\fam\itfam\tenit}%
\setfam\ttfam{10}88{tt}\def\tt{\fam\ttfam\tentt}%
\setfam\ssfam{10}88{ss}\def\ss{\fam\ssfam\tenss}%
\setfam\slfam{10}88{sl}\def\sl{\fam\slfam\tensl}%
\def\csc{\tencsc}%
\setdimens{12pt}{8.5pt}{3.5pt}}
%
% Switch to 11pt type
%
\message{11pt,}

%
% Switch to 12pt type
%
\message{12pt,}
\def\twelvept{\setfam1{12}97i\setfam2{12}97{sy}%
\setfam0{12}97{rm}\def\rm{\fam0\twelverm}%
\setfam\bffam{12}97{bf}\def\bf{\fam\bffam\twelvebf}%
\setfam\itfam{12}97{it}\def\it{\fam\itfam\twelveit}%
\setfam\ttfam{12}98{tt}\def\tt{\fam\ttfam\twelvett}%
\setfam\ssfam{12}98{ss}\def\ss{\fam\ssfam\twelvess}%
\setfam\slfam{12}98{sl}\def\sl{\fam\slfam\twelvesl}%
\def\csc{\twelvecsc}%
\setdimens{14pt}{10pt}{4pt}}
%
% Switch to 14pt type
%
\message{14pt,}

%
% Switch to 17pt type
%
\message{17pt.}

 at 11truept
 at 12truept
\font\bbbit=cmbxti10 at 17truept

% Redefine \item and \itemitem so as to distinguish between paragraph
% indentation size and item indentation size.

\newdimen\itemindent \itemindent=3em
\def\item#1{\par\hangindent=\itemindent\hangafter=1\hskip\itemindent
            \llap{#1\enspace}\ignorespaces}

% Automatically counting version of item.

\newcount\ic \ic=0 % Item counter
\def\itm{\advance\ic by 1 \japitem{\number\ic)}\def\ttt{\number\ic}}

% Big items - turn off hanging indentation after one line.

\let\em=\it % Set emphasis font to be italic

% Some common abbreviations:

\def\ds{\displaystyle}

\def\Hhundred{\hbox{$H_0=100\,\rm km\,s^{-1}\,Mpc^{-1}$}}
\def\phiunits{\hbox{$\times 10^{-2}\,h^3\,\rm Mpc^{-3}$}}

\def\BK{\hbox{$B-K$}}

\def\li#1#2{#1{\csc\lowercase{#2}}} % Abs/Em line e.g. \li{Fe}{II}

\def\HII{\li{H}{II}}

\def\OII{[\li{O}{II}]}

\def\slabel#1{\def\ttt{\number\Sec}\label{#1}\hglue -0.05cm}
\def\sslabel#1{\def\ttt{\number\Sec.\number\SecSec}\label{#1}\hglue -0.05cm}

% Open the .AUX file for tags:

\auxinit

% Figures

\expandafter \gdef \csname overlap.fig\endcsname {1}
\expandafter \gdef \csname StarGal.fig\endcsname {2}
\expandafter \gdef \csname StarGalvsK.fig\endcsname {3}
\expandafter \gdef \csname RefCounts.fig\endcsname {4}
\expandafter \gdef \csname RawCounts.fig\endcsname {5}
\expandafter \gdef \csname StarModel.fig\endcsname {6}
\expandafter \gdef \csname Knoev.fig\endcsname {7}
\expandafter \gdef \csname bjnoev.fig\endcsname {8}
\expandafter \gdef \csname NMLE.fig\endcsname {9}
\expandafter \gdef \csname NMME.fig\endcsname {10}
\expandafter \gdef \csname NzK.fig\endcsname {11}

% Tables

\expandafter \gdef \csname Fieldcentres.tab\endcsname {1}
\expandafter \gdef \csname EnormousWasteOfSpace.tab\endcsname {2}
\expandafter \gdef \csname CCDint.tab\endcsname {3}
\expandafter \gdef \csname Catalogue.tab\endcsname {4}
\expandafter \gdef \csname IRNumCounts.tab\endcsname {5}
\expandafter \gdef \csname GalaxyMix.tab\endcsname {6}

% \font\euler=eurm10
% \def\umu{\hbox{$\euler\mu$}} \def\upi{\hbox{$\euler\pi$}}
% Uncomment the above two lines if the Euler font is available

% \Referee   %  uncomment this for referee mode
\Autonumber  %  auto-number sections, subsections and subsubsections

% \pagerange, \pubyear and \volume are defined at the Journals office and
% not by an author.

%\pagerange{1--6}
%\pubyear{1989}
%\volume{226}
% \umufiche{}     % for articles with microfiche
% \authorcomment{}  % author comment for footline

\begintopmatter  %  start the two spanning material

\vglue-2.2truecm
%\centerline{\japit For submission to Monthly Notices of the R.A.S.}
\centerline{\japit Accepted for publication in Monthly Notices of the R.A.S.}
\vglue 1.7truecm

\title{An imaging {\bbbit K\/}-band survey --- I:
The catalogue, star and galaxy counts}

\author{Karl Glazebrook$^{1,3}$,
J.A. Peacock$^2$, C.A. Collins$^{2,4}$
 and L. Miller$^2$ }

\affiliation{$^1$Department of Astronomy, \bigstrut University of Edinburgh,
Blackford Hill, Edinburgh EH9 3HJ, UK \hfill\break
$^2$Royal Observatory, Blackford Hill, Edinburgh EH9 3HJ, UK\hfill\break
$^3$Present address: Institute of Astronomy, Madingley Road,
Cambridge CB3 0HA, UK\hfill\break
$^4$Present address: Department of Physics, South Road, Durham DH1 3LE, UK}

\shortauthor{K. Glazebrook, J.A. Peacock, C.A. Collins and L. Miller}

\shorttitle{Imaging $K$-band survey --- I.}

% \acceptedline is to be defined at the Journals office and not
% by an author.

\abstract
We present results from a large area (552\,\sqamin)
imaging $K$-band survey of faint objects.
The survey is a high galactic latitude
blank-field sample to a 5$\sigma$ limit of $K\simeq 17.3$. The
methods for constructing the infrared survey are described,
including flatfielding,
astrometry, mosaicing and photometry. Also described are optical CCD
observations
which cover the survey to provide optical-infrared colours of almost all
the objects in the sample.
Star-galaxy discrimination is performed and the results used to derive the
infrared
star and galaxy counts. $K$-band ``no-evolution'' galaxy-count
models are constructed and compared
with the observed data. In the infrared, there is no counterpart for the large
excess of faint galaxies over the no-evolution model seen in optical counts.
In contrast,
the $K$ counts require little or no evolution in the luminosity or space
density
of galaxies, although it is shown that the count predictions can be
remarkably insensitive to evolution
under certain reasonable assumptions. Finally, model predictions for
$K$-selected redshift surveys are derived.

\maketitle  %  finish the two spanning material

\section{Introduction}

\tx
The development of 2-dimensional near-infrared detectors has finally
made it possible to survey substantial areas of the
sky at these wavelengths to cosmologically interesting depths. This is
the first in a series
of papers describing
a project designed to
map several hundred square arcminutes of sky to a
depth of $K\simeq 17$. This paper in particular is concerned with
the details of the construction and calibration of this catalogue,
and the associated optical CCD imaging for all the objects.

Such a survey has a multitude of uses. Firstly, there are serendipitous
searches for populations of objects which might be quite bright in
the infrared but absent, or very faint, in the deepest
optical surveys. Such populations could include, for example,
protogalactic objects and low-luminosity stars and brown dwarfs.

It has been proposed  that protogalaxies
(hereafter PG's) should have flat spectra similar to giant \HII\ regions
(Koo 1986). The density of such objects on the sky
should be very high, at least $\sim 5/\sqamin$, from simple
estimates based on the space density of modern galaxies
(Collins \& Joseph 1988).
If such objects were formed at sufficiently high redshifts
then they would not be prominent in the optical, due to the redshifting
of the Lyman limit into the passbands. However the $K$-band ($2.2\micron$) is
sensitive to such objects out to $z\simeq 20$.

Brown dwarfs are commonly defined as sub-stellar objects formed from
the same fragmentation process as the visible stars (Black, 1986).
They may have been detected, via
infrared imaging, as companions
to white dwarfs by  Zuckerman
\& Becklin (1987) and Becklin \& Zuckerman (1988) and a systematic survey,
based
on $I$ plates, by  Hawkins \& Bessel (1988)
shows that low-luminosity stars are found
as an extension of the stellar luminosity function.
Because the blackbody flux from such objects peaks at $K$, and the objects
are additionally reddened by the presence of strong absorption bands in
the optical, a $K$-band survey opens up the possibility of revealing and
quantifying new populations of such objects.

Secondly the survey provides a catalogue of galaxies selected by
near-infrared flux, which allows a new approach to the study of galaxy
evolution.
It is now well established that the faint galaxy number-magnitude counts
in the optical
show a blue excess population over that predicted
without evolution (e.g. Peterson \etal. 1979, Kron 1980,
Hall \& Mackay
1984, Tyson 1988). Deep redshift surveys
by Broadhurst \etal. (1988), Colless \etal.
(1990) and Lilly \etal. (1991) are beginning to show that the excess is
not due to high redshift objects, as would be expected in simple models
of luminosity evolution in the stellar populations, but appears to be
due to lower-luminosity objects with a higher space density.

At redder wavelengths the
light becomes dominated by
stars with progressively longer lifetimes. The near-infrared light from
a galaxy is dominated by the ``old-stellar population''. This consists of old,
evolved, stars on the giant branch with lifetimes of gigayears. Thus the
relative contribution from on-going star-formation to the total light
will be much less.
In the $I$-band ($\sim 1\micron$)
the excess in the counts over no-evolution is already much less than in the
$B$-band (e.g. Tyson, 1988).
This leads us to expect that
the $K$-band light will be relatively
insensitive to star-formation. Thus the $K$ light should
provide an excellent tracer of direct evolution in galaxy density.

The paper is laid out as follows: Section 2 discusses the
construction of the infrared survey. The
choosing of field centres from Schmidt plates and the method adopted for
observing are discussed. The data reduction procedures peculiar to infrared
array detectors are detailed  and the methods of mosaicing and performing
photometric and astrometric calibrations are described.

Section 3  describes the CCD observations that were performed to
obtain accurate colours for all the infrared selected objects. Photometry,
astrometry and matching with the infrared data are detailed.

Section 4 derives the infrared star and galaxy counts for the survey
and corrects them for observed error rates, from the spectroscopy,  in
star-galaxy classification.

Section 5 derives galaxy count predictions from
literature spectral evolution models and compared in $K$ and $b_j$ with the
survey
data and other published data. The effect of clustering on the galaxy counts
are considered. More general parametric models of density and
luminosity evolution in the galaxy population are also developed and compared
with the data.

Finally the results and conclusions are summarized in Section 6.

\section{The Infrared Survey}

\tx
The design of an infrared survey is a trade off between area and depth
due to the limitations of observing time. It is also
heavily constrained
by the desire to obtain optical information on the same set of
objects.

On a 4\,m optical telescope,  a
spectrum good enough to yield a redshift from absorption features in a night's
observing (30\,000\,s) can be obtained
down to $B\simeq 23.5$ or even $B\simeq 24$ if the
object has very strong emission lines. In simple passively evolving
galaxy models (e.g. Rocca-Volmerange and Guideroni 1988) the galaxies
can be as red as $B-K\simeq 5-6$ at redshifts of order $0.5$--1. This
implies a limit of $K\simeq 18$ for complete samples. At this magnitude
a galaxy with the characteristic luminosity $L^*$ is seen at $z\sim 0.3$,
so the sample is deep enough to be cosmologically interesting.
Because of the desire to study a sample {\em selected\/} in the infrared
rather than make infrared observations of an optically limited sample,
it is necessary to include the reddest objects.
This sets the approximate limiting magnitude for the present survey.

In order to address the question of galaxy evolution
a fairly large sample of at least $\simeq 50$--100 galaxies would be
be required. More galaxies are observed by covering a larger area
rather than by integrating longer --- the
number of detected objects in a background
limited observation
rises only $\propto t^{3\over 4}$ for a uniform population
in a simple Euclidean universe, which
provides an upper bound to the observed number-magnitude optical counts.
Thus the estimated surface density at  $K=18$ of  $\sim 1$ galaxy$/\sqamin$
(based on optical surface densities and assumed colours) implies
several hundred
$\sqamin$ would have to be surveyed to provide such a large sample. Due
to the limited size of the detectors it was decided to survey as large
an area of sky as possible in one infrared band only, the 2.2\micron\
$K$-band, and supplement this with optical CCD imaging and spectroscopy.

\subsection{Observations}

\tx\sslabel{Obs.sec}
The infrared observations for this project were all made over the period
1987--1988 using the Infrared Camera IRCAM at the
3.8\,m United Kingdom Infrared Telescope on Mauna Kea, Hawaii.
IRCAM uses a $62\times 58$  indium antimonide (InSb) array for imaging
observations in the 1--5\,\micron\ band (McLean \etal. 1986). A pixel size of
$\rm1.2\asec/pixel$ was chosen to cover the largest area possible with
reasonable sampling and without field vignetting.

For IRCAM at $K$ in the rms noise per pixel in 1 second was equivalent to
$K=17.8$
(Casali \etal. 1987) which predicts a $5\,\sigma$
detection at the required depth of $K=18.0$ in 5 minutes. This Fig.~allows
for both shot and read noise making approximately equal contributions because
of
the exposure being broken down into
10 second segments between readouts. Any longer
would saturate the detector from the sky signal, causing non-linearity.

With such a small detector it is necessary to cover the sky in a mosaic
pattern, ideally with overlaps to cross-calibrate photometry and astrometry
between
different nights.
A simple rectilinear grid with  overlaps was ruled out as there would
be insufficient bright stars in the overlaps ---
stellar distribution models (Bahcall \& Soneira 1980)
predict only 0.3--1 stars/\sqamin\ in the fields  brighter than $B=20$.

Thus a sparse rather than a full mosaic
strategy was adopted. The images were taken in a $2\times 2$ pattern around
preselected
bright stars so that each frame would have the star in one of its four corners
--- about 10 pixels from the edge. Each quarter was
observed on different nights so that the reference star provided a
photometric and astrometric cross-calibration.
The disadvantage of this approach is
that the sky is filled in less efficiently with gaps in regions devoid
of bright stars. This introduces a slight bias towards bright stars
owing to the effects of star clusters and multiple-star systems.
Galaxy positions are not correlated
with those of bright stars so they are unbiased by this procedure.

Reference stars were selected from COSMOS scans of U.K. Schmidt plates  in
equatorial fields well seperated around the sky in RA. These were chosen either
from Kapteyn's Selected Areas (Kapteyn 1906) or from areas which were to be
surveyed  with the {\em LDSS-1\/} multislit spectrograph by Colless \etal.
(1990), obtained by private communication. Since the UKIRT mosaicing software
limits the maximum offset from a mosaic centre to
$\pm 500\asec$ the areas used were
further subdivided into $10\amin\times 10\amin$ zones, either by gridding the
Selected Areas or using the {\em LDSS-1\/} sub-field centres.

The candidate reference stars were selected from the plate to
a faint limit of $R=17$--18, and a bright limit 1--2 magnitudes brighter
so that a similar surface density was obtained in each field. The actual
stars used were chosen to be well seperated so as to minimise the overlap
between adjacent IRCAM fields centred on them. Typically there were 10--20
of these per $10\amin\times 10\amin$ zone.

The survey was carried out in three observing runs in October 1987, 1988 and
March 1988. The limited data obtained in October 1987  were not used
in the final survey due to problems with electronic artifacts on images and
with bad
weather.  In the
October 1988 run (RA $22$--$02\hrs$) a total of 662 frames, of suitable quality
for
the final survey, were obtained covering $\simeq 366\,\sqamin$. About 50\% of
these data were obtained in photometric conditions and the rest were taken in
slight cirrus. In the March 1988 run
(RA $09$--$17\hrs$) a total of 205 survey frames were obtained
covering  $\simeq 228\,\sqamin$. In the October run the frames consisted mainly
of $2\times 2.5$ minute integrations while in the March run single $2.5$
minutes
exposures were taken, thus going less deep. The individual field centers
of the $10\amin\times 10\amin$ zones are listed in
Table~\use{Fieldcentres.tab} along with their galactic latitude
and the centres of all the individual frames,
together with exposure times and reference star field IDs,
are listed in Table~\use{EnormousWasteOfSpace.tab}.

\subsection{Data Reduction}

\tx
As with CCD data, bias and dark frames have to be subtracted from the
IRCAM frames. These frames
contain considerably more structure than they do in the optical. The dark
current ($\simeq 100\rm\,e^-sec^{-1}$) is slightly non-linear with time,
so a dark frame of the same exposure as the data is subtracted, rather than
scaling a different exposure dark. There is also a small non-linearity
in the photon detection rate, the correction for which is well determined
and applied automatically at the telescope.

In the infrared the sky is much brighter and
the fractional $\sqrt{N}/N$ shot noise consequently much smaller. This
means the flatfields have to have higher fractional
accuracy to reach this limit. Also
the colour of the sky changes due to variable OH emission lines which
significantly alters the flatfield on timescales of
10--15 minutes. Because of this, the optimum strategy was to construct
a new flatfield every 4--6 frames by median filtering the data frames
to remove objects, excluding regions around known reference stars.
Using this procedure the final stacked
frames had a rms equal to that predicted from shot noise (from
sky level as measured off the frame) plus readout noise which
implies a flat-field accuracy of $\ls 5\times 10^{-4}$.

Once the flatfielding was complete a fixed set of bad pixels was removed
by interpolation.
IRCAM also suffers from intermittent ``hot''
pixels which vary in sensitivity and location from frame to frame.
These were removed in the image
detection stage of the reduction (see Section \use{ImagDet.sec}).

\subsection{Image detection}

\tx\sslabel{ImagDet.sec}
Image detection was done, following  Tyson 1988, by setting a threshold
above the sky background and joining up pixels higher than this
into discrete objects. Those objects with area greater than a certain
number of pixels are selected. Experimentation showed a threshold
of twice the global frame rms and an area cut of 3 pixels
to be optimal in selecting objects but rejecting bad pixels.

Because of the large pixel scale it was not found necessary to pre-smooth
the images before object detection.  However
some of the IRCAM images were slightly vignetted resulting in
a background gradient across the chip which was removed by applying
a low-pass  $17\times 17$ pixel median filter. This also removed
some background patterns due to the electronics (see Section
\use{MultCol.sec}).

\subsection{Photometry}

\tx\sslabel{IRphotom.sec}
It was decided to adopt a fixed aperture magnitude scheme for the
photometry, this having the advantages that it is simple to define
and easy for other observers to check.
Deriving total magnitudes using a growth curve and extrapolating to
infinite aperture is problematic owing to
sky-noise and contamination from other objects. Provided
the aperture is more than a few arcseconds the correction will be small
for point sources and the magnitude is well-defined and seeing independent
for galaxies.

The other common scheme, using
isophotal magnitudes, systematically  underestimates the total
flux of faint objects as the area above the threshold  isophote gets
smaller with magnitude. Another systematic effect which partially compensates
for this
is that an isophote can only include {\em positive\/} and never negative sky
noise. However isophotal magnitudes are difficult to check or interpret as,
unlike fixed aperture magnitudes, the position of the  isophote varies
depending on the noise.

Two sets of standard stars were used to calibrate the photometry. The UKIRT
standard star list provided a list of commonly used bright ($K\simeq
6\hbox{--}7$) standards. In addition, fainter
stars from Leggett \& Hawkins (1988) and from
Hawkins (private communication, 1988) were used which provided
stars in the North and
South Galactic Poles and Hyades cluster areas. These have
magnitudes in the range $K\simeq 10$--14 and using them meant that
the linearity of the detector could be checked to faint magnitudes.

In order to determine the aperture
size to be used for the photometry
extinction curves of zeropoint against airmass were plotted for a
series of increasing apertures (4\,\asec, 5\,\asec, 6\,\asec, 8\,\asec,
10\,\asec\
and 12\,\asec). As the apertures increase the rms scatter about the best fit
line tends to a constant due to the diminishing effect of
small variations in the seeing.

The infrared and optical data in the October fields had a range of seeing
values
of  $1.9\pm 0.16\asec$ $1.4\pm 0.25\asec$ FWHM respectively. Adopting a 4\asec\
diameter aperture for standards and objects
and taking a Gaussian PSF from the images this gives typical
zeropoint errors of
$0.03$ magnitudes from the range of seeing involved.
In the March data the seeing was a lot worse --- as bad as
3--4\,\asec\ in both the infrared and optical data and so an 8\,\asec\
aperture was used for the photometry.

The data taken in non-photometric conditions had to be calibrated against the
photometric data using the reference stars. In
some cases these  were not bright enough
to do this accurately as the original Schmidt plate selected stars were
too blue to be bright in $K$.

To calibrate these data it was decided to interpolate the zeropoints from
neighbouring frames with brighter reference stars.
The zeropoint was determined for each frame where the
reference star was brighter than $10\,\sigma$ using the median magnitude
of the star from photometric nights.
Next, the zeropoints were smoothed by taking the median value of each
group of 5 consecutive observations to reduce the noise further.
The typical scatter of the zeropoints about these median values
was $\simeq 0.1\,$ magnitudes as expected, thus the final zeropoint
is good to $0.1/\sqrt{5}=0.04$ magnitudes.
Finally, the zeropoints of frames with reference stars fainter than
$10\,\sigma$ were interpolated from the smoothed median values.

\subsection{Mosaicing}

\tx
Once the zeropoints were determined,
up to 8 sub-images had to be mosaiced together --- $2\times
2.5$ minute exposures  of each $2\times 2$ mosaic.
Using the reference star centroids, the sub-images were registered to
the nearest pixel. Higher accuracy was found not to be necessary as
the seeing was typically rather greater than 1 pixel.
The resultant flux in each mosaic
pixel was computed as the sum
of the counts in each sub-image weighted by the image rms and scaled
according to the difference in zeropoints.

Image detection and photometry for the mosaiced images
was similar to that for the individual sub-images.
One complication is
that some portions of the mosaic with different overlap factors have different
amounts of noise and this had to be allowed for in the image detection, though
this was not so important for the photometry.
This was done by diving each image by a ``mask'' image, whose pixel values
reflected the local noise average and then using the result for the image
detection. The photometry was performed on the
original image.
The mask was constructed by replacing each pixel with the square root
of the median value in a local box of the
squared flux --- because the background is
subtracted and the mean is zero this forms a robust estimator of the rms.
It was found by experimentation that a $9\times 9$ box worked best and was
least
sensitive to objects while correctly reflecting the local rms.
The typical rms on a twice-covered (i.e. 5 minutes total exposure)
portion of a mosaic was $\simeq 25\,\electron$ which was equivalent to a
$3\,\sigma$ limiting magnitude of $K=17.5$ in a 4\,\asec\ aperture.

As well as the overlaps internal to the 4-mosaics
there was a small amount
of overlap between some
adjacent 4-mosaics which resulted in the same object
appearing twice in the final catalogue.
To remove these duplicates (3\% of the objects),
objects within 3\,\asec\ of a given
object  on a {\em different\/} 4-mosaic were deleted from the
catalogue. It should be noted that
this means that pairs would not be detected across 4-mosaic boundaries.

\subsection{Astrometry}

\tx\sslabel{IRastro.sec}
The objects on each IRCAM image were matched up with objects detected on
COSMOS scans of UKSTU
plates to derive the coordinate transformation between the two systems.
As there were few objects on each IRCAM image only a 4 parameter orthogonal
transformation was used (rotation, scale, x and y offset). This is non-linear
but for small angles and scale changes is approximately linear, so a least
squares solution was fitted to the residuals and iterated.

Once the rotation and scale had been determined for each individual frame
with sufficient objects, global median values were determined for these
parameters and the fits repeated to determine the best fit offsets.
The rms position residuals which resulted from this process were typically
$0.3$--$0.7\asec$ as expected from the typical measurement accuracy of the
UKSTU plates and the 1 pixel registration.

\section{CCD Imaging Observations}

\tx\slabel{CCD.sec}
Optical CCD images were made of almost all the fields surveyed in the
infrared. This allows the study of the colours of the infrared-selected sample
and the relation to optically-selected samples.
Also optical imaging allows the possibility of more reliable star-galaxy
separation based on the relative sizes of star and galaxy profiles
(see Section \use{StarGal.sec}).

\subsection{Observations}

\tx
All the CCD observations were performed on the 2.5\,m Isaac Newton Telescope
(INT) on La Palma over the period in April 1988, September 1988 and April 1989
using the RCA $512\times 320$ chip at a $0.74\asec$ pixel scale.

Because of the high density of IRCAM fields in the October data it was decided
to cover the 9 existing $10\amin \times 10\amin$ fields using an overlapping
$2 \times 4$ grid
pattern.
In contrast the March fields had a much lower density and the optimum strategy
was to
place the CCD so as to maximize the number of IRCAM fields. The
overlaps were still quite large and could be used for cross-calibration.
Additionally a
number of  positions with  short exposures and large overlaps were taken
in photometric conditions  to facilitate calibration.

Complete coverage of all the IRCAM data was obtained in $R$, in addition
complete $B$ coverage was obtained in the March fields ($V$ and $I$ were also
observed for a limited subset of the March fields). A few of the October fields
were also covered in $B$. Note only the field 851STARS\_2 did not have any
optical
coverage, if this field is excluded the infrared survey area is
$551.9\,\sqamin$.
Table~\use{CCDint.tab} summaries the magnitude
limits reached and the equivalent noise levels.

The images were reduced following standard CCD procedures. They were bias and
dark frame subtracted and then divided by a flatfield. A master flatfield was
constructed for each night by median filtering the normalized data frames in
each band and the photon shot noise limit
was reached. The RCA chip had one bad column which was interpolated over.

\begingroup

\def\Bccd{\hbox{$B_{\rm CCD}$}}
\def\Rccd{\hbox{$R_{\rm CCD}$}}
\def\Blan{\hbox{$B_{\rm LAN}$}}
\def\Rlan{\hbox{$R_{\rm LAN}$}}

Standard stars  were selected from the list of Landolt (1983)
to match the colours
typical of faint galaxies. Fitting a linear extinction law gave a rms residual
of $0.01$--$0.05$ magnitudes for photometric nights, and the values of the
absolute zeropoint (extrapolated to zero airmass) and
extinction were consistent between
nights and runs.
A substantial colour term between \Blan\ and \Bccd\ was found,
but not in $V$, $R$ or $I$. The relation defined by:

$$ \eqalign{ \Blan &= \Bccd - \beta (B-R)_{LAN} \cr
             \Rlan &= \Rccd \cr}$$

where $\beta$ was consistent between nights and observing runs and has a mean
value of $-0.082$. Tabulated magnitudes of the catalogue objects are given as
\Bccd\ as not all objects were paired in $B$ and $R$.

\endgroup

\subsection{Image Detection and Photometry}

\tx
The image detection proceeded exactly as described in Section \use{ImagDet.sec}
for the infrared images. Because of the smaller pixels (0.74\asec)
the object profiles were better
sampled, so fewer real objects were lost and
fewer spurious features were picked
up as verified by visual checking of the images. Some of the frames
suffered from background variations induced by the presence of dust on the
camera window.  These were effectively removed by
subtraction of a $20\asec \times
20\asec$ median filtered frame.

Aperture photometry was performed on the objects using a 4\asec\ diameter
aperture for the October data and 8\asec\ for the March data and object
catalogues were constructed for each image. These were then paired with Schmidt
plate data to obtain the rotation and scale transformations as described in
Section \use{IRastro.sec} for the infrared data.

\subsection{Cross-calibration of CCD data --- a new non-iterative technique}

\tx
As only 30\% of the nights were photometric it was necessary to cross-calibrate
large parts of the photometry via objects in common between overlapping
frames.  It was found that the most robust method, of determining the magnitude
zeropoint offset was to take the median magnitude difference of the brightest
10 objects, matched with a 3\asec\ tolerance.  Using a median made this
insensitive to the presence of $\ls$1--2 matching errors per frame.  Cross
checking photometric data it was found that the typical shifts were of the
order $\pm 0.05$\,mags between frames.

For the October fields, which had a simple $2\times 4$ grid arrangement of
frames,
the non-photometric images were matched up with adjacent frames according
to the arrangement in the grid.
The zeropoint errors of each
matching were found to be of order $0.05$ magnitudes. Not every
frame was immediately adjacent to a photometric frame. The maximum number of
frames matched across to calibrate a particular image was 4 in a few extreme
cases --- this would give a systematic error of 0.1\,mags in the zeropoint at
most.

The geometry of the March fields was considerably more complex so a more
general method of cross-calibrating data was devised.
This involves allowing the zeropoints of all the non-photometric frames to be
free parameters and then finding a least-squares solution for them which
minimizes the magnitude overlap residuals via a matrix inversion.

\begingroup

\def\ds{\displaystyle}
\def\thetaij{\theta_{ij}}
\def\thetajk{\theta_{jk}}
\def\wij{w_{ij}}
\def\wjk{w_{jk}}
\def\Deltaij{\Delta_{ij}}
\def\kronij{\delta_{ij}}

Consider $n$ frames of which $(1 \ldots m)$ are uncalibrated and $(m+1 \ldots
n)$ are calibrated and let
$\Deltaij$ be the magnitude difference between frames $i$ and $j$.

$$\Deltaij = \rm \left<Mag\,_i-Mag\,_j\right>_{pairs}
\quad\quad(\hbox{Note:\ } \Deltaij=-\Delta_{ji})$$

Let $a_i$ be the floating zeropoint of frame $i$, with $a_i = 0$ if $i>m$.

If we define the overlap function:

$$ \thetaij =
\cases{ = 1, &if frames $i$ and $j$ overlap \cr
        = 0, &if no overlap \cr
        = 0, &if $i=j$ \cr} $$

Then the sum of squares to be minimized is:

$$ S = \sum_{i=1}^n \sum_{j=1}^n \wij\thetaij (\Deltaij + a_i - a_j)^2 $$

where $\wij$ are the weights used for each match. In this case they were set
equal to unity although they could, for example, be set according to the errors
on
each overlap.

Differentiating with respect to $a_i$ gives the matrix equation:

$$ \ds \sum_{j=1}^m A_{ij} a_j = b_i $$

where

$$ \eqalign{
   A_{ij}     &= \wij\thetaij - \kronij  \sum_{k=1}^n \wjk\thetajk \cr
   b_i        &= \sum_{j=1}^n \wij\thetaij\Deltaij   \cr} $$

This gives a single-step solution for the $a_i$.
It is not required that all the CCD frames
be contiguous. If there are disconnected groups
of frames then the matrix becomes block-diagonal and the different solutions
become independent of each other. If there are uncalibrated frames which do not
join up in any way onto calibrated frames then the matrix becomes singular so
it
cannot be inverted. These frames have to be removed from the list.

The overall rms residual ($\epsilon$) of the solution is given by:
$$ \epsilon = {\ds\sum_{i=1}^n \sum_{j=1}^n \wij\thetaij(\Deltaij+a_j-a_i)^2
   \over \ds\sum_{i=1}^n \sum_{j=1}^n \wij\thetaij }$$

\penalty-2000

A simple example is useful in order to clarify this technique. Consider the
arrangement of CCD images shown in Fig.~\use{overlap.fig},
where the $*$ marks the calibrated images.
The matrix equation in this case is:

$${\def~{\phantom{-}}
   \pmatrix{ -2 & ~1  & ~0 & ~0 \cr
             ~1 & -2  & ~0 & ~0 \cr
             ~0 & ~0  & -1 & ~1 \cr
             ~0 & ~0  & ~1 & -2 \cr} }
\pmatrix{a_1\cr a_2\cr a_3\cr a_4\cr} =
\pmatrix{\Delta_{12}+\Delta_{16}\cr
         \Delta_{21}+\Delta_{26}\cr
         \Delta_{34}\cr
         \Delta_{43}+\Delta_{45}\cr} $$

which results in the solutions:
$$ \eqalign{
 a_1 &= {2\over 3} \Delta_{61} + {1\over 3} (\Delta_{21}+\Delta_{62})\cr
 a_2 &= {2\over 3} \Delta_{62} + {1\over 3} (\Delta_{12}+\Delta_{61})\cr
 a_3 &= \Delta_{43} + \Delta_{54} \cr
 a_4 &= \Delta_{54} \cr} $$

These results make intuitive sense. In the
left-hand case the
offsets are linear combinations of the offsets along two different paths from
the calibrated image. In the right-hand
case there is only one path and the offsets
are just the sum of the individual offsets along that path. The
matrix has decomposed into block-diagonal form as the two groups of frames are
disconnected.

The rms magnitude residual on each image from {\em different\/}
overlaps can be computed as:
$$ \hbox{rms}\,_i = {\ds\sum_{j=1}^n \wij\thetaij(\Deltaij+a_j-a_i)^2
   \over \ds\sum_{j=1}^n \wij\thetaij }$$

The typical rms residual was found to be $0.02$--$0.04$ magnitudes, depending
on band. Histogram plots made for each band confirmed that there was no
problem with anomalously large values.

In conclusion it appears that the zeropoint errors on the non-photometric data
have been determined to $<0.05$\,mags by two similar methods in the October
and March fields.

\endgroup

\subsection{Multicolour Catalogues}

\tx\sslabel{MultCol.sec}
Multicolour catalogues were made by successively pairing the infrared
catalogue with the optical catalogues for the different bands, always based
on the infrared position. As the nearest object to the infrared position
was taken this automatically removed objects multiply covered
in the CCD data. The final paired infrared-optical magnitudes are shown in
Table~\use{Catalogue.tab}.
For the October data the optical coverage was complete in $R$ (except for
851STARS\_2) but only
partial in $B$.  Out of the 576 infrared objects, 167 of them (the ``$K$-only''
objects) were detected solely in the $K$-band and not in $R$ or $B$;
for some reason there were infrared detections which were not present at
$R<23$.
Of course if these were genuine galaxies they would have to be very red
($R-K\gs 6$) and so they would be very interesting as candidate protogalaxies
(see Introduction).

The images were examined visually to try to ascertain whether these were due
to any obvious flaw in the reduction procedures. Most of the
$K$-only objects ($\simeq 95\%$) were found {\em not} to be real,
but comprised a variety of image artifacts:

\ic=0

\itm Hot pixels manifest themselves as spikes consisting
of usually 1, but occasionally a group of 2--3,  bright pixels.
Despite the 3-pixel area cut
in the object detection many of them still make it into the catalogue.
Some frames seemed to have large-scale correlated
noise streaks caused by problems with the readout electronics.
These can be broken up
into a large number of spurious objects by the image detection
algorithm. At the detection stage this was partially compensated for by the
subtraction of the median-filtered sky background. The worst images were not
used.

\itm Due to the faint isophote scheme used in the
object detection a number of close objects would be separately
detected at $K$ but not in the optical. This is because for bright $K$
objects would be many magnitudes above the sky in the optical,
so merging into a single $R$ detection leaving one surplus $K$
detection.

\itm The other major cause of artifacts is multiple reflections within the
camera
which allows bright stars to produce secondary ghost images. This can be
checked
by re-observation with the image on a different portion of the detector. If it
was a ghost it will not have moved to the correct place.

{}From the most promising ``$K$-only'' objects seven were re-imaged in
service observing time to check on the reality of the detections,
--- none of these objects were re-confirmed so it appears most likely
that none of the $K$-only detections correspond to genuine objects.

Given this knowledge, for the March data it
was decided to cut deeper into the
signal/noise in object detection
and avoid the extra spurious objects by tightening the optical
pairing constraint. The
final catalogue was generated with detection thresholds $>2.5\sigma$
(equivalent
to $K=17.3$ in a 4\asec aperture) and
2\asec\ pairing tolerance and contained 663 objects of which 281 were detected
in
at least one of the optical bands. If the catalogue is restricted to $5\sigma$
then there are 312 objects of which 195 were paired in at least one optical
band. 74 of these $K$-only objects were checked visually (the other 43 being
due
to the lack of optical coverage in one field) and only 4 emerged as candidate
objects, the other being ghosts, hot pixels, etc.

It is possible to get an upper limit to the contamination of the paired
objects, if it is assumed that all the $K$-only objects are spurious. Given a
$R$-band optical density of $\simeq 8$ objects$/\,\sqamin$ down to the flux
limit, and the pairing box size, then the expected number of spurious $K$ pairs
is
$\simeq 0.04\times$ the number of spurious $K$-only objects. Thus in the
March field catalogue (256 $(K,R)$ pairs) there are $\simeq 15$ spurious {\em
pairs}. This can be checked by increasing the pair box to $3\asec$, as the
positional accuracy is good to 0.5\asec\ then the increase in $(K,R)$ pairs
should be predominately due to the random matching up of genuine $R$ objects
with spurious $K$-only objects. Thus $\simeq 19$ extra pairs are expected whist
17 are found, which confirms the estimate of the contamination rate.

\subsection{Star-Galaxy Classification}

\tx\sslabel{StarGal.sec}
Automated classification of objects into stars and galaxies was
performed using a
technique  based on image size.
The area, in pixels, above the
threshold isophote was determined for each catalogue object. At bright
magnitudes
galaxies cover a much larger area than stars, so  a histogram  of objects
against area is strongly bimodal. At fainter magnitudes
the galaxy peak moves down in area and merges with the stellar peak as the
galaxies become unresolved.
For each object, with magnitude $M$ and on frame $j$
a parameter $y$ was calculated:

$$ y \equiv \log_{10}\left[\hbox{Object area}\right] -
       \log_{10}\left[A_*(M,j)\right] $$

where $A_*$ is the averaged area of an object on the stellar locus at that
magnitude on the same image. Since the point-spread-function varies from
frame to frame this locus is calculated separately for each one.

The star-galaxy classification was performed using the paired $R$-band CCD data
rather than directly on the $K$ data as each CCD image had a much larger
numbers
of objects per frame (several hundred as opposed to 1--2) which made the
determination of the stellar locus less noisy. The CCD data also had a smaller
pixel scale and generally  better seeing.

The procedure adopted  was to read in all the optical objects ({\em not\/} just
those paired with {\em infrared} objects) and bin each image in magnitude and
$\log_{10}\left[\hbox{image area}\right]$. Next the binned data were smoothed
slightly along the magnitude axis by applying a top hat filter of width 1
magnitude. Then the stellar locus in each magnitude bin and image was
determined
from the objects with least area. It was taken as the area value of the second
smallest object if the total number of objects in the magnitude bin was $>10$
or
else the lowest object.

Fig.~\use{StarGal.fig} shows plots of $y$ vs $R$
magnitude.
Note that for $R>20$ {\em only a third of the points are plotted}
--- this is done so the structure in the plot can be more easily discerned
among the rapidly increasing number of objects.
It can be seen that there is a well defined stellar locus down to at
least $R\simeq 20$. Fainter than this the galaxies become unresolved (for
2-4\asec\ seeing) and the quantisation of area becomes apparent in the plots.

A conservative (in the sense of including stars rather than missing galaxies)
value of $y$=0.3 was chosen to separate stars from galaxies. Additionally in
the
final selection all objects which were saturated on the $R$-band CCD image were
classified as stars because galaxies do not have a high enough central surface
brightness to saturate for these observations.

Though this procedure is crude and rather semi-empirical it does serve as a
useful measure to ensure that not too many stars are observed in subsequent
spectroscopy at bright magnitudes ($R<20$). It is not intended to be, and does
not need to be an accurate measure, as the cut is conservative and a
representative fraction of objects classified as ``stars'' were included in the
spectroscopy in any case as a check.

\subsection{Summary Final Catalogue}

\tx
A total area of 552\,\sqamin\ (after correcting for overlaps) was surveyed with
826 IRCAM images.  Using a median filtering technique the images were
flatfielded to $\ls 1$ part in $10^4$ and all the frames were essentially
limited by sky shot and readout noise. The limiting depth was $K=17.5$
($3\,\sigma$ in a 4\,\asec\ aperture) over 64\% of the area and $K=17.1$ over
the rest. The $1\,\sigma$ surface brightness limit of the survey was thus
$19.7$--$20.1\,\rm mags/\,\sqasec$.

634 objects (including 179 reference objects) were detected and matched up with
optical images.  These are listed in Table~\use{Catalogue.tab} together with
the optical magnitudes and star-galaxy classification derived in section
\use{CCD.sec}.  Using overlaps between photometric and non-photometric data the
magnitude zeropoints on all the frames were determined to $\ls 0.1\,$mag and
photometric catalogues were constructed with a positional accuracy of $\simeq
0.5\,\asec$.

\section{K-band star and galaxy counts}

\tx
Galaxy counts are a classical tool
of cosmology. The determination of the 2\,\micron\  counts is important
because,
in principle, they are much less sensitive to the evolution of stellar
populations than the optical galaxy counts.
This allows them to be used as a direct test of cosmological models as well as
to probe galaxy evolution. There has been little work done in
this area until recently, because  the recent arrival of infrared array
detectors only now allows a large area to be surveyed quickly.

There are several important points addressed directly by the $K$-band
counts:

\ic=0

{\pretolerance 10000

\itm {\bf Geometry.}\quad
The insensitivity to evolution means the
counts are an excellent probe of the geometry of the Universe. It may
be possible to obtain new constraints on $\Omega_0$, and also
test non-standard world
models such as those including the Cosmological Constant.

}

\itm {\bf Normalization.}\quad
One problematical issue in galaxy count studies
is the true local space density of galaxies. The
range of estimates from differing surveys
typically vary by a factor of two (Efstathiou \etal., 1988).
For number-magnitude count predictions this
normalization is usually allowed to float --- instead the models are normalized
to the data at some intermediate magnitude. This of course affects the size of
the excess over no-evolution and, hence, the amount of evolution required. The
problem is that if the models are normalized at faint enough $B$ magnitudes to
avoid problems from local fluctuations in galaxy density, then one is
sufficiently
deep that spectral evolution is already important. As there should be less
spectral
evolution in $K$ it would be possible to obtain a more reliable normalization,
on
larger scales, and transfer this to the $B$-band to determine just how much
evolution is actually needed.

\itm {\bf Extinction.}\quad Light at
$2\micron$ is  relatively
insensitive to dust extinction  --- $A_K\simeq 0.1 A_V$. And unlike longer
{\em far-infrared\/} wavelengths,
such as those surveyed by the IRAS satellite, there is little thermal
{\em emission\/} from warm dust $\rm\ls 100\,K$ as seen in many
starburst galaxies (Rowan-Robinson \& Crawford 1991).

\itm {\bf Luminosity Evolution.}\quad
The $K$ data should allow the quantification of just how much
luminosity evolution
in the $B$ light of galaxies, is allowed. This should manifest itself in a
greater excess over no-evolution in the $B$ number-magnitude counts than in
$K$,
and, equivalently, in the evolution of the \BK\ colours. Also by selecting
galaxies in a band insensitive to star-formation it should be possible to
find out the true fraction of objects with strong emission lines such as \OII,
which make up such a large proportion of the Broadhurst  \etal. (1988) and
Colless
\etal. (1990) samples.

\itm {\bf Density evolution.}\quad
The $K$ data also provide a probe of
the amount of evolution
in the space density of galaxies. This will cause an excess in the
number-magnitude counts over no-evolution which, unlike luminosity evolution,
would be the {\em same\/} in $B$ and $K$.

\subsection{Observed $\bf K$-band counts and corrections.}

\tx\sslabel{RawCounts.sec}
The raw counts of galaxies and stars are listed in
Table~\use{IRNumCounts.tab}.  In total there
are 287 galaxies and 167 stars in the all the fields.
These are counts for objects paired in the $K$ and
$R$ bands; as is discussed in  Section \use{MultCol.sec} there is no evidence
for any of the unpaired $K$ detections being real. The counts are given
in bins of 1 magnitude, the completeness is discussed below.
Poisson errors are assumed for these --- the real error will, of course, be
somewhat larger due to the effect of clustering but this is not large
compared to the Poisson error. This is discussed
further in section ~\use{clust.sec}).

The objects used as astrometric and photometric
references in the  construction of the infrared survey (column
``R'' in Table~\use{Catalogue.tab})
present a significant difficulty in the count
analysis. Clearly, as a result of using references in this
way, the survey is biased to contain too many bright
objects. However, it is not correct simply to exclude these
objects from the analysis: in the limit that the mosaic survey
became a completely filled area, all the reference objects
would clearly be retained as genuine survey members.
In general, for a survey with a filling factor $f$, it is
obvious that the the reference objects should be given a
weight which is on the order of $f$ ($\sim 20$ per cent
in our case). We now show how to make a precise estimate
of the weight to apply in practice.

For this analysis, we need to know three things:
(i) the surface density of candidate reference objects,
from which we can estimate the number of objects of this
sort expected in an unbiased survey, $\bar N$;
(ii) the number of such objects actually used as references, $N_R$;
(iii) the number of such objects which were serendipitously
included in the survey, in addition to the preselected references, $N_S$.
The correct weight to give to the reference objects is then clearly
just the fraction of the reference objects needed to supply the
difference between the expected number of reference-type objects
and the number included serendipitously:
$$
w={\bar N - N_S \over N_R}.
$$
The relevant numbers are as follows: the surface density of
candidate references gives $\bar N=218$.
$N_R$ was 207 and $N_S$ was 144, yielding an estimate of
$w=0.36$. The correction must be applied using this value of
$w$ for all $K$ bins, although the actual
value of the correction to each $K$ bin
obviously depends on the number of reference stars and
galaxies in the bin, which is listed
in Table~\use{IRNumCounts.tab}.

The magnitude of these upward corrections is highest
at intermediate $K$ magnitudes as expected from the selection
criteria used for the stars. It peaks at $+2.2\sqrt{N}$
(where $\sqrt{N}$ is the random Poisson error) in the
brightest bin containing any galaxies ($14\le K<15$) - this is
where the most reference objects lie compared to the rapidly diminishing
number of field galaxies.
In all other bins the systematic reference object correction is $\ls$
the random error so the effect on the number-magnitude relation
is small.

Star-galaxy separation is vital to the counts as stars outnumber galaxies for
$K<16$.  The reference stars provide a useful independent check on the $R$
CCD star-galaxy classification as they  were independently classified by the
COSMOS image analysis software based on the Schmidt plate data. There were  37
``galaxies'' out of 116 reference objects in the October fields, of which
$3/21$
were later identified spectroscopically as stars. In the
March fields the galaxy fraction was 11/63; none of these ``galaxies'' were
examined spectroscopically. It appears that the misclassification rate of the
CCD star-galaxy separation is consistent between the reference objects and the
rest (see below), and that the COSMOS classification is much less reliable than
the CCD classification.

The CCD classification
failure rate can be estimated from early spectroscopic results, to
be fully presented in Paper II.
We can define parameters $\alpha$ as the fraction of
``stars'' which turn out to be really galaxies and $\beta$ as the fraction of
``galaxies'' which turn out to be really stars. For the entire identified
spectroscopic sample ($R\ls 21$) $\alpha = 4\pm 2\%$  and $\beta = 20\pm 6\%$,
averaging over all $K$ magnitudes, the errors
being based on Poisson statistics. $\beta$ is much larger than $\alpha$ because
a deliberately conservative star-galaxy cut was chosen for completeness
reasons,
i.e. to include stars rather than lose galaxies.
The effect of this error on the normalization of
the galaxy counts is about half that of the Poisson errors in each magnitude
bin. However what is most important is the {\em differential} change in
star-galaxy misclassification with $K$ magnitude --- the slope of the galaxy
relation is much more important than the normalization. This misclassification
can be corrected for by using the relations:

$$\eqalign{ N'_g &= N_g(1-\beta)  + \alpha N_s \cr
            N'_s &= N_s(1-\alpha) + \beta  N_g \cr }$$

where $N_s$ and $N_g$ are the original star and galaxy numbers in each
magnitude
bin and $N'_g$ and $N'_s$ are the corrected values (note $N'_s+N'_g=N_s+N_g$ of
course). The errors $\Delta\alpha$, $\Delta\beta$,  $\sqrt{N_s}$ and
$\sqrt{N_g}$ can all be folded in to give errors on $N'_g$ and $N'_s$ which are
now slightly in excess of their Poisson values. (In the figures that follow the
corrected error bars appear smaller. This is not in fact the case as this is a
log plot: the actual absolute errors increase).

$\alpha$ and $\beta$ should be strong functions of $R$, the band used to do
the separation. To correct the counts we need estimates of average values
in the different $K$ ranges.
It can be seen from Fig.~\use{StarGalvsK.fig}, which shows the classification
image-area parameter $y$ against $K$ magnitude, that the star and galaxy loci
are
well separated down to $K\simeq 15$ and that the misclassification should
apply mostly in the $K>15$ regime. In the spectroscopic sample 12 out of 15
of the misclassified objects had $K>15$. For the final corrected counts all
objects with $K<15$ (94) were inspected visually to check for errors and
only 4 were found. The $K>15$ bins were corrected using the values
$\alpha=7.7\pm 7.7\%$ (1 object) and $\beta=23.3 \pm 8.8\%$ derived from the
spectroscopy in this magnitude range.
These values are consistent with the $K<15$ values, and those
for the entire sample, indicating that any systematics in $\alpha$ and
$\beta$ are of lesser importance than the random error, which is
incorporated into the final error.

Another issue to be considered with regard to the reference objects is
the effect of galaxy clustering --- since 27\% of our reference objects
are unintentionally galaxies clustering will introduce an excess of
nearby objects. This can be easily checked by comparing the raw counts
in fields with and without galaxies as references --- this is shown in
Fig.~\use{RefCounts.fig}. It can be seen there is no significant clustering
effect.

Table~\use{IRNumCounts.tab} shows the final counts of stars and galaxies
after applying these corrections, together with the propagated errors.
These counts are plotted in Fig.~\use{RawCounts.fig} which compares
stars and galaxies.

There are also, in principle two more corrections to be made:

\ic=0

\itm The true counts are mutiplied by a selection function which represents the
probability of detection as a function of magnitude. Because the objects are
$2\,\sigma$ detections in at least 3 pixels this produces a set of peak flux
selected objects.

\itm Once the objects are detected then the magnitude is measured in a larger
aperture introducing a magnitude scatter that introduces a Malmquist-like
correction (Eddington 1940) that biases the observed distribution away from
the true one.

Because our detections have high significance we expect the size of these
corrections to be small for $K<17$. Because they are small we can estimate
their amplitude by using the noise and detection parameters for the different
parts of the survey and making a rough estimate for the number-magnitude
relation fainter than $K=17$.

At $K=17$ the $\log$ number-magnitude slope is close to $\simeq 0.45$; if we
allow a generous range of slope from $0.3$ to $0.6$ we get a range of
integrated correction for the final $16<K<17$ bin from $+2\%$ to $-1\%$. Since
these corrections are negligible compared to the random error we do not apply
them. For $K>17$ the incompleteness rapidly becomes very large as the survey
limits are approached and we do not plot the counts at these magnitudes.

The final corrected star counts are also shown in Fig.~\use{StarModel.fig}
split according to the different galactic latitudes in the survey.
Although the reference stars have been
allowed for it should be born in mind
there could still be a residual
bias in these counts from stellar binaries
and clusters.
Nevertheless, the star counts can be compared with the model of
Bahcall \& Soneira (1980), these were
modified to the infrared by I.N. Reid (private
communication, 1990). Two models are shown both consisting
of various stellar components:

\ic=0

\itm ``Normal model'' ---
an old disk stellar with age 2--10\,Gyr, scale height of
350\,pc, luminosity function from Reid (1987) and weight at zero scale
height ($w$) $=0.80$.
Plus an intermediate age disk (0.3--2\,Gyr, 250\,pc, $w=0.17$),
plus an old extended thick disk (2--3\,Gyr, 1000\,pc, $w=0.015$) and
plus a halo component
with local density 0.15\% of the disk and disk-shaped luminosity function.

\itm ``Extended flat LF'' ---
the same as (1) except that the luminosity function is assumed to
be flat beyond $M_V\ge+13$ (Dahn \etal., 1986) which gives the
maximum possible contribution of low-mass late-type red M-stars.

These are also shown in Fig.~\use{StarModel.fig} and it can be seen
that in most cases
the agreement is good though the normalization appears to be
slightly too high in some cases. This is worse in the 09\hrs\ field which is
also
the field at lowest galactic latitude. Though clearly these first direct $K$
star counts
have insufficient objects to distinguish these two models it is encouraging
that they agree so well with predictions made by an extrapolation of optical
stellar
data.

\section{Constraints on Geometry and
Galaxy Evolution from the K Counts}

\tx
Fig.~\use{Knoev.fig} shows the $K$-band counts determined from this survey
together with those
of Cowie \etal. (1990) and Cowie (1991).  Cowie \etal's strategy is
complementary
to ours as their program was to go very deep ($\sim 50$ hours integration) in
tiny
fields ($\sim 4\,\sqamin$). Together, our surveys define the galaxy counts very
well over 8 magnitudes. Cowie \etal. perform star-galaxy separation based on
colours,
but their stellar contribution  is not significant in this magnitude range:
the star counts reach a plateau at $K=16$ while the galaxies continue to
increase.
Also shown is the
number-magnitude determination of Jenkins \& Reid (1991) based on a
confusion limited survey of
113\,\sqamin\ to $K=19$ using the single channel photometer UKT9 on UKIRT
with a 19.6\asec\ diameter aperture.  Jenkins \& Reid analysed the flux density
distribution and fitted a parametric form to $n(m)$ (after allowing for
predicted star counts), $\pm1\sigma$ errors being determined from Monte-Carlo
simulations.

Their final results, with errors, are shown in Fig.~\use{Knoev.fig} as dotted
lines. Note there is a very satisfactory agreement between the 3 independent
determinations in the slope. The survey area  of
Jenkins \& Reid is 113\,\sqamin, this is distributed over 11 widely
separated patches of sky in an attempt to get a fair sample. The work presented
here is based on 552\,\sqamin\ and is almost equally  distributed over 6
patches
spaced around the celestial equator at high galactic latitude.
For the present purposes the normalization
based on the larger area will be adopted.

The optical galaxy counts are shown in Fig.~\use{bjnoev.fig}. This is a
new compilation of the most recently available data and matches well previously
published compilations. The faint points are taken from the deep CCD counts of
Tyson (1988), Lilly \etal. (1991) and Metcalfe \etal. (1991).  At intermediate
magnitudes are the counts of Jones \etal. (1991) based on deep AAT plates. The
brightest points are taken from the two largest surveys in existence, both
based on southern UK Schmidt plate data, the 946\,\sqdeg\ Edinburgh-Durham
Southern Galaxy Catalogue (Collins \& Nichol, 1991) and the  4300\,\sqdeg\ APM
Galaxy Survey (Maddox \etal. 1990). Note these two surveys are not independent:
the EDSGC sky coverage is a smaller subset of the APM survey, analysed
independently.

The errors bars are determined from the observed field-field rms scatter where
given. All the counts have been transformed to the Schmidt $b_j$-band using the
transformations given in the references. The plate data are for isophotal
magnitudes (see references for surface brightness limits) while the CCD data
represent total magnitudes.
The agreement between the normalization
of different determinations is now of the order of $\ls 50\%$, much better than
in previous compilations of older data.

\subsection{Non-evolving prediction}

\tx\sslabel{NoEv.sec}
The first step is to compare the counts to the prediction of a model
which assumes an unchanging  galaxy
population. For this calculation two ingredients are required: the
K-correction and the local galaxy luminosity function.

The K-correction corrects luminosity distances for the redshifting
of a galaxy's spectrum through the fixed observed passband. It is
given by:
{\def\ss{\scriptstyle}
$$ K(z) = -2.5\log_{10}\left[ {1\over(1+z)}
{\ds\int_{-\infty}^{\infty} T_{\lambda} f_{{\ss\lambda\over\ss(1+z)}}
d\lambda
\over
 \ds\int_{-\infty}^{\infty} T_{\lambda} f_{\lambda} d\lambda } \right] $$}
where $T_{\lambda}$ is the bandpass transmission as a function of wavelength
and $f_{\lambda}$ is the galaxy's spectrum.
This spectrum was taken from the final epoch of the evolving galaxy
models generated by
Rocca-Volmerange and Guideroni (1988). At late times these approximately match
the optical and near-infrared colours of local galaxies and provide a
similar optical K-correction to that of Broadhurst \etal\ (1988).  In the
$K$-band the
K-corrections of the different types are very similar and the no-evolution
model
is thus insensitive to the exact mixture.

The type-dependent  luminosity function was taken from King \& Ellis (1985)
with the relative mix of morphological types calculated from that observed in
the local population ($b_j<16.75$)  by Shanks \etal.  (1984). This is
summarized in Table~\use{GalaxyMix.tab}. This luminosity function is very
similar to that used by Broadhurst \etal.
and those of Efstathiou \etal. (1988) and Loveday \etal. (1992)
though they found less type dependence. The $K$-band luminosity
function was constructed by applying model \BK\ colours to the $B$-band
luminosity function --- this is very similar to the $K$ luminosity derived from
the Durham-AAT redshift survey (R.M Sharples, private communication, 1990).

The normalization, $\phi^*$, is usually derived from local redshift surveys
along with the rest of the luminosity function, and Efstathiou \etal. (1988) do
this for five different surveys.  They gave a mean $\phi^*$ of $(1.56\pm
0.34)\phiunits$ although the dispersion among estimates from the differing
surveys amounts to a factor of two.  Additionally from a
large APM-selected redshift survey Loveday \etal. (1992)
find $\phi^*=(1.4\pm 0.17)\phiunits$. An alternative approach is to choose a
value of $\phi^*$ which normalizes the prediction to the bright end of the
counts. Note that the $b_j$ and $K$ normalisations are, in principle,  not
independent --- the same total number of galaxies
integrated down to zero flux must be seen in $b_j$ and $K$, this requires
the {\em same} value of $\phi^*$ to be used.

Taking $\phi^*=1.5\phiunits$ the $K$ and $B$-band no-evolution models are
plotted with the counts in Fig.~\use{Knoev.fig} and Fig.~\use{bjnoev.fig}.
The $b_j$ counts show
the well-known result of a huge galaxy excess compared to the non-evolving
prediction. If the curves are normalized as shown then this excess is a factor
of 3 at $b_j=22$ and 10 at $b_j=28$. It can be seen at once that the observed
$K$ counts are a much better match to the non-evolving prediction than are the
counts in $b_j$. This follows the trend seen in optical counts in other bands
(e.g. Tyson   1988) where the galaxy excess is greatest in $b_j$ and
progressively less in $R$ and $I$ filters.

It can be seen that changing to a non-standard world model --- e.g. introducing
a cosmological constant or using a ``Tired light'' model of the redshift
would not simultaneously rectify both the $K$ and $B$-band counts. While
increasing
the volume element would increase the number of faint $B$ galaxies,
this would have the same effect in $K$ and
thus overpredict the counts in this band.

We shall not consider the blue counts in any detail in the remainder of this
paper.
It is clear that any explanation for the excess of blue galaxies must involve
some star formation to give the galaxies additional ultraviolet output at
moderate
to high redshifts. Only a small number of high-mass stars are required to
accomplish
this. However such an event will have very little effect on the properties of
galaxies in the infrared. The $K$-band counts are therefore much better suited
to setting constraints on cosmological geometry or cosmological evolution of
the
mass function. We consider these topics below.

\subsection{Galaxy Clustering and count normalization}

\tx\sslabel{clust.sec}
So far, we have not considered the extent to which our
counts (and those of others) could be in error owing to
galaxy clustering; different areas of sky will have
counts which differ from the global average, and we need
to estimate how large these fluctuations can be.
This is relatively straightforward in principle,
given some knowledge of the power spectrum for galaxy
clustering. The process of making a magnitude-limited
survey over some area of the sky corresponds to a
convolution of the galaxy density field with some window
function. Hence, the fractional variance ($\sigma^2$) in the
resulting number of galaxies is given by an integral
over the power spectrum times the
azimuthally averaged squared transform of
the window function determined the volume sampled:
$$
\sigma^2=\int \Delta^2(k)\; {dk\over k}\; \langle |W_k|^2\rangle.
$$
We have used a dimensionless notation for the power
spectrum: $\Delta^2(k)$ is the contribution to the
fractional density variance per $\ln k$ (see Peacock 1991).
The window function is a product of radial and angular selection
functions: $W({\bf r})=\phi(r) f(\theta)$, so that the
$k$-space window is
$$
W_k= { \int\!\!\!\int r^2\phi(r)\;  f(\theta)\, e^{i{\bf k\cdot r}}
  \; dr\; d^2\theta \over
 \int f(\theta)\; d^2\theta\quad \int r^2\phi(r)\; dr }.
$$
We assume here an $\Omega=1$ model and take $r$ to be the
comoving radius, so that the comoving spatial geometry
is Euclidean.
To evaluate the necessary integrals for our exact angular
selection and realistic radial selection is a tedious
exercise. Fortunately, insight into what is happening
can be combined with acceptable accuracy in a simple
analytical model.

For the radial selection, $\phi(r)\propto r^{-1/2}\exp(-[r/r^*]^2)$
is often adopted as a reasonable approximation to the
integral Schechter function.
For the final $k$-space result, it is adequate to model
the angular selection as convolution with an angular
Gaussian: $f(\theta)\propto \exp(-\theta^2/2 R^2)$,
where $R^2\simeq A/12$ approximates the effect
of a square of area $A$ (this relation
between $R$ and survey area is almost independent of
the exact survey geometry in simple cases).
We consider the appropriate value of $R$ for our data below.
If we further assume that the angle $R$ is small, then
an excellent approximation
($\ls 1$ per cent error)
to the azimuthally-averaged value of $|W_k|^2$ is
$$
\langle |W_k|^2\rangle=
\left[ 1+(kr^*R)^2/2 \right]^{-2} \left[ 1+ (kr^*/2)^3 \right]^{-1/3}.
$$
Even for an exact evaluation of the window for a square
survey region, this formula is within about
10 per cent of the correct answer.
The above expression shows that the power is reduced by one
power of $k$ for wavelengths less than the depth of
the survey, and by four additional powers of $k$
if the wavelength is also smaller than the typical
transverse size of the survey cone. For realistic
power spectra, the dominant fluctuations will
be those where the second type of filtering starts to
become important.

We have applied the above expression to a simple
model for the clustering power spectrum, which corresponds
to a power-law correlation function $\xi(r)=(r/r_0)^{-\gamma}$:
$$
\Delta^2(k)=0.903 (kr_0)^{\gamma}.
$$
We take a slope of $\gamma=1.8$ and a normalization of $r_0=5\,h^{-1}\,$Mpc.
Empirically it appears that the observed power spectrum is
convex on large scales (Peacock 1991), so using a power law
will be conservative in the sense of overestimating
the effect of clustering on the galaxy counts.
Provided $R$ is small, the integral can be performed
to yield
$$
\sigma^2=2.36\left({r_0\over r^*}\right)
\left({r_0\over Rr^*}\right)^{0.8}.
$$

We can now put some numbers into this equation:
the bright blue counts are based on
the APM survey of 4300 square degrees, which corresponds to
$R=0.33$ radians.
Our present survey lies in 6 zones, for each of which
the area is on average $99\,\sqamin$, corresponding to
$R=8.4\times 10^{-4}$ radians.
However, each zone consists of a sparsely-filled
mosaic with a filling factor of about 0.25; it turns
out that the transform of the angular selection function
is dominated by this larger area (see
equation 4.5 of Kaiser \& Peacock 1991); we therefore
adopt a value of $R$ twice the above figure.
Furthermore, since we have 6 widely-spaced zones,
the overall rms in the counts is a factor $\sqrt{6}$
smaller than the single-field figure.
Lastly, the deep $K$-band counts are based on
4 frames each of area 1.4 arcmin$^2$,
corresponding to $R=1.0\times 10^{-4}$.
Inserting these figures gives our final results:
$$
\eqalign{
\sigma &\simeq (r^*/13h^{-1}{\rm Mpc})^{-0.9}\quad{\rm (APM)}\cr
\sigma &\simeq (r^*/51h^{-1}{\rm Mpc})^{-0.9}\quad{\rm (This\ paper)}\cr
\sigma &\simeq (r^*/224h^{-1}{\rm Mpc})^{-0.9}\quad{\rm (Cowie)}\cr
}
$$

A more practical measure of depth than $r^*$ is the median
redshift of a survey. For this selection function,
the median comoving distance is $0.968 r^*$, and so
the critical depths above correspond to median redshifts
of 0.0042, 0.017, and 0.08 respectively. More interesting are
perhaps the redshifts where the rms uncertainty reaches
about 20 per cent; this represents the
bright limit for useful conclusions given our limited
survey areas. For APM this is a median redshift of 0.026;
for our survey it is 0.11.
According to our no-evolution models, such median
redshifts correspond to magnitude limits of
$b_j=15$ and $K=13.7$.

These calculations have interesting implications
for the normalization, $\phi^*$, and for conclusions
concerning evolution in general.
We need to ignore the brighter parts of the counts
in setting the normalization: clearly the adopted
is a compromise between normalizing at low redshift
so as to avoid evolutionary effects
but wanting to be certain that the effects of clustering
are completely negligible. If we use the magnitudes
calculated above where the clustering uncertainty
is small, the implied
value of the normalization is (consistent between
both wavebands) is
$$
\phi^*=(1.5 \pm 0.2) \times 10^{-2}h^3\; {\rm Mpc}^{-3}.
$$

This agrees with the empirical value of $\phi^*$ adopted in Section
\use{NoEv.sec} and provides additional support for the values found by
Efstathiou \etal. (1988) and Loveday \etal. (1992) in optical redshift surveys.
It is encouraging that $\phi^*$ is the same in both $b_j$ and $K$, this
indicates
that our assumption of negligible evolution to the depth where clustering
becomes unimportant is valid.
The normalization is at a higher redshift
in $K$ ($0.11$ vs $0.026$ in $b_j$) but because
the slope of the no-evolution model is so
closer in $K$ to the data than it
is in $b_j$ the effects of evolution will be much
less. The normalization would also match up with that of future larger area
infrared surveys unless the $K$ count slope changed markedly
at brighter magnitudes.

Shanks (1990) and Maddox \etal. (1990) noted that the slope of the $b_j$ counts
between $b_j=15$ and $b_j=20$ was sufficiently steep that either there was very
strong evolution or a large local hole in the galaxy distribution. Loveday
\etal.
(1992) directly estimated the radial density variation from $bj=15$ to $b_j=17$
from their redshift survey and find no evidence for a hole in this range.
It appears
from the above argument that a hole is inconsistent with the observed
large-scale
power in galaxy clustering and that this is additionally supported by the
infrared data.
We can turn the argument around and ask what values of $\gamma$ and $r_0$
are required to cause a 100\% effect at $b_j=15$? It turns out that either
$\gamma\sim 0.3$ or $r_0\sim 40\,h^{-1}\,$Mpc (separately)
is required --- both of these would
increase the power to $\Delta^2(k)\sim0.3$ on scales of $\sim 80\,h^{-1}\,$Mpc
(the depth at $b_j=15$)
compared to the  $\Delta^2(k)\ls0.01$ observed (Peacock 1991).

Thus it appears that it is difficult to reconcile the steep slope of the
optical counts with observed galaxy clustering and we must consider the
possibility of strong evolution. Obviously the infrared data will
provide new constraints on evolutionary models, so it is this we must
consider next.

\subsection{Luminosity evolution}

\tx
We first consider the limits set by the $K$-counts to models of luminosity
evolution in which galaxies were brighter in the past. Lilly \& Longair (1984)
suggested from a study of radio galaxies that the brightening amounted to
1 magnitude by $z=1$. It is convenient
to parameterise the luminosity evolution
by a simple functional form.
Many schemes have been used, for example Broadhurst \etal. (1992) who choose
a function which is exponential with time. However the exact form of the
relation is unimportant, as the counts are mainly sensitive to the first order
linear term with redshift. We choose to investigate this directly using:

$$ L^*\propto 1 + b z$$

For $z<<1$ this is exactly equivalent to the functional form of Broadhurst
\etal., the higher order terms in $z^2$ etc. being small. However we are
applying this  in the infrared not the optical and do not wish to attribute
physical significance to $b$. Since we wish to investigate what minimum
departure from the no-evolution model is allowed this seems to us a  sensible
course.

In $K$, how big a value of $b$ is required to match {\em passive} evolution of
the galaxy population, as the stars get younger with increasing redshift? The 1
magnitude brightening of Lilly \& Longair   can be regarded as an upper limit
--- the effect of increased star-formation in $K$ is in the opposite sense to
the passive evolution, {\em i.e.} to {\em decrease} the mean luminosity with
redshift rather than increase it as it does in $B$.  This may seem
counter-intuitive but does make sense --- the $K$ light is approximately
proportional to the number of stars in the galaxy because the light is
dominated
by long-lived stellar populations. Passive evolution makes these stars younger
and brighter in the past, but the effect of star-formation is to reduce the
{\em
number} of stars in the past and hence make the galaxy dimmer. This was checked
quantitatively by running some simple Bruzual (1983) models --- an upper bound
of $\simeq 1$ magnitude by $z=1$ was found for a passively evolving model in an
$\Omega_0=0.1$ cosmology. Increasing $\Omega_0$ or the star-formation rate
reduces this and makes $b$ smaller.
This upper bound corresponds to $b=1.5$, and is plotted in
Fig.~\use{NMLE.fig} along with $b=0.6$ (0.5 magnitudes to $z=1$) and $b=6.3$
(1.5
magnitudes) to compare the effect of other amounts of nett positive
luminosity evolution on the number-magnitude counts.
Note that Fig.~\use{NMLE.fig} has a slope of $0.6$ (the Euclidean
slope) subtracted from the data and the models so as to reduce the number of
orders of magnitude plotted and to allow small discrepancies to be more easily
discerned.

Galaxies to $z\simeq 1$ are seen approximately to $K\simeq 18$ in these
models, so attention should be focused on this region,
rather than the deeper counts of Cowie. It can be seen that only very
mild amounts of luminosity evolution are consistent with the
data: $b\ls 0.6$ (0.5 magnitudes) for $0\le\Omega\le1$. Larger values of $b$
can
be tolerated only if normalization is regarded as a free parameter; however,
the
normalization ($\phi^*=1.5\phiunits$) is already at the bottom end of the range
of values allowed by the optical data. Thus it appears that the result of
Lilly \& Longair is not applicable to the field galaxy population.

\subsection{Merger evolution}

\tx\sslabel{MergEv.sec}
The infrared counts provide, in principle, an approximate
measure of the mass function of galaxies. As the old stars which dominate the
$K$
light evolve only very slowly,  $K$
light should trace the galactic stellar mass and be insensitive to small
amounts of star formation. The $K$ light is also insensitive to dust extinction
($A_{K}\simeq 0.1 A_{V}$) so all the light is being seen, even if the
interstellar medium is heavily disrupted by a galaxy merger (e.g. Thronson
\etal. 1990). The $K$ counts thus in principle set limits on changes with epoch
to the mass function of galaxies.

The evolution in the optical luminosity function is best described
empirically by pure number-density
evolution, {\it i.e.} evolution in the value of $\phi^*$.
(Lilly, Cowie \& Gardner 1991). This is because the
number-magnitude counts increase over no-evolution while the mean redshift
remains unevolved. Functional forms of
number-density evolution have been used to fit the optical counts, e.g.
Koo (1990) who
using $\phi^*\propto (1+z)^m$ found $m=1.5$--2.
Such models are inconsistent since what they
assume is merging plus a no-evolution K-correction. However,
galaxies with free gas
could not fail to form stars during a merger, which would change
the luminosity, especially in $b_j$. Since the faint
galaxies are indeed bluer, i.e. the excess is less in redder bands, a proper
model
of number-density evolution should seek to account for this.

At first sight, the $K$ counts strongly constrain such models:
a pure change in the number density of galaxies
quickly overpredicts the $K$ counts.  This is a
general effect --- if a particular model of pure density evolution,
which fits the $b_j$ data, is simply translated to $K$ then it produces
too many faint galaxies in that band.

If any change in galaxy number density is due to  mergers,
then it is
possible to construct a more realistic model by requiring that the
total amount of light in the $K$ luminosity function be conserved.
The key assumption
here is that
the combined $K$ light from two systems before a merger is  approximately the
same as the $K$ light afterwards;  $K$
light is thought of as tracing
stellar {\em mass}. Clearly this will be false at some level, but it
is a more natural starting point than pure number-density evolution where
one effectively assumes that the light {\em doubles}.
For a Schechter function
$$ L_{\rm TOT} = \int_{-\infty}^\infty L \,\phi(L)\, dL \propto \phi^* L^*  $$
so any formalism which conserves $\phi^* L^*$ as a function of
$z$ will do. To investigate evolution in the number density the following
parameterisation was adopted:
$$\eqalign{\phi^* &\propto 1 + Q z \cr
L^*&\propto {1\over 1 + Q z}\cr}$$

Note that this actually
implements luminosity {\em devolution} as the galaxies break up into smaller
units at high $z$. This is similar to the approach of
Guideroni \& Rocca-Volmerange (1991) except that they modelled the optical
$b_j$
counts rather than $K$
--- to get from the mass function to the $b_j$ light requires complex
models of spectral evolution and an assumption for the star-formation rate
which
may depend in a complicated way on the merger history. By modelling $K$ we
reduce
the importance of such effects and just consider the mass function (in stars).

Fig.~\use{NMME.fig} shows the results of such models
with $Q=2$, 4 and 7 and $\phi^*=1.5\phiunits$,
the normalization used before.
Note that the effect is to bend up the faint end in
such a way that the slope is a better match to the data. The normalization is a
little low, but this could be rectified as we do have some freedom to normalize
{\em up}. Note also that the number-magnitude counts are {\em insensitive to
large values of $Q$}. This is because the effect of such merging is to
stretch the model along a line of slope 0.4 in the $\log n-m$ plane --- i.e.
the
surface brightness of the population on the sky is conserved.
Because this is close to the observed count slope, a wide range of $Q$
can be consistent with the count data.

Of course such models are extremely crude: all that is specified is that
galaxies of all masses double their mass in similar times. It does not
specify how many mergers actually happen in that time --- solutions where
galaxies always merged with units of the same mass or digested
many much smaller subunits can not be distinguished.
But it does demonstrate an important
point, namely that from the count slope alone it is not easy to tell whether
galaxies at $z=1$ exist in many subunits.

The existence of radio-galaxies at high-$z$ provides a {\em lower} limit on the
amount of merging if the $K$ luminosity function evolves in a self-similar way.
Radio-galaxies have a comoving space density at $z\gs 1$ of $10^{-6}\,h^3\,\rm
Mpc^{-3}$ (Dunlop \& Peacock 1990). Their mean absolute $K$ magnitude is
$-24.0$, $\simeq 0.5$ magnitudes brighter than the $M^*$ for normal galaxies
with a scatter of $0.5$ magnitudes (Lilly \& Longair 1984). If we assume half
the radio-galaxies are brighter than the mean this places an observational
lower limit on the number density of galaxies with $M<-24.0$.

If there is too
much self-similar evolution there will not be enough galaxies of any kind
brighter than $M=-24$ to account for the observed radio-galaxies.
This is a strong constraint due to the rapid exponential fall-of in the
Schechter luminosity function at $M<M^*$.
Putting in the figures
this  constrains the self-similar evolution in $\phi^*$ and $L^*$ to be less
than a factor of $\sim 6$ at $z\sim 1$.  For our linear form this corresponds
to
$Q\ls 5$, for the exponential form of Broadhurst \etal. (1992) this is $Q\ls 4$
since this increases faster at higher $z$. This implies that if the preferred
amount of evolution in Broadhurst \etal. ($Q=4$) is allowed all luminous
galaxies at high-$z$ would have to be radio-galaxies, and radio-galaxies would
just be defined by a mass cutoff at early times. Given the peak in radio AGN
activity at $z\simeq 2$ (Dunlop \& Peacock 1990) it is certainly not
implausible to suggest that all massive galaxies would be active if we are
seeing them soon after their assembly.

A direct way to measure the evolution in the $K$ luminosity function is to
measure the redshift distribution.
Fig.~\use{NzK.fig} shows the predicted redshift distributions for
a $16<K<17$ selected galaxy sample for increasing values of $Q$.
Such merging models
predict that the mean redshift of {\em $K$ selected} samples will be {\em
less\/} than the no-evolution prediction because galaxies become less luminous
in the past. Such a comparison, with actual survey data, will be made in
Paper II.

Such a merging model can naturally explain the $b_j$ counts and redshift
distributions if there is enough enhancement to the $b_j$ light to compensate
for the decrease in galaxy mass, thus keeping the mean redshift at the
no-evolution value. This would have to be via some merger-driven starburst
mechanism, which would also explain the galaxies in the $b_j$ excess being
spectroscopically evolved (e.g. Broadhurst, Ellis \& Glazebrook (1992)).

\section{Conclusions}

\tx
We have presented a $K$-band survey complete to $K\simeq 17.3$ over
552\,\sqamin,
and have constructed galaxy counts from a sample of 481 galaxies and stars.

It is possible to explain the infrared number-magnitude galaxy counts in a
model
based upon local data on
optical galaxies. Unlike the blue counts, the $K$ counts are well
matched by a non-evolving galaxy model. If the models are normalized at bright
magnitudes using our data, the deep $K$ counts of Cowie are
only a factor of $\sim 2$ at most (for $\Omega=1$)
in excess of the non-evolving model, in contrast with
the factor of $\sim 10$ excess seen in $b_j$. This follows the trend previously
seen as number-magnitude counts were extended to redder optical bands.

The bright counts exclude pure $K$ luminosity evolution in excess of $0.5$
magnitudes to $z=1$.
However this does not rule out models with active
star-formation in the past --- these predict {\em less}
luminosity evolution then simple passive stellar evolution.
Models of density evolution, and models with non-standard volume
elements, which explain the $b_j$ number counts and redshift distributions by
increasing the number of galaxies in the past,
generally overpredict the faint $K$
counts. It is possible to match the slope by invoking a merging type
evolutionary model, but only if the powerful assumption is made that merging
conserves $K$ light (i.e. $\simeq$ stellar mass).
It is argued that this is a more logical
starting point than empirical number-density evolution.
In this case the $K$ counts are
insensitive to the amount of merging. This can be tested by an
infrared-selected
redshift distribution, such a test will be carried out in Paper II.

The infrared counts favour a local space density of galaxies
$\phi^*=1.5\pm 0.2\phiunits$. This value is in good
agreement with the value determined by Efstathiou \etal. (1988).  This
normalization is much better defined than
in the optical as the counts are very close to the no-evolution prediction
and so the value is insensitive to evolution, provided the density and
luminosity scale as $\phi^*\propto 1/L^*$.

\section{Acknowledgements}

\tx
We acknowledge the generous allocations of telescope time on the U.K Infrared
Telescope, operated by the Royal Observatory Edinburgh, and the Isaac Newton
Telescope, operated by the Royal Greenwich Observatory in the Spanish
Observatorio del Roque de Los Muchachos of the Instituto de Astrofisica de
Canarias.  We also thank the staff and telescope operators of these telescopes
for their enthusiasm and competent support.
The photographic photometry was performed on plates supplied by the
U.K. Schmidt Telescope Unit using the COSMOS measuring machine at the Royal
Observatory, Edinburgh.  The computing and data reduction was carried out on
STARLINK which is funded by the SERC. Special thanks go to James Dunlop
for allowing us to use his software to
compute the K-corrections from galaxy templates.
KGB acknowledges the support of a SERC
research studentship.

\section*{References}

\tx
\ref Bahcall, J.N. \& Soneira, R.M., 1980. \name{\ApJ\ Suppl.}, \vol{44}, 73.
\ref Becklin, E.E. \& Zuckerman, B., 1988. \name{Nature}, \vol{336}, 656.
\ref Black, D.C., 1985. In: \name{Astrophysics of Brown Dwarfs}, ed. Kafatos,
M.C.,
Harrington, R.S. and Maran, S.P., Cambridge University Press, Cambridge, p.
139.
\ref Broadhurst, T.J., Ellis, R.S. \& Shanks, T., 1988. \name{\MNRAS},
\vol{235}, 827.
\ref Broadhurst, T.J., Ellis, R.S., and Glazebrook, K., 1992.
\name{Nature}, \vol{355}, 55.
\ref Bruzual, G., 1983. \name{\ApJ}, \vol{273}, 105.
\ref Casali, M.M., Aspin, C.A., McLean, I.S., 1987. \name{IRCAM users guide}.
\ref Colless, M., Ellis, R.S., Taylor, K. and Hook, R.N., 1990. \name{\MNRAS},
\vol{244}, 408.
\ref Collins, C.A., and Joseph, R.D., 1988. \name{\MNRAS}, \vol{235}, 209.
\ref Collins, C.A. \& Nichol, R.C., 1991. \name{\MNRAS}, submitted.
\ref Cowie, L.L., Gardner, J.P., Lilly, S.J. and McLean, I.S., 1990.
       \name{\ApJ}, \vol{360}, L1.
\ref Cowie, L. L., 1991. In: \name{Observational Tests of Inflation},
ed. Shanks, T. , Kluwer,  Dordrecht, p.257.
\ref Dahn, C. C.,  Liebert, J. and Harrington, R.S., 1986.  \name{Astr. J.},
\vol{91}, 621.
\ref Dunlop, J.S. \& Peacock, J.A., 1990. \name{\MNRAS}, \vol{247}, 19.
\ref Eddington, A.S., 1940. \name{\MNRAS}, \vol{100}, 352.
\ref Efstathiou, G., Ellis, R.S. and Peterson, B.A., 1988. \name{\MNRAS},
\vol{232},
431.
\ref Guideroni, B. \& Rocca-Volmerange, B.,1991. \name{\AstrAst},
\vol{252}, 435.
\ref Hall, P., \& Mackay, C.D, 1984. \name{\MNRAS}, \vol{210}, 979.
\ref Hawkins, M.R.S. \& Bessell, M.S., 1988. \name{\MNRAS}, \vol{234}, 177.
%Heydon-Dumbleton, N.H, Collins, C.A. and MacGillivray, H.T, 1989.
%\name{\MNRAS}, \vol{238}, 379.
\ref Jenkins, C.R. \& Reid, I.N., 1991. \name{\AstrJ}, \vol{101}, 1595.
\ref Jones, L.R., Fong, R., Shanks, T., Ellis, R.S. \& Peterson, B.A., 1991.
\name{\MNRAS}, \vol{249}, 481.
\ref Kaiser, N. \& Peacock, J.A., 1991. \name{\ApJ}, \vol{379}, 482.
\ref Kapteyn, J.C, 1906. \name{Plan of Selected Areas.}
\ref King C.R. \& Ellis, R.S., 1985. \name{\ApJ}, \vol{288}, 456.
\ref Koo, D.C., 1986. In: \name{Proceedings of the Erice Workshop on The
Spectral
Evolution of Galaxies}, ed. Chiosi, C. and Renzini, A.,  Reidel, Dordrecht, p.
419.
\ref Koo, D.C., 1990. In: \name{The Evolution of the Universe of Galaxies: The
Edwin
Hubble Centennial Symposium, Astronomical Soc. of the
Pacific Conference Series Vol. 10}, ed. Kron, R.G., p. 268.
\ref Kron, R.G.,, 1980. \name{\ApJ\ Suppl.}, \vol{43}, 305.
\ref Leggett, S.K. \& Hawkins, M.R.S., 1988. \name{\MNRAS}, \vol{234}, 1065.
\ref Lilly, S.J. \& Longair, M.S., 1984. \name{\MNRAS}, \vol{211}, 833.
\ref Lilly, S.J., Cowie, L.L. and Gardner, J.P., 1991. \name{\ApJ},
\vol{369}, 79.
\ref Loveday, J., Peterson, B.A., Efstathiou, G. and Maddox, S.J., 1992.
\name{\ApJ}, \vol{390}, 338.
\ref Maddox, S.J., Sutherland, W.J., Efstathiou, G., Loveday, J. and Peterson,
B.A.,
1990. \name{\MNRAS}, \vol{247}, 1P.
\ref McLean, I.S., Chuter, T.C., McCaughrean, M.J. \& Rayner, J.T., 1986.
In \name{Instrumentation in Astronomy VI}, ed. D.L.Crawford (Proc SPIE
Vol 627), p.430.
\ref Metcalfe, N., Shanks, T., Fong, R. \& Jones L.R., 1991. \name{\MNRAS},
\vol{249}, 498.
\ref Peacock, J.A., 1991. \name{\MNRAS}, \vol{253}, 1P.
\ref Peterson, B.A., Ellis, R.S., Kibblewhite, E.J., Bridgeland, M., Hooley, T.
and
Horne, D., 1979. \name{\ApJ}, \vol{233}, L109.
\ref Reid, I. N., 1987. \name{\MNRAS}, \vol{225}, 873.
\ref Rocca-Volmerange, B. \& Guideroni, B., 1988. \name{\AstrAstSup},
\vol{75}, 93.
\ref Rowan-Robinson, M. \& Crawford, J., 1991. \name{\MNRAS}, \vol{238}, 523.
\ref Shanks, T.S., Stevenson, P.R., Fong, R., MacGillivray, H.T., 1984.
\name{\MNRAS}, \vol{206}, 767.
\ref Shanks, T.S., 1990. In: \name{The Galactic and Extragalactic Background
Radiation}, ed. Bowyer, S. \& Leinert, C., p.269.
\ref Thronson, H.A., Jr., Majewski, S., Descartes, L. and Hereld M., 1990.
\name{\ApJ}, \vol{364}, 456.
\ref Tyson, J.A., 1988. \name{\AstrJ}, \vol{96}, 1.
\ref Zuckerman, B. \& Becklin E.E., 1987. \name{Nature}, \vol{330}, 138.

\vfill\eject
\strut\vfill\eject
\bgroup

\itemindent=5em

\def\item#1{\par\noindent{#1} \hangindent\parindent \quad\ignorespaces }

\parskip=4ex

\section*{Figures}

\tx
\item{\bf Fig.~\use{overlap.fig}} Example geometry of CCD frames for
the least-square cross-calibration technique.

\item{\bf Fig.~\use{StarGal.fig}} The star-galaxy image classification
parameter $y$
(see text) plotted against CCD $R$ magnitude. Note for $R>20$ only a third
of the points are plotted.

\item{\bf Fig.~\use{StarGalvsK.fig}} The star-galaxy image classification
parameter $y$
(see text) plotted against IRCAM $K$ magnitude.

\item{\bf Fig.~\use{RefCounts.fig}} Raw galaxy counts in fields with and
without galaxies as references (renormalised according to area).  The reference
star and reference galaxy points are shown slightly offset from the raw points
for clarity.

\item{\bf Fig.~\use{RawCounts.fig}} Final star and galaxy counts for the
survey.

\item{\bf Fig.~\use{StarModel.fig}} The star counts plotted for different
galactic
coordinates and compared with models of I.N. Reid (see text).

\item{\bf Fig.~\use{Knoev.fig}} The $K$ galaxy counts compared with those of
Cowie \etal. (1990, 1991) and Jenkins \& Reid (1991). Note the latter are shown
by dotted lines marking the $\pm 2\sigma$ limits from their statistical
analysis
of sky fluctuations. No-evolution predictions are shown.

\item{\bf Fig.~\use{bjnoev.fig}} The $b_j$ galaxy counts from a number of
authors together with no-evolution predictions.

\item{\bf Fig.~\use{NMLE.fig}} Parametric models of luminosity evolution
compared with
the $K$ galaxy counts.

\item{\bf Fig.~\use{NMME.fig}} Parametric models of merger evolution compared
with
the $K$ galaxy counts.

\item{\bf Fig.~\use{NzK.fig}} The predicted number-redshift distributions for a
$16<K<17$
selected sample for no-evolution and increasing amounts of merging. Note the
shift
of the mean redshift to lower values.

\vfill\eject

\section*{Tables}

\tx
\item{\bf Table~\use{Fieldcentres.tab}} Centres of the $10\amin\times 10\amin$
survey
zones.

\item{\bf Table~\use{EnormousWasteOfSpace.tab}} The centres of all individual
IRCAM frames together with exposure times in seconds and an ID corresponding to
the field --- in the sense that there is one unique ID per reference
star/galaxy
if they were all detected. Because many IRCAM frames have the same
referance object they have the same field ID (see section~\use{Obs.sec} for
details).

{

\pretolerance 10000

\item{\bf Table~\use{CCDint.tab}} Magnitude limits reached,
sky brightness (magnitudes$/\,\sqasec$) observed and equivalent noise levels
for the optical  CCD observations.

}

\item{\bf Table~\use{Catalogue.tab}} The final catalogue, in RA order, of all
detected $K$ sources together with optical magnitudes were available. Column
``G'' is whether or not the object is a galaxy and column ``R'' is whether or
not the object was selected in as a reference (see text for details). The ID
number is unique for each object, the field ID numbers (FID) are the same as in
Table~\use{EnormousWasteOfSpace.tab}.

\item{\bf Table~\use{IRNumCounts.tab}} The raw counts of galaxies ($N_g$),
stars ($N_s$), reference galaxies ($N_{gr}$), reference stars ($N_{sr}$) and
final corrected counts galaxies ($N'_g$) and stars ($N'_s$) with errors.

{\pretolerance 10000

\item{\bf Table~\use{GalaxyMix.tab}} Luminosity function used in the
no-evolution models.

}

\egroup

\vfill\eject

% starting tables

{

%======= Additional active Characters ===========
% ALL ACTIVE CHARACTERS MUST BE DECLARED AND DEFINED AT THE VERY
% BEGINNING OR THERE IS A DANGER OF INCORRECT TOKENIZATION
% ======== THESE ARE USED IN THE TABLE MACROS =========
\catcode`\|= \active
\def|{\ifmmode \vert\else \char`\|\fi} % effectively undoes activeness
\def\q@m{\string"}
\catcode`\"=\active \def"{\char`\"}
%=== these are redefined inside the table macros =====

%============= Table making macros ============
% the basic format is
% \begintable
% <special definitions for this table>
% \begintableformat
%           format as in halign except it needs ## instead of #
%           in tableformat, " means strut column
%           spacing controlled by \left, \center, \right
%           can use \math or \displaymath in conjuction with spacing
% \endtableformat
% \br{<struts>} .... | .... | .... " ..... | .... \er{<stuff>}  (rows)
% \-                                                (horizontal rules)
% ....etc
% \endtable
%
% each row has format
% \br{<..>}  <item> | <item} " .... | <item> \er{<..>}
%                   \br, \er mean beginning of row, end of row
%                   in table | means rule, " means no rule in strut column
%                   \: is standard strut, | is standard vrule
%
% \tablespread {to <dimen>}  width of table
% \tr=width of rules (default .4pt)
% \midtabglue sets glue in table (default 0pt plus 1fill)
% also can set explicit hrules and vrules
% tokens (e.g. \tablespread, \tr, \midtabglue, etc.) can be set in \everytable

% these macros require that | and " be active during the entire
% document to work correctly. ... although some definitions specifically
% set them active

\def\hssf{\hskip 0pt plus 1fill minus 1fill}
\def\n@ewaligndefs{\def\center##1{\hssf ##1\hssf\null}
          \def\left##1{##1\hssf\null}
          \def\right##1{\hssf ##1\null}}
\newdimen\trulesize
\let\tr=\trulesize
\trulesize = .4pt
\def\zerocenteredbox#1{\ifmmode \ifinner \setbox2 =\hbox{$#1$}\else
                           \setbox2 =\hbox{$\displaystyle#1$}\fi
                     \else \setbox2 =\hbox{#1}\fi
      \setbox0=\hbox{\lower.5ex\hbox{$\vcenter{\box2}$}}\ht0=0pt\dp0 =0pt\box0}

%this macro creates a strut with the that is higher by #2 and deeper than #3
% than the natural size of #1 ... the sizes may be negative
\def\modifystrut#1#2#3{\setbox4=\hbox{#1}\dimen0=\ht4
             \advance \dimen0 by #2 \dimen2 = \dp4
             \advance  \dimen2 by #3
            \vrule width 0pt height \dimen0 depth \dimen2}
\let\mst=\modifystrut
%general math form

%
\newskip\tcs
\newtoks\tablespread
\newskip\midtabglue \midtabglue = 0pt plus 1fill
\newtoks\everytable  \everytable = {\relax}

{\catcode`\|=\active   \catcode`\" = \active
  \gdef\begintable{\vbox\bgroup \tcs=.5em % uses font in force when entering
                  \catcode`\|=\active
                  \catcode`\"=\active
                   \def\:{\relax \vrule height 2.5ex depth .9ex width 0pt}
          \def\-{\ifcase\a@lignstate \fulltablerule{\tr}
                  \else
                  \thrule{\tr}\fi}
          \let\t@xx =\relax % for premature expansions
          \everycr={\noalign{\global\a@lignstate=0}}
          \def\fulltablerule##1{\noalign{\hrule height
                   ##1}}
          \def\thrule##1{\omit\leaders\hrule height ##1\hfill}
          \def\center{\hskip\tcs\hss ########\hss\hskip\tcs}
          \def\left{\hskip\tcs ########\hss\hskip\tcs}
          \def\right{\hskip\tcs\hss ########\hskip\tcs}
          \def\sprule{\tvrule{2.5\tr}}
          \def|{\ifcase\a@lignstate \def\t@xx{\tvrule{\tr}}\or
                            \def\t@xx{\tvrule{\tr}}\or
                             \def\t@xx{\unskip&\tvrule{\tr}&}\else
                             \def\t@xx{\tvrule{\tr}}\fi\t@xx}
          \def\|{\ifcase\a@lignstate \def\t@xx{\sprule}\or
                            \def\t@xx{\sprule}\or
                             \def\t@xx{\unskip&\sprule&}\else
                              \def\t@xx{\sprule}\fi\t@xx}
          \def"{&########&} % for table format
          \def\br##1{\global\a@lignstate=1 ##1\unskip\global\a@lignstate=2&}
          \def\er##1{\global\a@lignstate=3\unskip&##1\unskip
                      \global\a@lignstate=0\cr}
          \def\tvrule##1{\hss\vrule width ##1\hss}
          \def~{\penalty\@M \hphantom{0}}
%                               % changes ~ to be phantom of width .5em
          \tablespread = {}
          \the\everytable
               }
  \gdef\begintableformat #1\endtableformat{\offinterlineskip \tabskip = 0pt
       \edef\t@blform{####\tabskip =\midtabglue &#1&####\tabskip=0pt\cr}
%                                               % adds rules front and back
                                  \n@ewaligndefs
                                 \def"{\ifcase\a@lignstate \def\t@xx{\relax}\or
                                          \def\t@xx{\relax}\or
                                        \def\t@xx{\unskip&&}\else
                                        \def\t@xx{\relax}\fi\t@xx}
                                \edef\h@align{\halign \the\tablespread}
                                  \h@align\bgroup\span\t@blform}
      }% " and | are always active in INRSTEX

\def\use#1{\omit\mscount=#1 \advance\mscount by -1\multiply\mscount by2
                \loop\ifnum\mscount>1 \sp@n\repeat
                \ifnum\mscount>0 \span \else \relax \fi}

\def\sa#1{\setbox0=\hbox{#1}\hbox to \wd0{}}
\def\endtable{\crcr\egroup\egroup}

\catcode`\@=12

%
% Extra quick macros by KGB (Use \tab{format} .... \etab) :
%
\newdimen\nicetabruleskip
\nicetabruleskip=0.2em
\def\tabtoprule{\-\br{\ustrut|}}

\def\tab#1{\begintable                        % Begin table and draw top rule
\def\dstrut{\mst{\:}{0pt}{\the\nicetabruleskip}}  % Extra struts on hrules
\def\ustrut{\mst{\:}{\the\nicetabruleskip}{0pt}}  %
\def\n{\er{|}\br{\:|}}                        % Next row
\def\r{\er{|\dstrut}\-\br{\ustrut|}}          % Next row + rule
%
% Next row + double rule. The argument gives the internal rules, e.g. for a
% 4 column table \dr{|||} or \dr{"""} if thats what is preferred.
%
\def\dr##1{\er{|\dstrut}\-\br{\vrule height\nicetabruleskip width0pt depth0pt|}
         ##1\er{|}\-\br{\ustrut|}}
\def\L{\quad\left\quad}                 % Nice left justified
\def\R{\quad\right\quad}                % Nice right justified
\def\C{\quad\center\quad}               % Nice center justified
\def\para##1{\vtop{\noindent\hsize=##1########\dstrut}}
                                        % (######## --> ## in actual format.)
\def\P##1{\quad\para{##1}\quad}         % Nice paragraphs (e.q. \P{4in}
                                        % in format)
\begintableformat #1\endtableformat\tabtoprule}
         % End table and draw bottom rule
                                        % (use instead of \n or \r on
                                        % last line)
  % Alias for \use to make multicolumn entries
%
\nopagenumbers

\noindent
{\bf Table 1}
\bigskip

\halign{
 \quad#\quad\hfill & \hfill\quad$#$\quad\hfill & \hfill\quad$#$\quad\hfill &
\hfill\quad$#$\quad\hfill  \cr
 \ Field   & $RA$                   & $DEC$  &   b \cr
\noalign{\medskip}
 SA114\_3  & 22\hrs 38\mins 56.0\secs & +00\deg 36\mins 00\secs&   -48\deg\cr
 SA114\_4  & 22\hrs 40\mins 16.0\secs & +00\deg 26\mins 00\secs&   -48\deg\cr
 SA114\_5  & 22\hrs 39\mins 36.0\secs & +00\deg 26\mins 00\secs&   -48\deg\cr
 SA114\_6  & 22\hrs 38\mins 56.0\secs & +00\deg 26\mins 00\secs&   -48\deg\cr
 SA92\_2   & 00\hrs 52\mins 20.0\secs & +00\deg 20\mins 00\secs&   -62\deg\cr
 SA92\_4   & 00\hrs 53\mins 00.0\secs & +00\deg 10\mins 00\secs&   -62\deg\cr
 SA93\_1   & 01\hrs 53\mins 16.0\secs & +00\deg 45\mins 00\secs&   -58\deg\cr
 SA93\_5   & 01\hrs 52\mins 36.0\secs & +00\deg 35\mins 00\secs&   -58\deg\cr
 SA93\_8   & 01\hrs 52\mins 36.0\secs & +00\deg 25\mins 00\secs&   -58\deg\cr
 851STARS\_1  & 09\hrs 18\mins 24.5\secs & -00\deg 21\mins 55\secs&
+32\deg\cr
 851STARS\_2  & 09\hrs 18\mins 39.0\secs & -00\deg 03\mins 03\secs&
+33\deg\cr
 855LDSS\_1   & 10\hrs 43\mins 55.4\secs & +00\deg 05\mins 33\secs&
+49\deg\cr
 855LDSS\_2   & 10\hrs 43\mins 06.2\secs & -00\deg 08\mins 15\secs&
+49\deg\cr
 855LDSS\_4   & 10\hrs 44\mins 03.3\secs & -00\deg 20\mins 26\secs&
+49\deg\cr
% 855STARS\_1  & 10\hrs 41\mins 59.2\secs & -00\deg 13\mins 26\secs&
%%+49\deg\cr
 864LDSS\_1   & 13\hrs 41\mins 57.7\secs & +00\deg 06\mins 25\secs&
+60\deg\cr
 864LDSS\_2   & 13\hrs 41\mins 09.1\secs & -00\deg 00\mins 16\secs&
+60\deg\cr
 864LDSS\_3   & 13\hrs 41\mins 08.8\secs & +00\deg 10\mins 18\secs&
+60\deg\cr
% 864STARS\_1  & 13\hrs 39\mins 20.8\secs & +00\deg 18\mins 48\secs&
%%+60\deg\cr
% 864STARS\_2  & 13\hrs 40\mins 10.8\secs & -00\deg 02\mins 57\secs&
%%+60\deg\cr
% 875STARS\_5  & 17\hrs 19\mins 59.3\secs & +00\deg 10\mins 19\secs&
%%+19\deg\cr
}

\vskip 2.5truecm

\noindent
{\bf Table 3}
\bigskip

\noindent
{\bf\quad September 1988 data}
\bigskip
\halign{
\hfil\quad#\quad\hfil & \hfil\quad#\quad\hfil
& \hfil\quad#\quad\hfil & \hfil\quad#\quad\hfil & \hfil\quad#\quad\hfil \cr
\noalign{\medskip}
Band &Sky brightness    &$1\rm\sigma 1s$& $t_{int}$  &  Mag. limit     \cr
B    &  21.2            &  23.5         & ~600\,s    &   23.9          \cr
R    &  20.6            &  23.2         & ~600\,s    &   23.8          \cr
}
\bigskip
\bigskip
\noindent
{\bf\quad April 1989 data}
\bigskip
\halign{
\hfil\quad#\quad\hfil & \hfil\quad#\quad\hfil
& \hfil\quad#\quad\hfil & \hfil\quad#\quad\hfil & \hfil\quad#\quad\hfil \cr
Band &Sky brightness    &$1\rm\sigma 1s$& $t_{int}$  &  Mag. limit     \cr
\noalign{\medskip}
B    &  22.0            &  23.9         & ~600\,s    &   24.3          \cr
V    &  21.0            &  23.6         & ~600\,s    &   24.1          \cr
R    &  20.5            &  23.2         & ~600\,s    &   23.8          \cr
I    &  19.2            &  22.4         & 2000\,s    &   23.6        \cr
}

\vfill\eject
\strut\vfill\eject

\noindent
{\bf Table 5}
\bigskip

\def\pmb#1{ \setbox0=\hbox{#1}%
   \kern-.025em\copy0\kern-\wd0
   \kern.05em\copy0\kern-\wd0
   \kern-.025em\raise.0433em\box0 }

\medskip
\halign{
\hfil\quad$#$\quad & \hfil\quad$#$\quad & \hfil\quad$#$\quad &
\hfil\quad$#$\quad & \hfil\quad$#$\quad & \hfil\quad$#$\quad &
\hfil\quad$#$\quad & \hfil\quad$#$\quad & \hfil\quad$#$\quad \cr
     $K range$\hfil &    N_g&    N_s&   N_{gr}&   N_{sr}&
    N_g'&  \Delta N_g'&   N_s'&  \Delta N_s'\cr
\noalign{\smallskip}
     10\le K<11 &     0 &     6 &     0 &     0 &   0.0 &   0.0 &   6.0 &   2.4
\cr
     11\le K<12 &     0 &    10 &     0 &     0 &   0.0 &   0.0 &  10.0 &   3.2
\cr
     12\le K<13 &     0 &    16 &     0 &     9 &   0.0 &   0.0 &  19.2 &   4.1
\cr
     13\le K<14 &     3 &    23 &     1 &    32 &   3.4 &   1.8 &  34.5 &   5.2
\cr
     14\le K<15 &     6 &    31 &    15 &    49 &  11.4 &   2.8 &  48.6 &   6.1
\cr
     15\le K<16 &    57 &    40 &    18 &    32 &  52.7 &   9.1 &  62.3 &   9.4
\cr
     16\le K<17 &   166 &    31 &    14 &     9 & 133.8 &  18.2 &  71.5 &  16.4
\cr
     17\le K<18 &    55 &    10 &     0 &     0 &  43.0 &   7.5 &  22.0 &   6.0
\cr
\noalign{\smallskip}
       $Total$\hfil&287 &   167 &    48 &   131 & 244.2 &  22.0 & 274.2 &  22.2
\cr
}

\vskip 2.5truecm

\noindent
{\bf Table 6}
\bigskip

\halign{\hfil\quad#\quad\hfil & \hfil\quad#\quad\hfil &
\hfil\quad$#$\quad\hfil & \hfil\quad$#$\quad\hfil \cr
Type   & Fraction &   {M^*_{b_j}}^{\dagger} & b_j-K \cr
\noalign{\smallskip}
E/SO   &  0.35    &   -19.59 & 3.99 \cr
Sa     &  0.07    &   -19.39 & 3.79 \cr
Sb     &  0.18    &   -19.39 & 3.61 \cr
Sc     &  0.17    &   -19.39 & 3.26 \cr
Sd     &  0.15    &   -18.94 & 2.85 \cr
Im     &  0.08    &   -18.94 & 2.22 \cr
}
\quad\quad\tenpt$\dagger$
\quad $\alpha=-1.00$, \Hhundred.

}  % end tables

\vfill\eject
\strut\vfill\eject

\nopagenumbers
\baselineskip=0.85\baselineskip
\noindent
{\bf Table 2: Frame centres}
\bigskip
\tabskip 0.1truecm
%% FOLLOWING LINE CANNOT BE BROKEN BEFORE 80 CHAR
\halign{#&\hfill#&\hfill#&\quad\hfill#&\hfill#&\hfill#&\quad#&\quad\hfill#\hfill\cr
&  \llap{\rlap{$\alpha(1950)$}\quad$\;\;$} &       &    &
\llap{\rlap{$\delta(1950)$}\quad\quad}&       &\hfill $\!\!\!t_{\rm exp}$/s
\hfill    &field id \cr
\cr
   00 &52 & 5.12  & 00 &15 &12.69  & 300    &021201 \cr
   00 &52 & 6.66  & 00 &21 & 7.69  & 300    &021203 \cr
   00 &52 & 8.32  & 00 &15 &54.69  & 300    &021201 \cr
   00 &52 & 9.59  & 00 &24 & 9.69  & 300    &021202 \cr
   00 &52 & 9.86  & 00 &21 &49.69  & 300    &021203 \cr
   00 &52 &10.52  & 00 &17 &34.69  & 300    &021204 \cr
   00 &52 &12.79  & 00 &23 &27.69  & 300    &021202 \cr
   00 &52 &13.72  & 00 &16 &52.69  & 300    &021204 \cr
   00 &52 &15.26  & 00 &18 &53.69  & 300    &021207 \cr
   00 &52 &18.46  & 00 &19 &35.69  & 300    &021207 \cr
   00 &52 &18.59  & 00 &23 &53.69  & 300    &021205 \cr
   00 &52 &21.79  & 00 &24 &35.69  & 300    &021205 \cr
   00 &52 &22.59  & 00 &20 &50.69  & 300    &021206 \cr
   00 &52 &25.72  & 00 &24 &59.69  & 300    &021208 \cr
   00 &52 &25.79  & 00 &20 & 8.69  & 300    &021206 \cr
   00 &52 &28.32  & 00 &16 &28.69  & 300    &021211 \cr
   00 &52 &28.79  & 00 &21 &27.69  & 150    &021209 \cr
   00 &52 &28.92  & 00 &24 &17.69  & 150    &021208 \cr
   00 &52 &31.52  & 00 &17 &10.69  & 150    &021211 \cr
   00 &52 &31.99  & 00 &22 & 9.69  & 300    &021209 \cr
   00 &52 &32.46  & 00 &19 & 2.69  & 150    &021210 \cr
   00 &52 &33.12  & 00 &19 &58.69  & 150    &021213 \cr
   00 &52 &35.66  & 00 &18 &20.69  & 300    &021210 \cr
   00 &52 &35.79  & 00 &24 &16.69  & 300    &021212 \cr
   00 &52 &36.12  & 00 &15 &58.69  & 150    &021214 \cr
   00 &52 &36.32  & 00 &20 &40.69  & 300    &021213 \cr
   00 &52 &38.99  & 00 &23 &34.69  & 150    &021212 \cr
   00 &52 &39.32  & 00 &15 &16.69  & 300    &021214 \cr
   00 &52 &43.46  & 00 & 7 &16.91  & 300    &021403 \cr
   00 &52 &43.46  & 00 & 7 &58.91  & 300    &021403 \cr
   00 &52 &44.73  & 00 &10 &24.91  & 300    &021402 \cr
   00 &52 &44.73  & 00 & 9 &42.91  & 300    &021402 \cr
   00 &52 &45.46  & 00 &11 &52.91  & 300    &021401 \cr
   00 &52 &45.46  & 00 &11 &10.91  & 300    &021401 \cr
   00 &52 &46.66  & 00 & 7 &58.91  & 300    &021403 \cr
   00 &52 &46.66  & 00 & 7 &16.91  & 300    &021403 \cr
   00 &52 &47.93  & 00 &10 &24.91  & 300    &021402 \cr
   00 &52 &47.93  & 00 & 9 &42.91  & 300    &021402 \cr
   00 &52 &48.66  & 00 &11 &52.91  & 300    &021401 \cr
   00 &52 &48.66  & 00 &11 &10.91  & 300    &021401 \cr
   00 &52 &49.59  & 00 & 6 & 6.91  & 300    &021404 \cr
   00 &52 &49.59  & 00 & 5 &24.91  & 300    &021404 \cr
   00 &52 &52.59  & 00 & 8 & 7.91  & 300    &021406 \cr
   00 &52 &52.59  & 00 & 7 &25.91  & 300    &021406 \cr
   00 &52 &52.79  & 00 & 5 &24.91  & 300    &021404 \cr
   00 &52 &52.79  & 00 & 6 & 6.91  & 300    &021404 \cr
   00 &52 &53.46  & 00 & 9 &11.91  & 300    &021405 \cr
   00 &52 &53.46  & 00 & 9 &53.91  & 300    &021405 \cr
   00 &52 &55.79  & 00 & 7 &25.91  & 300    &021406 \cr
   00 &52 &55.79  & 00 & 8 & 7.91  & 300    &021406 \cr
   00 &52 &56.66  & 00 & 9 &11.91  & 300    &021405 \cr
   00 &52 &56.66  & 00 & 9 &53.91  & 300    &021405 \cr
   00 &53 & 1.86  & 00 &10 &15.91  & 300    &021407 \cr
   00 &53 & 1.86  & 00 &10 &57.91  & 300    &021407 \cr
   00 &53 & 5.06  & 00 &10 &57.91  & 300    &021407 \cr
   00 &53 & 5.06  & 00 &10 &15.91  & 300    &021407 \cr
   00 &53 & 6.86  & 00 & 7 &33.91  & 300    &021408 \cr
   00 &53 & 6.86  & 00 & 6 &51.91  & 300    &021408 \cr
   00 &53 &10.06  & 00 & 7 &33.91  & 300    &021408 \cr
   00 &53 &10.06  & 00 & 6 &51.91  & 300    &021408 \cr
   00 &53 &11.13  & 00 &10 &26.91  & 300    &021410 \cr
   00 &53 &11.13  & 00 &11 & 8.91  & 300    &021410 \cr
   00 &53 &11.19  & 00 & 5 &51.91  & 300    &021411 \cr
   00 &53 &11.19  & 00 & 5 & 9.91  & 300    &021411 \cr
   00 &53 &14.33  & 00 &10 &26.91  & 300    &021410 \cr
   00 &53 &14.33  & 00 &11 & 8.91  & 300    &021410 \cr
   00 &53 &14.39  & 00 & 5 & 9.91  & 300    &021411 \cr
   00 &53 &14.39  & 00 &12 &24.91  & 300    &021409 \cr
   00 &53 &14.39  & 00 & 5 &51.91  & 300    &021411 \cr
   00 &53 &14.39  & 00 &13 & 6.91  & 300    &021409 \cr
   00 &53 &17.59  & 00 &12 &24.91  & 300    &021409 \cr
   00 &53 &17.59  & 00 &13 & 6.91  & 300    &021409 \cr
   01 &52 &19.06  & 00 &26 &23.80  & 450    &031802 \cr
   01 &52 &19.08  & 00 &34 &40.67  & 450    &031503 \cr
   01 &52 &21.33  & 00 &38 &29.00  & 150    &030501 \cr
   01 &52 &21.39  & 00 &28 &30.80  & 450    &031801 \cr
   01 &52 &21.68  & 00 &36 &38.67  & 450    &031502 \cr
   01 &52 &22.26  & 00 &25 &41.80  & 300    &031802 \cr
   01 &52 &22.46  & 00 &22 &24.80  & 300    &031804 \cr
   01 &52 &22.52  & 00 &19 &58.80  & 450    &031805 \cr
   01 &52 &23.32  & 00 &23 &22.80  & 300    &031803 \cr
   01 &52 &24.59  & 00 &29 &12.80  & 300    &031801 \cr
   01 &52 &25.35  & 00 &39 &27.67  & 450    &031501 \cr
   01 &52 &25.66  & 00 &21 &42.80  & 450    &031804 \cr
   01 &52 &25.72  & 00 &20 &40.80  & 300    &031805 \cr
   01 &52 &26.13  & 00 &30 &56.00  & 150    &030502 \cr
   01 &52 &26.52  & 00 &24 & 4.80  & 450    &031803 \cr
   01 &52 &29.08  & 00 &31 & 2.67  & 300    &031507 \cr
   01 &52 &29.61  & 00 &33 & 8.67  & 300    &031506 \cr
   01 &52 &30.06  & 00 &26 &39.80  & 300    &031807 \cr
   01 &52 &30.35  & 00 &37 &30.67  & 450    &031504 \cr
   01 &52 &30.39  & 00 &29 &18.80  & 450    &031806 \cr
   01 &52 &30.80  & 00 &33 & 2.00  & 150    &030503 \cr
   01 &52 &31.68  & 00 &35 &30.67  & 300    &031505 \cr
   01 &52 &33.26  & 00 &27 &21.80  & 450    &031807 \cr
   01 &52 &33.59  & 00 &28 &36.80  & 300    &031806 \cr
   01 &52 &35.19  & 00 &24 &52.80  & 450    &031809 \cr
   01 &52 &37.01  & 00 &36 &46.67  & 300    &031508 \cr
   01 &52 &37.46  & 00 &20 &10.80  & 300    &031811 \cr
   01 &52 &38.15  & 00 &29 &58.67  & 150    &031511 \cr
   01 &52 &38.19  & 00 &29 &26.80  & 300    &031808 \cr
   01 &52 &38.39  & 00 &25 &34.80  & 300    &031809 \cr
   01 &52 &39.55  & 00 &34 &53.67  & 300    &031509 \cr
   01 &52 &39.80  & 00 &39 &34.00  & 150    &030504 \cr
   01 &52 &40.66  & 00 &20 &52.80  & 450    &031811 \cr
   01 &52 &41.39  & 00 &28 &44.80  & 450    &031808 \cr
   01 &52 &41.66  & 00 &23 &21.80  & 450    &031810 \cr
   01 &52 &41.81  & 00 &32 &27.67  & 300    &031510 \cr
   01 &52 &44.12  & 00 &27 &24.80  & 450    &031813 \cr
   01 &52 &44.86  & 00 &22 &39.80  & 300    &031810 \cr
   01 &52 &44.93  & 00 &31 &36.00  & 150    &030507 \cr
   01 &52 &46.95  & 00 &36 & 3.67  & 150    &031513 \cr
   01 &52 &47.28  & 00 &40 &12.67  & 150    &031512 \cr
   01 &52 &47.32  & 00 &28 & 6.80  & 300    &031813 \cr
   01 &52 &47.60  & 00 &34 &48.00  & 150    &030505 \cr
   01 &52 &47.60  & 00 &32 &16.00  & 150    &030506 \cr
   01 &52 &48.46  & 00 &26 &51.80  & 450    &031814 \cr
   01 &52 &50.46  & 00 &30 & 6.80  & 300    &031812 \cr
   01 &52 &51.66  & 00 &26 & 9.80  & 300    &031814 \cr
   01 &52 &53.32  & 00 &21 & 5.80  & 300    &031816 \cr
   01 &52 &53.59  & 00 &25 &40.80  & 300    &031815 \cr
   01 &52 &53.66  & 00 &29 &24.80  & 450    &031812 \cr
   01 &52 &56.52  & 00 &20 &23.80  & 450    &031816 \cr
   01 &52 &56.61  & 00 &33 &30.67  & 150    &031514 \cr
   01 &52 &56.79  & 00 &26 &22.80  & 450    &031815 \cr
   01 &52 &59.49  & 00 &41 &30.72  & 300    &031101 \cr
   01 &53 & 0.89  & 00 &44 &57.72  & 300    &031105 \cr
   01 &53 & 1.22  & 00 &49 &47.72  & 300    &031102 \cr
   01 &53 & 2.69  & 00 &42 &12.72  & 300    &031101 \cr
   01 &53 & 3.73  & 00 &42 &11.00  & 150    &030102 \cr
   01 &53 & 4.09  & 00 &45 &39.72  & 300    &031105 \cr
   01 &53 & 4.42  & 00 &49 & 5.72  & 300    &031102 \cr
   01 &53 & 4.69  & 00 &47 &15.72  & 300    &031103 \cr
   01 &53 & 7.89  & 00 &47 &57.72  & 300    &031103 \cr
   01 &53 & 8.09  & 00 &46 &31.72  & 300    &031104 \cr
   01 &53 &10.82  & 00 &40 &56.72  & 300    &031108 \cr
   01 &53 &11.29  & 00 &45 &49.72  & 300    &031104 \cr
   01 &53 &11.89  & 00 &44 &20.72  & 300    &031107 \cr
   01 &53 &12.40  & 00 &49 &22.00  & 150    &030101 \cr
   01 &53 &14.02  & 00 &40 &14.72  & 150    &031108 \cr
   01 &53 &15.09  & 00 &45 & 2.72  & 300    &031107 \cr
   01 &53 &15.73  & 00 &47 &58.00  & 150    &030103 \cr
   01 &53 &17.20  & 00 &42 &50.00  & 150    &030104 \cr
   01 &53 &17.69  & 00 &49 &54.72  & 300    &031106 \cr
   01 &53 &19.29  & 00 &42 &18.72  & 300    &031112 \cr
   01 &53 &19.75  & 00 &46 &42.72  & 150    &031110 \cr
   01 &53 &20.89  & 00 &49 &12.72  & 300    &031106 \cr
   01 &53 &21.07  & 00 &39 &53.00  & 150    &030106 \cr
   01 &53 &22.49  & 00 &41 &36.72  & 150    &031112 \cr
   01 &53 &22.95  & 00 &46 & 0.72  & 300    &031110 \cr
   01 &53 &23.09  & 00 &44 &18.72  & 300    &031111 \cr
   01 &53 &24.62  & 00 &47 &54.72  & 150    &031109 \cr
   01 &53 &26.29  & 00 &45 & 0.72  & 150    &031111 \cr
   01 &53 &27.82  & 00 &48 &36.72  & 300    &031109 \cr
   01 &53 &28.55  & 00 &42 &10.72  & 150    &031113 \cr
   01 &53 &30.20  & 00 &46 &53.00  & 150    &030105 \cr
   01 &53 &30.87  & 00 &47 &48.00  & 150    &030108 \cr
   01 &53 &31.75  & 00 &42 &52.72  & 300    &031113 \cr
   01 &53 &32.69  & 00 &40 &49.72  & 150    &031114 \cr
   01 &53 &33.47  & 00 &46 &11.00  & 150    &030109 \cr
   01 &53 &34.34  & 00 &50 &15.00  & 150    &030107 \cr
   01 &53 &35.89  & 00 &40 & 7.72  & 300    &031114 \cr
   09 &18 & 3.71  &$-$00 &19 &45.82  & 300    &110101 \cr
   09 &18 & 3.71  &$-$00 &19 & 3.82  & 300    &110101 \cr
   09 &18 & 6.57  &$-$00 &19 &59.82  & 300    &110103 \cr
   09 &18 & 6.57  &$-$00 &20 &41.82  & 300    &110103 \cr
   09 &18 & 6.91  &$-$00 &19 &45.82  & 300    &110101 \cr
   09 &18 & 6.91  &$-$00 &19 & 3.82  & 300    &110101 \cr
   09 &18 & 8.78  &$-$00 &19 &48.82  & 300    &110102 \cr
   09 &18 & 8.78  &$-$00 &19 & 6.82  & 300    &110102 \cr
   09 &18 & 9.78  &$-$00 &20 &41.82  & 300    &110103 \cr
   09 &18 & 9.78  &$-$00 &19 &59.82  & 300    &110103 \cr
   09 &18 &10.04  &$-$00 &25 & 6.82  & 300    &110104 \cr
   09 &18 &10.04  &$-$00 &25 &48.82  & 300    &110104 \cr
   09 &18 &11.98  &$-$00 &19 &48.82  & 300    &110102 \cr
   09 &18 &11.98  &$-$00 &19 & 6.82  & 300    &110102 \cr
   09 &18 &13.24  &$-$00 &25 & 6.82  & 300    &110104 \cr
   09 &18 &13.24  &$-$00 &25 &48.82  & 300    &110104 \cr
   09 &18 &13.84  &$-$00 &24 & 2.82  & 150    &110106 \cr
   09 &18 &13.84  &$-$00 &23 &20.82  & 300    &110106 \cr
   09 &18 &15.24  &$-$00 &25 &13.82  & 300    &110107 \cr
   09 &18 &15.24  &$-$00 &25 &55.82  & 300    &110107 \cr
   09 &18 &15.44  &$-$00 &18 & 9.82  & 300    &110105 \cr
   09 &18 &15.44  &$-$00 &18 &51.82  & 300    &110105 \cr
   09 &18 &17.04  &$-$00 &23 &20.82  & 300    &110106 \cr
   09 &18 &17.04  &$-$00 &24 & 2.82  & 300    &110106 \cr
   09 &18 &18.44  &$-$00 &25 &55.82  & 300    &110107 \cr
   09 &18 &18.44  &$-$00 &25 &13.82  & 300    &110107 \cr
   09 &18 &18.64  &$-$00 &18 &51.82  & 300    &110105 \cr
   09 &18 &18.64  &$-$00 &18 & 9.82  & 300    &110105 \cr
   09 &18 &20.31  &$-$00 &25 &55.82  & 300    &110108 \cr
   09 &18 &20.31  &$-$00 &26 &37.82  & 300    &110108 \cr
   09 &18 &23.31  &$-$00 &22 &16.82  & 300    &110109 \cr
   09 &18 &23.31  &$-$00 &22 &58.82  & 300    &110109 \cr
   09 &18 &23.51  &$-$00 &25 &55.82  & 300    &110108 \cr
   09 &18 &23.51  &$-$00 &26 &37.82  & 300    &110108 \cr
   09 &18 &26.24  &$-$00 & 7 & 5.89  & 150    &110202 \cr
   09 &18 &26.51  &$-$00 &22 &58.82  & 300    &110109 \cr
   09 &18 &26.51  &$-$00 &22 &16.82  & 300    &110109 \cr
   09 &18 &28.71  &$-$00 &24 &36.82  & 300    &110110 \cr
   09 &18 &28.71  &$-$00 &23 &54.82  & 300    &110110 \cr
   09 &18 &29.44  &$-$00 & 7 &47.89  & 150    &110202 \cr
   09 &18 &29.44  &$-$00 & 0 & 7.89  & 150    &110201 \cr
   09 &18 &31.91  &$-$00 &23 &54.82  & 300    &110110 \cr
   09 &18 &31.91  &$-$00 &24 &36.82  & 300    &110110 \cr
   09 &18 &32.64  & 00 & 0 &34.11  & 150    &110201 \cr
   09 &18 &32.64  &$-$00 & 3 &49.89  & 150    &110206 \cr
   09 &18 &35.50  &$-$00 & 3 &39.89  & 150    &110205 \cr
   09 &18 &35.84  &$-$00 & 4 &31.89  & 150    &110206 \cr
   09 &18 &36.77  &$-$00 & 0 &29.89  & 150    &110204 \cr
   09 &18 &38.04  &$-$00 &21 &53.82  & 300    &110111 \cr
   09 &18 &38.04  &$-$00 &22 &35.82  & 300    &110111 \cr
   09 &18 &38.70  &$-$00 & 2 &57.89  & 150    &110205 \cr
   09 &18 &38.97  &$-$00 & 0 & 0.89  & 150    &110203 \cr
   09 &18 &39.97  &$-$00 & 1 &11.89  & 150    &110204 \cr
   09 &18 &40.24  &$-$00 &19 & 8.82  & 300    &110113 \cr
   09 &18 &40.24  &$-$00 &19 &50.82  & 300    &110113 \cr
   09 &18 &41.24  &$-$00 &22 &35.82  & 300    &110111 \cr
   09 &18 &41.24  &$-$00 &21 &53.82  & 300    &110111 \cr
   09 &18 &41.65  &$-$00 &18 &43.82  & 300    &110112 \cr
   09 &18 &41.65  &$-$00 &18 & 1.82  & 300    &110112 \cr
   09 &18 &42.17  & 00 & 0 &41.11  & 150    &110203 \cr
   09 &18 &42.44  &$-$00 & 1 &50.89  & 150    &110208 \cr
   09 &18 &42.84  &$-$00 & 1 & 8.89  & 150    &110207 \cr
   09 &18 &43.44  &$-$00 &19 &50.82  & 300    &110113 \cr
   09 &18 &43.44  &$-$00 &19 & 8.82  & 300    &110113 \cr
   09 &18 &44.84  &$-$00 &18 & 1.82  & 300    &110112 \cr
   09 &18 &44.84  &$-$00 &18 &43.82  & 300    &110112 \cr
   09 &18 &45.64  &$-$00 & 2 &32.89  & 150    &110208 \cr
   09 &18 &46.04  &$-$00 & 0 &26.89  & 150    &110207 \cr
   09 &18 &57.04  &$-$00 & 6 &18.89  & 150    &110209 \cr
   09 &19 & 0.24  &$-$00 & 5 &36.89  & 150    &110209 \cr
   10 &42 &47.61  &$-$00 &11 &30.82  & 150    &120201 \cr
   10 &42 &50.81  &$-$00 &10 &48.82  & 150    &120201 \cr
   10 &42 &55.34  &$-$00 &10 & 6.82  & 150    &120203 \cr
   10 &42 &56.07  &$-$00 & 5 &55.82  & 150    &120202 \cr
   10 &42 &58.54  &$-$00 & 9 &24.82  & 150    &120203 \cr
   10 &42 &59.28  &$-$00 & 6 &37.82  & 150    &120202 \cr
   10 &43 & 8.94  &$-$00 &11 &56.82  & 150    &120206 \cr
   10 &43 &10.74  &$-$00 & 9 &31.82  & 150    &120205 \cr
   10 &43 &12.14  &$-$00 &12 &38.82  & 150    &120206 \cr
   10 &43 &13.95  &$-$00 & 8 &49.82  & 150    &120205 \cr
   10 &43 &16.14  &$-$00 & 3 &39.82  & 150    &120204 \cr
   10 &43 &19.34  &$-$00 & 4 &21.82  & 150    &120204 \cr
   10 &43 &19.48  &$-$00 & 9 &45.82  & 150    &120208 \cr
   10 &43 &22.48  &$-$00 & 7 & 6.82  & 150    &120207 \cr
   10 &43 &22.68  &$-$00 &10 &27.82  & 150    &120208 \cr
   10 &43 &23.48  &$-$00 &12 &49.82  & 150    &120209 \cr
   10 &43 &25.68  &$-$00 & 6 &24.82  & 150    &120207 \cr
   10 &43 &26.68  &$-$00 &12 & 7.82  & 150    &120209 \cr
   10 &43 &34.67  & 00 & 6 &34.63  & 300    &120101 \cr
   10 &43 &34.67  & 00 & 5 &52.63  & 300    &120101 \cr
   10 &43 &37.07  & 00 & 3 &42.63  & 300    &120103 \cr
   10 &43 &37.07  & 00 & 3 & 0.63  & 300    &120103 \cr
   10 &43 &37.53  & 00 & 9 &21.63  & 300    &120102 \cr
   10 &43 &37.53  & 00 &10 & 3.63  & 300    &120102 \cr
   10 &43 &37.87  & 00 & 5 &52.63  & 300    &120101 \cr
   10 &43 &37.87  & 00 & 6 &34.63  & 300    &120101 \cr
   10 &43 &40.26  & 00 & 3 & 0.63  & 300    &120103 \cr
   10 &43 &40.26  & 00 & 3 &42.63  & 300    &120103 \cr
   10 &43 &40.73  & 00 &10 & 3.63  & 300    &120102 \cr
   10 &43 &40.73  & 00 & 9 &21.63  & 300    &120102 \cr
   10 &43 &41.26  & 00 & 1 &44.63  & 300    &120104 \cr
   10 &43 &41.26  & 00 & 2 &26.63  & 300    &120104 \cr
   10 &43 &44.46  & 00 & 2 &26.63  & 300    &120104 \cr
   10 &43 &44.46  & 00 & 1 &44.63  & 300    &120104 \cr
   10 &43 &46.13  & 00 & 7 &47.63  & 300    &120105 \cr
   10 &43 &46.13  & 00 & 7 & 5.63  & 300    &120105 \cr
   10 &43 &46.77  &$-$00 &16 &30.63  & 300    &120401 \cr
   10 &43 &47.97  &$-$00 &22 &15.63  & 150    &120403 \cr
   10 &43 &49.33  & 00 & 7 & 5.63  & 300    &120105 \cr
   10 &43 &49.33  & 00 & 7 &47.63  & 300    &120105 \cr
   10 &43 &53.97  &$-$00 &18 &56.63  & 300    &120402 \cr
   10 &43 &55.40  & 00 & 4 & 5.63  & 150    &120106 \cr
   10 &43 &55.40  & 00 & 4 &47.63  & 300    &120106 \cr
   10 &43 &58.60  & 00 & 4 &47.63  & 300    &120106 \cr
   10 &43 &58.60  & 00 & 4 & 5.63  & 300    &120106 \cr
   10 &43 &59.03  &$-$00 &21 &20.63  & 300    &120405 \cr
   10 &44 & 2.97  &$-$00 &19 &31.63  & 300    &120404 \cr
   10 &44 & 4.26  & 00 & 3 &55.63  & 300    &120108 \cr
   10 &44 & 4.26  & 00 & 4 &37.63  & 300    &120108 \cr
   10 &44 & 5.26  & 00 & 9 &15.63  & 150    &120107 \cr
   10 &44 & 5.26  & 00 & 8 &33.63  & 300    &120107 \cr
   10 &44 & 6.90  &$-$00 &18 &25.63  & 300    &120406 \cr
   10 &44 & 7.46  & 00 & 4 &37.63  & 150    &120108 \cr
   10 &44 & 7.46  & 00 & 3 &55.63  & 300    &120108 \cr
   10 &44 & 8.46  & 00 & 8 &33.63  & 300    &120107 \cr
   10 &44 & 8.46  & 00 & 9 &15.63  & 300    &120107 \cr
   10 &44 & 9.90  &$-$00 &22 &47.63  & 300    &120407 \cr
   10 &44 &12.60  & 00 & 6 & 5.63  & 300    &120109 \cr
   10 &44 &12.60  & 00 & 5 &23.63  & 300    &120109 \cr
   10 &44 &15.80  & 00 & 5 &23.63  & 150    &120109 \cr
   10 &44 &15.80  & 00 & 6 & 5.63  & 300    &120109 \cr
   10 &44 &18.97  &$-$00 &18 &28.63  & 300    &120409 \cr
   10 &44 &21.36  &$-$00 &15 &33.63  & 300    &120408 \cr
   10 &44 &22.16  &$-$00 &22 &13.63  & 300    &120410 \cr
   13 &40 &53.32  & 00 & 3 &36.81  & 150    &130201 \cr
   13 &40 &55.12  &$-$00 & 4 & 2.19  & 150    &130202 \cr
   13 &40 &55.48  & 00 &12 &47.50  & 150    &130303 \cr
   13 &40 &55.48  & 00 &10 &12.50  & 150    &130301 \cr
   13 &40 &57.52  &$-$00 & 1 &25.19  & 150    &130204 \cr
   13 &40 &58.08  & 00 &10 & 3.50  & 150    &130304 \cr
   13 &40 &58.85  & 00 & 1 &37.81  & 150    &130203 \cr
   13 &41 & 1.69  & 00 & 5 &57.50  & 150    &130306 \cr
   13 &41 & 4.42  & 00 & 8 &34.50  & 150    &130305 \cr
   13 &41 & 5.88  & 00 &14 &18.50  & 150    &130302 \cr
   13 &41 & 6.52  &$-$00 & 3 & 2.19  & 150    &130205 \cr
   13 &41 & 8.59  &$-$00 & 3 &41.19  & 150    &130208 \cr
   13 &41 &10.25  &$-$00 & 0 & 8.19  & 150    &130207 \cr
   13 &41 &11.62  & 00 &11 &32.50  & 150    &130307 \cr
   13 &41 &13.39  &$-$00 & 5 &18.19  & 150    &130211 \cr
   13 &41 &13.92  & 00 & 3 &17.81  & 150    &130206 \cr
   13 &41 &22.92  & 00 & 2 & 7.81  & 150    &130209 \cr
   13 &41 &23.05  &$-$00 & 3 & 0.19  & 150    &130210 \cr
   13 &41 &26.05  & 00 & 1 &25.81  & 150    &130212 \cr
   13 &41 &27.72  &$-$00 & 3 & 2.19  & 150    &130213 \cr
   13 &41 &41.07  & 00 & 9 & 2.62  & 150    &130102 \cr
   13 &41 &44.27  & 00 & 8 &20.62  & 150    &130102 \cr
   13 &41 &45.81  & 00 & 7 &39.62  & 150    &130103 \cr
   13 &41 &48.27  & 00 & 4 &44.62  & 150    &130104 \cr
   13 &41 &48.61  & 00 & 2 & 8.62  & 150    &130105 \cr
   13 &41 &48.67  & 00 &10 &16.62  & 150    &130101 \cr
   13 &41 &49.01  & 00 & 8 &21.62  & 150    &130103 \cr
   13 &41 &51.47  & 00 & 4 & 2.62  & 150    &130104 \cr
   13 &41 &51.81  & 00 & 2 &50.62  & 150    &130105 \cr
   13 &41 &51.87  & 00 &10 &58.62  & 150    &130101 \cr
   13 &41 &54.87  & 00 & 4 &49.62  & 150    &130108 \cr
   13 &41 &56.61  & 00 & 2 &41.62  & 150    &130109 \cr
   13 &41 &56.94  & 00 & 7 &44.62  & 150    &130107 \cr
   13 &41 &57.07  & 00 &10 &52.62  & 150    &130106 \cr
   13 &41 &58.07  & 00 & 4 & 7.62  & 150    &130108 \cr
   13 &41 &59.81  & 00 & 3 &23.62  & 150    &130109 \cr
   13 &42 & 0.14  & 00 & 8 &26.62  & 150    &130107 \cr
   13 &42 & 0.27  & 00 &10 &10.62  & 150    &130106 \cr
   13 &42 & 0.54  & 00 & 6 & 6.62  & 150    &130112 \cr
   13 &42 & 3.61  & 00 & 7 &28.62  & 150    &130111 \cr
   13 &42 & 3.74  & 00 & 5 &24.62  & 150    &130112 \cr
   13 &42 & 3.81  & 00 &10 &29.62  & 150    &130110 \cr
   13 &42 & 6.81  & 00 & 8 &10.62  & 150    &130111 \cr
   13 &42 & 7.01  & 00 & 9 &47.62  & 150    &130110 \cr
   13 &42 & 7.81  & 00 & 1 &10.62  & 150    &130113 \cr
   13 &42 & 8.87  & 00 & 3 & 2.62  & 150    &130115 \cr
   13 &42 &10.21  & 00 & 9 &17.62  & 150    &130114 \cr
   13 &42 &11.01  & 00 & 1 &52.62  & 150    &130113 \cr
   13 &42 &12.07  & 00 & 3 &44.62  & 150    &130115 \cr
   13 &42 &13.41  & 00 & 8 &35.62  & 150    &130114 \cr
   22 &38 &35.33  & 00 &35 &55.29  & 300    &011309 \cr
   22 &38 &35.33  & 00 &35 &13.29  & 300    &011309 \cr
   22 &38 &36.33  & 00 &32 &57.29  & 300    &011304 \cr
   22 &38 &36.33  & 00 &33 &39.29  & 300    &011304 \cr
   22 &38 &37.00  & 00 &29 &20.00  & 150    &010612 \cr
   22 &38 &38.33  & 00 &31 &44.29  & 300    &011303 \cr
   22 &38 &38.33  & 00 &31 & 2.29  & 300    &011303 \cr
   22 &38 &38.53  & 00 &35 &13.29  & 300    &011309 \cr
   22 &38 &38.53  & 00 &35 &55.29  & 300    &011309 \cr
   22 &38 &38.80  & 00 &40 &41.29  & 300    &011308 \cr
   22 &38 &38.80  & 00 &41 &23.29  & 300    &011308 \cr
   22 &38 &39.14  & 00 &25 & 2.00  & 150    &010611 \cr
   22 &38 &39.53  & 00 &33 &39.29  & 300    &011304 \cr
   22 &38 &39.53  & 00 &32 &57.29  & 300    &011304 \cr
   22 &38 &40.53  & 00 &37 &12.29  & 300    &011306 \cr
   22 &38 &40.53  & 00 &37 &54.29  & 300    &011306 \cr
   22 &38 &40.71  & 00 &26 &28.48  & 300    &011606 \cr
   22 &38 &40.71  & 00 &27 &10.48  & 300    &011606 \cr
   22 &38 &41.53  & 00 &31 & 2.29  & 300    &011303 \cr
   22 &38 &41.53  & 00 &31 &44.29  & 300    &011303 \cr
   22 &38 &42.00  & 00 &41 &23.29  & 300    &011308 \cr
   22 &38 &42.00  & 00 &40 &41.29  & 300    &011308 \cr
   22 &38 &42.06  & 00 &33 &50.29  & 300    &011305 \cr
   22 &38 &42.06  & 00 &33 & 8.29  & 300    &011305 \cr
   22 &38 &42.98  & 00 &29 &33.48  & 300    &011607 \cr
   22 &38 &42.98  & 00 &28 &51.48  & 300    &011607 \cr
   22 &38 &43.60  & 00 &21 & 1.00  & 150    &010610 \cr
   22 &38 &43.73  & 00 &37 &54.29  & 300    &011306 \cr
   22 &38 &43.73  & 00 &37 &12.29  & 300    &011306 \cr
   22 &38 &43.91  & 00 &27 &10.48  & 300    &011606 \cr
   22 &38 &43.91  & 00 &26 &28.48  & 300    &011606 \cr
   22 &38 &44.19  & 00 &41 &16.29  & 300    &011307 \cr
   22 &38 &44.19  & 00 &40 &34.29  & 300    &011307 \cr
   22 &38 &45.26  & 00 &33 & 8.29  & 300    &011305 \cr
   22 &38 &45.26  & 00 &33 &50.29  & 300    &011305 \cr
   22 &38 &46.18  & 00 &28 &51.48  & 300    &011607 \cr
   22 &38 &46.18  & 00 &29 &33.48  & 300    &011607 \cr
   22 &38 &47.40  & 00 &40 &34.29  & 300    &011307 \cr
   22 &38 &47.40  & 00 &41 &16.29  & 300    &011307 \cr
   22 &38 &49.38  & 00 &26 &23.48  & 300    &011605 \cr
   22 &38 &49.38  & 00 &25 &41.48  & 300    &011605 \cr
   22 &38 &50.80  & 00 &28 &42.00  & 150    &010608 \cr
   22 &38 &51.60  & 00 &22 &55.00  & 150    &010607 \cr
   22 &38 &52.58  & 00 &25 &41.48  & 300    &011605 \cr
   22 &38 &52.58  & 00 &26 &23.48  & 300    &011605 \cr
   22 &38 &54.64  & 00 &24 & 6.48  & 300    &011602 \cr
   22 &38 &54.64  & 00 &24 &48.48  & 300    &011602 \cr
   22 &38 &55.67  & 00 &20 &59.00  & 150    &010604 \cr
   22 &38 &56.71  & 00 &26 &15.48  & 300    &011603 \cr
   22 &38 &56.71  & 00 &25 &33.48  & 300    &011603 \cr
   22 &38 &56.87  & 00 &31 & 5.00  & 150    &010609 \cr
   22 &38 &57.84  & 00 &24 &48.48  & 300    &011602 \cr
   22 &38 &57.84  & 00 &24 & 6.48  & 300    &011602 \cr
   22 &38 &57.93  & 00 &37 &17.29  & 300    &011302 \cr
   22 &38 &57.93  & 00 &37 &59.29  & 300    &011302 \cr
   22 &38 &59.38  & 00 &29 &44.48  & 300    &011604 \cr
   22 &38 &59.38  & 00 &30 &26.48  & 300    &011604 \cr
   22 &38 &59.91  & 00 &25 &33.48  & 300    &011603 \cr
   22 &38 &59.91  & 00 &26 &15.48  & 300    &011603 \cr
   22 &39 & 1.13  & 00 &37 &59.29  & 300    &011302 \cr
   22 &39 & 1.13  & 00 &37 &17.29  & 300    &011302 \cr
   22 &39 & 2.20  & 00 &22 &44.00  & 150    &010605 \cr
   22 &39 & 2.27  & 00 &26 &20.00  & 150    &010606 \cr
   22 &39 & 2.64  & 00 &30 &26.48  & 300    &011604 \cr
   22 &39 & 2.64  & 00 &29 &44.48  & 300    &011604 \cr
   22 &39 & 4.73  & 00 &35 &44.29  & 300    &011301 \cr
   22 &39 & 4.73  & 00 &35 & 2.29  & 300    &011301 \cr
   22 &39 & 6.25  & 00 &24 & 2.48  & 300    &011601 \cr
   22 &39 & 6.25  & 00 &23 &20.48  & 300    &011601 \cr
   22 &39 & 7.93  & 00 &35 & 2.29  & 300    &011301 \cr
   22 &39 & 7.93  & 00 &35 &44.29  & 300    &011301 \cr
   22 &39 & 9.45  & 00 &23 &20.48  & 300    &011601 \cr
   22 &39 & 9.45  & 00 &24 & 2.48  & 300    &011601 \cr
   22 &39 &12.47  & 00 &26 & 2.00  & 150    &010603 \cr
   22 &39 &12.67  & 00 &30 & 0.00  & 150    &010602 \cr
   22 &39 &15.76  & 00 &25 &50.59  & 300    &011508 \cr
   22 &39 &15.76  & 00 &26 &32.59  & 300    &011508 \cr
   22 &39 &17.34  & 00 &22 &57.00  & 150    &010601 \cr
   22 &39 &18.96  & 00 &26 &32.59  & 300    &011508 \cr
   22 &39 &18.96  & 00 &25 &50.59  & 300    &011508 \cr
   22 &39 &22.16  & 00 &22 &27.59  & 300    &011505 \cr
   22 &39 &22.16  & 00 &21 &45.59  & 300    &011505 \cr
   22 &39 &25.36  & 00 &21 &45.59  & 300    &011505 \cr
   22 &39 &25.36  & 00 &22 &27.59  & 300    &011505 \cr
   22 &39 &25.96  & 00 &25 & 6.59  & 300    &011507 \cr
   22 &39 &25.96  & 00 &24 &24.59  & 300    &011507 \cr
   22 &39 &29.15  & 00 &24 &24.59  & 300    &011507 \cr
   22 &39 &29.15  & 00 &25 & 6.59  & 300    &011507 \cr
   22 &39 &30.29  & 00 &22 & 2.59  & 300    &011506 \cr
   22 &39 &30.29  & 00 &22 &44.59  & 300    &011506 \cr
   22 &39 &32.09  & 00 &30 &40.59  & 300    &011504 \cr
   22 &39 &32.09  & 00 &31 &22.59  & 300    &011504 \cr
   22 &39 &33.29  & 00 &22 & 5.59  & 300    &011501 \cr
   22 &39 &33.29  & 00 &21 &23.59  & 300    &011501 \cr
   22 &39 &33.49  & 00 &22 &44.59  & 300    &011506 \cr
   22 &39 &33.49  & 00 &22 & 2.59  & 300    &011506 \cr
   22 &39 &35.29  & 00 &31 &22.59  & 300    &011504 \cr
   22 &39 &35.29  & 00 &30 &40.59  & 300    &011504 \cr
   22 &39 &35.96  & 00 &24 &39.59  & 300    &011502 \cr
   22 &39 &35.96  & 00 &25 &21.59  & 300    &011502 \cr
   22 &39 &36.49  & 00 &21 &23.59  & 300    &011501 \cr
   22 &39 &36.49  & 00 &22 & 5.59  & 300    &011501 \cr
   22 &39 &37.69  & 00 &28 &42.59  & 300    &011503 \cr
   22 &39 &37.69  & 00 &28 & 0.59  & 300    &011503 \cr
   22 &39 &39.16  & 00 &24 &39.59  & 300    &011502 \cr
   22 &39 &39.16  & 00 &25 &21.59  & 300    &011502 \cr
   22 &39 &40.89  & 00 &28 & 0.59  & 300    &011503 \cr
   22 &39 &40.89  & 00 &28 &42.59  & 300    &011503 \cr
   22 &39 &57.57  & 00 &20 &37.07  & 300    &011410 \cr
   22 &40 & 0.23  & 00 &24 &16.07  & 300    &011409 \cr
   22 &40 & 0.77  & 00 &21 &19.07  & 300    &011410 \cr
   22 &40 & 0.77  & 00 &20 &37.07  & 300    &011410 \cr
   22 &40 & 3.43  & 00 &23 &34.07  & 300    &011409 \cr
   22 &40 & 3.43  & 00 &24 &16.07  & 300    &011409 \cr
   22 &40 & 8.17  & 00 &29 & 2.07  & 300    &011407 \cr
   22 &40 & 8.17  & 00 &28 &20.07  & 300    &011407 \cr
   22 &40 &10.90  & 00 &21 &46.07  & 300    &011404 \cr
   22 &40 &10.90  & 00 &22 &28.07  & 300    &011404 \cr
   22 &40 &11.37  & 00 &28 &20.07  & 300    &011407 \cr
   22 &40 &13.30  & 00 &24 &15.07  & 300    &011405 \cr
   22 &40 &14.10  & 00 &22 &28.07  & 300    &011404 \cr
   22 &40 &15.31  & 00 &29 &37.07  & 300    &011408 \cr
   22 &40 &15.31  & 00 &30 &19.07  & 300    &011408 \cr
   22 &40 &16.03  & 00 &24 &30.07  & 300    &011406 \cr
   22 &40 &16.50  & 00 &23 &33.07  & 300    &011405 \cr
   22 &40 &16.50  & 00 &24 &15.07  & 300    &011405 \cr
   22 &40 &18.51  & 00 &30 &19.07  & 300    &011408 \cr
   22 &40 &19.03  & 00 &28 &32.07  & 300    &011403 \cr
   22 &40 &19.03  & 00 &27 &50.07  & 300    &011403 \cr
   22 &40 &19.23  & 00 &25 &12.07  & 300    &011406 \cr
   22 &40 &19.23  & 00 &24 &30.07  & 300    &011406 \cr
   22 &40 &22.23  & 00 &27 &50.07  & 300    &011403 \cr
   22 &40 &24.17  & 00 &20 &56.07  & 300    &011402 \cr
   22 &40 &26.90  & 00 &26 &58.07  & 300    &011401 \cr
   22 &40 &27.37  & 00 &21 &38.07  & 300    &011402 \cr
   22 &40 &27.37  & 00 &20 &56.07  & 300    &011402 \cr
   22 &40 &30.10  & 00 &26 &16.07  & 300    &011401 \cr
   22 &40 &30.10  & 00 &26 &58.07  & 300    &011401 \cr
}
\bye
\bye

\end